\documentclass[aps, preprint, nofootinbib,preprintnumbers,eqsecnum,superscriptaddress,sort]{revtex4}
\pdfoutput=1

\usepackage[
      colorlinks=true,
      linkcolor=blue,
      urlcolor=blue,
      filecolor=black,
      citecolor=red,
      pdfstartview=FitV,
      pdftitle={},
        pdfauthor={Oscar Dias, Gavin Hartnett, Jorge Santos},
        pdfsubject={},
        pdfkeywords={},
        pdfpagemode=None,
        bookmarksopen=true
      ]{hyperref}

\usepackage[normalem]{ulem}
\usepackage{amsmath}
\usepackage{enumerate}
\usepackage{amsfonts}
\usepackage{yfonts}

\usepackage{subfigure}
\usepackage{psfrag}

\usepackage{xcolor}
\usepackage{epsfig}
\usepackage[latin1]{inputenc}
\usepackage{float}
\usepackage{graphicx}
\usepackage{cancel}
\usepackage{mathrsfs}
\usepackage{amssymb}
\usepackage{amsfonts}
\usepackage{amsmath}
\usepackage{slashed}

\usepackage{graphicx}
\usepackage{bm}

\def\({\left(}
\def\){\right)}
\def\[{\left[}
\def\]{\right]}
\def\<{\langle}
\def\>{\rangle}
\newcommand{\be}{\begin{equation}}
\newcommand{\ee}{\end{equation}}
\newcommand{\bea}{\begin{eqnarray}}
\newcommand{\eea}{\end{eqnarray}}
\newcommand{\bwt}{\begin{widetext}}
\newcommand{\ewt}{\end{widetext}}

\newcommand{\bi}{\begin{itemize}}
\newcommand{\ei}{\end{itemize}}
\newcommand{\ben}{\begin{enumerate}}
\newcommand{\een}{\end{enumerate}}
\newcommand{\bca}{\begin{cases}}
\newcommand{\eca}{\end{cases}}
\newcommand{\bln}{\begin{align}}
\newcommand{\eln}{\end{align}}
\newcommand{\bst}{\begin{split}}
\newcommand{\est}{\end{split}}

\begin{document}

\title {Quasinormal modes of asymptotically flat rotating black holes}

\author{\'Oscar J. C. Dias}
\email{oscar.dias@ist.utl.pt}
\affiliation{Center for Mathematical Analysis, Geometry, \& Dynamical Systems, \\
Departamento de Matem\'atica and LARSyS, Instituto Superior T\'ecnico, 1049-001 Lisboa, Portugal}
\affiliation{STAG research centre and Mathematical Sciences, University of Southampton, UK}
\affiliation{Institut de Physique Theorique, CEA Saclay, CNRS URA 2306, F-91191 Gif-sur-Yvette, France}

\author{Gavin S. Hartnett}
\email{hartnett@physics.ucsb.edu}
\affiliation{Department of Physics, UCSB, Santa Barbara, CA 93106, USA}

\author{Jorge E. Santos}
\email{jss55@stanford.edu}
\affiliation{Department of Physics, Stanford University, Stanford, CA 94305-4060, U.S.A.}
\affiliation{Department of Applied Mathematics and Theoretical Physics, University of Cambridge, Wilberforce Road, Cambridge CB3 0WA, UK \\ \vspace{1 cm}\,}

\begin{abstract}
We study the main properties of general linear perturbations of  rotating black holes in asymptotically flat higher-dimensional spacetimes. In particular, we determine the  quasinormal mode (QNM) spectrum of  singly spinning and equal angular momenta Myers-Perry black holes (MP BHs). Emphasis is also given to the timescale of the ultraspinning and bar-mode instabilities in these two families of MP BHs. For the bar-mode instabilities in the singly spinning MP BH, we find excellent agreement with our linear analysis and the non-linear time evolution of Shibata and Yoshino for $d=6,7$ spacetime dimensions. We find that $d=5$ singly spinning BHs are linearly stable. In the context of studying general relativity in the large dimension limit, we obtain the QNM spectrum of Schwarzschild BHs and rotating MP BHs for large dimensions. We identify two classes of modes. For large dimensions, we find that in the limit of zero rotation, unstable modes of the MP BHs connect to a class  of Schwarzschild QNMs that saturate to finite values.
\end{abstract}

\today

\maketitle

\tableofcontents

%\newpage

%%%%%%%%%%%%%%%%%%%%%%%%%%%%%%
%%%%%%%%%%%%%%%%%%%%%%%%%%%%%%
\section{Introduction\label{sec:Intro}}
%%%%%%%%%%%%%%
The understanding of asymptotically flat higher-dimensional black holes (BHs) in classical general relativity has seen remarkable progress in recent years (see \cite{2012bhhd.book} for an excellent review on the subject). We have learned that BHs in $d\geq 5$ spacetime dimensions have markedly different properties from their four-dimensional counterparts: several asymptotically flat higher-dimensional BHs exist for the same set of asymptotic charges \cite{PhysRevLett.88.101101}, horizons can have distinct topologies \cite{PhysRevLett.88.101101,Emparan:2009at}, solutions with disconnected horizons exist \cite{Elvang:2007rd}, and cosmic censorship does not hold in $d\geq 5$\footnote{Strictly speaking, a black string is not asymptotically flat. To date, there is no proof that a violation of cosmic censorship occurs for asymptotically flat BHs in $d\geq 5$, even though very thin black rings provide an excellent candidate for such a phenomenon.} \cite{Lehner:2010pn}, to name but a few. Given the plethora of higher-dimensional BHs and the intricate nature of the phase space of BH solutions, it is interesting to study their stability. This paper addresses this question for a specific class of higher-dimensional BHs.

The most general exact solution known in closed form for an arbitrary number of dimensions is the Myers-Perry BH (MP BH) \cite{Myers:1986un}. In many ways, this is the natural generalization of the Kerr solution to higher dimensions. The spatial cross section of the horizon has topology $S^{d-2}$, and it is uniquely characterized by its mass and $\lfloor (d-1)/2\rfloor$ angular momenta. The properties of these solutions strongly depend on the spacetime dimension, as well as the angular momenta. In $d\geq 6$, if all angular momenta are non-vanishing, there is an extremal limit, meaning that all angular momenta are bounded above. If, on the other hand, at least one of the angular momenta vanishes, there is no upper bound on any of the remaining angular momenta and MP BHs are allowed to rotate arbitrarily fast. This fact was picked out by Emparan and Myers in \cite{Emparan:2003sy}, who conjectured rapidly spinning MP BHs to be unstable. When $\tilde{n}$ of the angular momenta are taken to be arbitrarily large, the horizon ``pancakes'' out near the poles, and acquires an almost exact $\mathbb{R}^{2\tilde{n}}\times S^{d-2\tilde{n}-2}$ topology, \emph{i.e.} they look like black branes. Since black branes are known to be unstable \cite{Gregory:1993vy,Lehner:2010pn}, they conjectured these highly deformed BHs to be unstable as well. The argument given above, strictly speaking, only works if the rotation parameters are taken to be infinitely large. However, Emparan and Myers went beyond this and estimated the rotation for which the instability should appear for singly spinning BHs. In order to do this, they used two distinct arguments that each give different estimates. In the first argument, the critical rotation was estimated by the condition that thermodynamic quantities, such as the temperature, behave similarly to those of a black brane. The other argument directly compares the entropy of a singly spinning BH to that of two disjoint BHs with the same total energy and angular momentum. The latter process necessarily breaks the rotation symmetry of the original BH. The critical rotation extracted from the former argument is systematically larger than that of the latter. Moreover, the first argument only applies to $d\geq6$, whereas the second gives a positive result for $d\geq5$. The first type of instability was coined the ultraspinning instability, and the second, rotational symmetry-breaking instability was dubbed the $m-$bar or bar-mode instability.

It took six years to know whether this conjecture was correct, and to detail its properties. In \cite{Dias:2009iu,Dias:2010maa} it was shown that singly spinning MP BHs in $d\geq 6$ are unstable to perturbations that do not break rotational symmetry, and in \cite{Shibata:2009ad,Shibata:2010wz} it was found that, at the full non-linear level, MP BHs are also unstable to perturbations that do break rotational symmetry, but this time for $d\geq 5$. The rotation required to herald the second type of instability was found to be much smaller than the first, in agreement with the conjecture of  \cite{Emparan:2003sy}. None of these results shed any light on what happens in the general case, where all angular momenta are non-vanishing. The arguments given by Emparan and Myers are also much weaker in this case, because there is an upper bound on the angular momenta and the horizons cannot get very distorted. The reason why singly spinning BHs were easier to study is related to the fact that they preserve a much larger symmetry group than the general case. There is however, one case that is amenable to a more systematic study, and that has an upper bound on the rotation. In odd spacetime dimensions, the equal angular momenta (EAM) MP solution exhibits a large symmetry group that can be employed to reduce the study of general linearized perturbations to a system of coupled ordinary differential equations. This was used in \cite{Dias:2010eu} to show that equal angular momenta MP solutions in odd $d\geq 7$ have ultraspinning instabilities \footnote{\label{foot:ultra}This instability was first found in $d \geq 9$ \cite{Dias:2010eu}. However, upon further investigation requested by the authors of Ref. \cite{Durkee:2010qu}, the instability was found to also exist in $d = 7$.}, very much like the singly spinning case. It was further shown in \cite{Dias:2011jg}, that the ultraspinning instability found in the equal angular momenta MP is connected to the one found in the singly spinning MP. More recently, in \cite{Hartnett:2013fba}, the $m-$bar instability was also shown to exist in equal angular momenta MP BHs.

However, to date, the study of quasinormal modes (QNMs) of higher-dimensional rotating BHs in asymptotically flat space is very meager. The main reason for this is the absence of a master equation that governs how generic gravitational perturbations propagate on such backgrounds. For the Kerr geometry such equation exists since Teukolsy's seminal paper \cite{Teukolsky:1973ha}. Partial progress was made for the $d=5$ equal angular momenta MP, where a clever decomposition was used \cite{Murata:2008yx}. Moreover, certain simple gravitational perturbations of higher-dimensional BHs have been studied \cite{Kodama:2009bf,Kunduri:2006qa}, but only for very special perturbation sectors where no instabilities were found. Another approach, pursued in \cite{Durkee:2010ea}, focused the study of near horizon geometries and their stability properties. The idea is simple: the authors prove that, for certain matter fields, and for axisymmetric perturbations, an instability of the near horizon geometry implies an instability of the full extremal geometry. However, it is much easier to study the stability of near horizon geometries, which often exhibit a large isometry group that reduces the problem to an algebraic calculation. Nevertheless, none of the studies described above gave a complete description of the behavior of QNMs in higher-dimensional rotating backgrounds, since their main goal was to study stability of such geometries. This article is entirely devoted to complete that gap, and presents a first exhaustive study of QNMs of the most  representative cases of higher-dimensional asymptotically flat rotating BHs.  Furthermore, the instability found in \cite{Shibata:2009ad,Shibata:2010wz} has never been reproduced with a linear calculation, which we attempt in this manuscript. 

Although in this paper we are mostly interested in the QNMs of rotating BHs, we begin by reviewing QNMs of the Schwarzschild-Tangherlini BH (henceforth referred to as simply Schwarzschild). There are two motivations for doing so. First, the rotating BHs considered in later sections reduce to Schwarzschild in the zero rotation limit, and in this limit the QNMs of these rotating BHs should reduce to those of Schwarzschild. A second motivation for studying Schwarzschild is the recent interest in studying general relativity in the large-$d$ limit \cite{Emparan:2014cia,Emparan:2013xia,Emparan:2013moa}. This is a very interesting limit to consider. The equations simplify drastically, and yet enough structure is preserved such that the physics does not become trivial. For example, the physical picture associated with the infinite-$d$ limit is that the space outside a BH is completely flat, BHs have zero cross-section, and can be modelled as dust. From this picture one might conclude that the infinite-$d$ limit is too strong to be useful, and that it has erased all the interesting features of BHs. However, this is certainly not the case, and our analysis of Schwarzschild QNMs and their connection with instabilities of rotating BHs adds to the interesting results of the infinite-$d$ limit. 

We conclude this section with a brief outline of the paper. In Section \ref{sec:IntroMP} we introduce the general metric of MP BHs, showing that under appropriate limits it reduces to the Schwarzschild, singly spinning MP and equal angular momenta MP BHs. In Section \ref{sec:Schw} we briefly review the Kodama-Ishibashi formalism, and present the QNMs of the Schwarzschild BH as a function of the spacetime dimension $d$. Section \ref{sec:MPEAM} studies generic gravitational perturbations, including scalar, vector and tensor-type perturbations, of the equal angular momenta MP BH and in Section \ref{sec:IntrossMP} we present gravitational QNMs of the singly spinning MP BH. We then conclude in Section \ref{sec:Conc}.

{\it Note added.} As this work was nearing completion, we learned from Emparan, Suzuki, and Tanabe of their recent work \cite{Emparan:unpub}  studying the (in)stability of odd-dimensional rotating black holes with equal angular momenta in the large dimension limit. When a comparison is possible, our numerical results agree with the analytical findings of \cite{Emparan:unpub}.

%%%%%%%%%%%%%%%%%%%%%%%%%%%%%%
%%%%%%%%%%%%%%%%%%%%%%%%%%%%%%
\section{General Myers-Perry black holes \label{sec:IntroMP}}
%%%%%%%%%%%%%%

%%%%%%%%%%%%%%%%%%%%%%%%%%%%%%
\subsection{Most general Myers-Perry family \label{sec:IntroMP0}}
%%%%%%%%%%%%%%

In this section we present the general MP solution \cite{Myers:1986un}, for general $d\geq 4$, and as a function of their mass $M$ and $\lfloor(d-1)/2\rfloor$ rotation parameters $\{a_i\}$. Its line element, in Boyer-Lindquist coordinates, takes the following form
\begin{multline}
ds^2 = -dt^2+\frac{r_0^{d-3}}{U(r,\mu_{1},\ldots,\mu_{\tilde{N}})}\left(dt-\sum_{i=1}^{\tilde{N}}a_i \mu_i^2\,d\varphi_i\right)^2+\\
\sum_{i=1}^{\tilde{N}}\mu_i^2(r^2+a_i^2)d\varphi_i^2+\frac{U(r,\mu_{1},\ldots,\mu_{\tilde{N}})\,dr^2}{F(r)-r_0^{d-3}}+\sum_{i=1}^{\tilde{N}+\varepsilon}(r^2+a_i^2)d\mu_i^2\,
\label{eq:MPgeneral}
\end{multline}
where
\begin{align}
U(r,\mu_{1},\ldots,\mu_{\tilde{N}})=r^{\varepsilon}\sum_{i=1}^{\tilde{N}+\epsilon}\frac{\mu_i^2}{r^2+a_i^2}\prod_{j=1}^{\tilde{N}}(r^2+a_j^2)\,,\quad\text{and}\quad F(r)=r^{\varepsilon-2}\prod_{j=1}^{\tilde{N}}(r^2+a_j^2)\,.
\end{align}
We have also defined $d = 2\tilde{N}+1+\varepsilon$ (where $\varepsilon = 1$ for even $d$, and $\varepsilon = 0$ otherwise) and for even $d$, $a_{N+1}=0$. Finally, the coordinates $\mu_i$ are not independent, and must satisfy
\begin{equation}
\sum_{i=1}^{\tilde{N}+\epsilon}\mu_i^2=1\,.
\end{equation}

The energy and angular momenta of this solution can be readily computed using Komar integrals, and yield:
\begin{equation}
M = \frac{(d-2)\,\omega_{d-2}}{16\pi G_d}\,r_0^{d-3}\,,\quad\text{and}\quad J_i=\frac{2}{d-2}M\,a_i\,,
\end{equation}
where $\omega_{d-2}$ is the area of a $(d-2)$ round sphere with unit radius and $G_d$ is the $d-$dimensional Newton's constant. The event horizon, located at $r=r_+$, is defined as the largest real root of $F(r)-r_0^{d-3}$.

Finally, for the sake of completeness, we also present the remaining thermodynamic quantities, such as the entropy $S_H$, Hawking temperature $T$ and angular velocities $\Omega_i$:
\begin{align}
S_H = \frac{\omega_{d-2}}{4\,r_+^{1-\varepsilon}\,G_d}\prod_{i=1}^{\tilde{N}}(r_+^2+a_i^2)\,,\quad T = \frac{r_+}{2\pi}\left[\sum_{i=1}^{\tilde{N}}\frac{1}{r_+^2+a_i^2}-\frac{1}{(1+\varepsilon)r_+^2}\right]\,\quad\text{and}\quad \Omega_i = \frac{a_i}{r_+^2+a_i^2}\,.
\label{eq:thermojj}
\end{align}

It is clear from the line element (\ref{eq:MPgeneral}) that, unless $d=4$ or $d=5$, studying a generic gravitational perturbation of a general MP solution in a given dimension, will necessarily involve solving PDEs in more than two variables. This makes the problem quite difficult to study. For this reason, we focus on particular configurations of the angular momenta parameters that result in solutions with enhanced symmetry, which we describe next.

%%%%%%%%%%%%%%%%%%%%%%%%%%%%%%
%%%%%%%%%%%%%%%%%%%%%%%%%%%%%%
\subsection{Schwarzschild, singly spinning and equal angular momenta MP black holes\label{subsec:Intro}}
%%%%%%%%%%%%%%
The simplest solution corresponds to the case where all of the angular momenta are set to zero. This is the Schwarzschild-Tangherlini solution first presented in \cite{Tangherlini:1963bw}, and whose line element is given by
\begin{equation}
ds^2 = -f(r)dt^2+\frac{dr^2}{f(r)} +r^2d\Omega^2_{d-2}, \qquad f(r) =   1-\frac{r_0^{d-3}}{r^{d-3}},
\label{eq:tangherlini}
\end{equation}
where $d\Omega^2_n$ is the line element of an $n-$dimensional unit round sphere. The spatial symmetry group of Schwarzschild is enhanced to $SO(d-1)$, which we will use in Section \ref{sec:Schw}.

Another simplifying limit occurs if we set all but one angular momenta to zero. This BH is often denominated singly spinning MP BH. In this case, the line element (\ref{eq:MPgeneral}) reduces to:
\begin{eqnarray}
\label{SMP:ds2orig}
&& ds^2=\frac{\left(1-\widetilde{x}^2\right)}{\Sigma }\,\left[d\phi \left(a^2+r^2\right)-a dt\right]^2 
-\frac{\Delta}{\Sigma }\,\left[dt-a d\phi \left(1-\widetilde{x}^2\right)\right]^2\nonumber\\
&& \hspace{1cm} +\frac{\Sigma }{\Delta }\,dr^2 +\frac{ \Sigma }{1-\widetilde{x}^2}\,d\widetilde{x}^2+r^2 \widetilde{x}^2\,d\Omega_{(d-4)}^2\,,
\end{eqnarray} 
where
\begin{equation}
\label{SMP:ds2origfunc}
\Delta=  r^2+a^2-\frac{r_0^{d-3}}{r^{d-5}}\,, \qquad   \Sigma= r^2+a^2  \widetilde{x}^2 \,\qquad\text{and}\qquad \widetilde{x}=\cos\theta \in(0,1)\,.
\end{equation} 
The symmetry group is now enhanced to $SO(d-3)$, which means that generic perturbations will reduce to a set of coupled PDEs in two variables, $r,\widetilde{x}$ (after Fourier mode decomposition both in time and rotational angle $\phi$). This line element, in many ways, is the simplest generalization of the Kerr BH to higher dimensions. Note that in this case, the temperature defined in Eq.~(\ref{eq:thermojj}) cannot be made zero for $d\geq6$, and as such no upper bound on the rotation exists. For the special case of $d=5$, there is an upper bound on the rotation, corresponding to $a = r_0$, but this corresponds to a naked singularity, i.e. $r_+ = 0$. Despite the many efforts over the past years, no decoupled equation describing how gravitational perturbations propagate on the singly spinning MP BH has been found, see for instance \cite{Godazgar:2011sn} for a recent effort in this direction. We shall proceed in Section \ref{sec:IntrossMP} by considering the full PDE problem.

A even more drastic simplification, which is less obvious, occurs when all the angular momenta are equal, i.e. $a_i = a$, and when $d$ is odd, as was first noted in \cite{Gibbons:2004uw}. In this case the general line element (\ref{eq:MPgeneral}), reduces to
\begin{equation}
ds^2 = -\frac{p(\hat{r})}{h(\hat{r})} dt^2 +\frac{d\hat{r}^2}{p(\hat{r})} + \hat{r}^2\left[h(\hat{r})^2\left(d\psi + A_a dx^a - u(\hat{r}) dt\right)^2 + \hat{g}_{ab}dx^a dx^b\right]
\label{eq:EMP}
\end{equation}
where $d=2N+3$, and the metric functions are defined as follows:
\begin{equation}
p(\hat{r}) = 1 - \frac{r_0^{2N}}{\hat{r}^{2N}} + \frac{r_0^{2N}a^2}{\hat{r}^{2(N+1)}}, \quad u(\hat{r}) = \frac{r_0^{2N}a}{\hat{r}^{2(N+1)}h(\hat{r})} \quad \text{and}\quad h(\hat{r}) = 1+\frac{r_0^{2N} a^2}{\hat{r}^{2(N+1)}}\,.
\end{equation}
Here $\hat{g}_{ab}$ is the Fubini-Study metric on $\mathbb{CP}^N$ and $A$ is related to its K\"{a}hler form by $J=dA/2$.  Some comments concerning this line element are in order. First, we note that the radial coordinate $\hat{r}$ is related to the general Boyer-Lindquist coordinates coordinate as $\hat{r}^2 = r^2+a^2$. Second, this line element has a much larger isometry group than the singly spinning MP solution, namely $\mathbb{R}\times U(1) \times SU(N+1)$. Finally, in passing from Eq.~(\ref{eq:MPgeneral}) to Eq.~(\ref{eq:EMP}) we have used the fact that any round $S^{2N+1}$ sphere can be written as a Hopt fibration over $\mathbb{CP}^N$, i.e.
\begin{equation}
d\Omega_{2N+1}^2 = (d\psi + A_a dx^a)^2 + \hat{g}_{ab}dx^a dx^b\,.
\end{equation}
These equal angular momenta BHs cannot rotate arbitrarily fast, and in fact have an extremal bound, \begin{equation} 
a_{\text{ext}} = \sqrt{ \frac{N}{N+1}} r_+. 
\end{equation}
A remarkable property about the line element (\ref{eq:EMP}) is that it is cohomogeneity-one, which is to say that it only depends non trivially on one coordinate, namely $\hat{r}$. Its large symmetry group will allow us to study how generic gravitational perturbations propagate on such background, by studying a system of coupled ODEs. This procedure was first used in \cite{Dias:2010eu},\cite{Kunduri:2006qa}, and shall be reviewed in Section \ref{sec:MPEAM}.

%%%%%%%%%%%%%%%%%%%%%%%%%%%%%%
%%%%%%%%%%%%%%%%%%%%%%%%%%%%%%
\section{Schwarzschild black holes
\label{sec:Schw}} \noindent
%%%%%%%%%%%%%%%%%%%%%%%%%%%%%%
%%%%%%%%%%%%%%%%%%%%%%%%%%%%%%

In this section we review gravitational perturbations of Schwarzschild and study the QNM spectrum in the limit of large dimensions. 

%%%%%%%%%%%%%%%%%%%%%%%%%%%%%%
\subsection{Review of the Kodama-Ishibashi formalism} \noindent
%%%%%%%%%%%%%%%%%%%%%%%%%%%%%%

Here we briefly review the Kodama-Ishibashi  (KI) master variable formalism \cite{Kodama:2003jz}, which has proven to be  an invaluable tool for the study of the linear stability of BH spacetimes. The KI  formalism exists for  spacetimes for which the line element can be written as
\be ds^2 = h_{AB}(y)dy^A dy^B + r^2(y) \hat{g}_{ab} dx^a dx^b. \ee
Here $h_{AB}$ is the Lorentzian metric of a two-dimensional orbit spacetime, and $\hat{g}_{ab}$ is the metric for a $n=d-2$ dimensional Euclidean signature space which (in our study) is restricted to be maximally symmetric with constant sectional curvature, normalized to be $0$ or $\pm 1$. In what follows, we will restrict ourselves to Schwarzschild BHs, for which the line element is given by \eqref{eq:tangherlini}.
%\be ds^2 = -f(r) dt^2 + \frac{dr^2}{f(r)} + r^2 d\Omega_n^2, \qquad f(r) = 1 - \Big(\frac{r_0}{r}\Big)^{D-3}, \ee

We wish to consider linearized gravitational perturbations of this spacetime. Although a general metric perturbation will depend on all the coordinates, the angular and time dependence can be separated out using spherical tensor harmonics on $S^{d-2}$ and complex exponentials of the form $e^{-i\omega t}$. The linearized Einstein equations then become a system of ordinary differential equations. Three types of harmonic tensors will be needed to construct the most generic perturbation: scalar, vector, and tensor. The scalar harmonics $S$ are both the most familiar and the simplest, and satisfy
\be (\hat{\nabla}^2 + \lambda_S) S = 0, \ee
where the requirement of regularity quantizes the eigenvalue as $\lambda_S = \widetilde{\ell}_S(\widetilde{\ell}_S+n-1)$, $\widetilde{\ell}_S = 0, 1, 2, ..$, and $\hat{\nabla}_a$ is the covariant derivative on the $n$-sphere. Less familiar are vector $V$ and tensor $T$ harmonics which satisfy similar equations:
\be (\hat{\nabla}^2 + \lambda_V) V_a = 0, \qquad \hat{\nabla}^a V_a = 0 \,; \ee
\be (\hat{\nabla}^2 + \lambda_T)T_{ab} = 0, \qquad \hat{\nabla}^a T_{ab}= 0, \qquad \hat{g}^{ab} T_{ab} = 0, \qquad T_{ab} = T_{(ab)}. \ee
The eigenvalues of these regular harmonics are quantized as $\lambda_V = \widetilde{\ell}_V(\widetilde{\ell}_V+n-1)-1$, $\widetilde{\ell}_V = 1,2,..$, and $\lambda_T = \widetilde{\ell}_T(\widetilde{\ell}_T+n-1)-2$, $\widetilde{\ell}_T = 1,2,..$ \footnote{\label{foot:L01}
Gravitational scalar perturbations with $\widetilde{\ell}_S = 0,1$ or vector perturbations with $\widetilde{\ell}_V = 1$ do not represent local degrees of freedom. Scalar perturbations with $\widetilde{\ell}_S = 0$ or vector perturbations with $\widetilde{\ell}_V = 1$ describe simply a variation of the mass and angular momentum parameters of the solution, respectively, and $\widetilde{\ell}_S=1$ corresponds to a pure gauge mode \cite{Kodama:2003jz,Dias:2013hn}. We do not consider these special modes further.
}. In terms of these harmonics, the most general metric perturbation can be constructed from superpositions of scalar, vector, and tensor perturbations, which take the form
\be
\delta g_{AB} = f_{AB}^{(S)} S, \quad \delta g_{Aa} = r f_A^{(S)} S_a, \quad \delta g_{ab} = r^2 (H_T^{(S)} S_{ab} + H_L \hat{g}_{ab}S) \qquad \text{(scalar)},
\ee
\be
\delta  g_{AB} =0,\quad  \delta g_{Aa} = r f_A^{(V)} V_a, \quad \delta g_{ab} = r^2  H_T^{(V)} V_{ab} \qquad \text{(vector)},
\ee
\be
\delta g_{AB} =0,\quad  \delta g_{Aa} =0,\quad \delta g_{ab} = r^2 H_T^{(T)} T_{ab} \qquad \text{(tensor)}.
\ee
In the above, $S_a, S_{ab}$,  and $V_{ab}$ are derived harmonics which can be written as derivatives of the more fundamental harmonic. For example, $S_a = - \lambda^{-1/2}_S \hat{\nabla}_a S $ is a scalar-derived vector harmonic. We refer the reader to the original paper for more details. The functions $f_{AB}, f_A$ are functions of the orbit coordinates $(t,r)$ only, and the time dependence can be trivially separated into Fourier modes, as in $ f_{AB} \propto e^{-i\omega t}$, as mentioned above.

With this parametrization, the linearized equations become a system of ODE's. At this point the gauge redundancy has not been taken into account, and in fact the above decomposition is not gauge invariant; under a linearized gauge transformation, many of the above functions will shift. Kodama and Ishibashi were able to combine the perturbation functions for each of the sectors into a single function called the master variable which is a gauge invariant quantity \footnote{The tensor sector is trivially put into master variable form, since there is only one perturbation function and it is already gauge invariant.}. A differential map acting on this master variable reconstructs the metric perturbation in a given gauge. The remarkable end result is that one can study gravitational perturbations of Schwarzschild by simply solving a Schr\"{o}dinger equation for each sector. The master equations are of the form:
\be - f \partial_r (f \partial_r \phi_I) + (V_I -\omega^2)\phi_I = 0,  \ee
where $I$ is an index that runs over scalar, vector, or tensor perturbations and $\phi_I$ is the master variable. The expressions for the potentials $V_I$ are rather lengthy and we will not reproduce them here; they can be found in the original Kodama Ishibashi paper \cite{Kodama:2003jz}. QNMs are then solutions to these equations with appropriate boundary conditions: the perturbations should be ingoing at the horizon and outgoing at infinity.

%%%%%%%%%%%%%%%%%%%%%%%%%%%%%%
\subsection{Gravitational QNMs of Schwarzschild in higher dimensions} \noindent
%%%%%%%%%%%%%%%%%%%%%%%%%%%%%%

The QNM  spectrum of the four dimensional Schwarzschild BH has been well understood for many years, and there are also many studies of the spectrum in higher dimensions. For a detailed review, see \cite{Berti:2009kk}. Here, we are interested in the the spectrum not at any one particular value of $d$, but as a function of $d$. In particular, we are interested in the behaviour of the frequencies as $d$ tends toward infinity. Although Schwarzschild BHs exist for integer $d \ge 4$, we will find it useful to consider $d$ a \emph{continuous parameter} and to study the spectrum as $d$ increases. The Kodama-Ishibashi (KI) equations were derived analytically for arbitrary integer $d$, and we will simply use these equations but allow $d$ to vary continuously. Here a comment is in order. Considered as a function of $d$, the potentials are not analytic, and one should use care in considering the QNM frequencies as continuous functions of $d$. Our main motivation for considering non-integer $d$ is twofold: it will make clearer the pattern of the QNM behavior as $d$ is increased, and it will allow us to use a certain powerful numerical method which utilizes a continuous parameter. Of course, when $d$ is an integer, we will check that our results for the continuous-$d$ code agree with the results of the integer-$d$ code.

As mentioned above, the KI  formalism allows the complicated equations for gravitational perturbations to be reduced to simple Schr\"{o}dinger-type ODE's. A simple Frobenius analysis yields the following behaviour near the two boundaries \footnote{The case $d=4$ is special and has a slightly different fall-off near infinity.} (here we suppress the $I$-index),
\begin{eqnarray}
\phi =  && e^{i \omega r_{\star}} \phi^{\text{out}}_{\infty} + e^{-i \omega r_{\star}} \phi^{\text{in}}_{\infty} \qquad \text{(infinity)}, \\
\phi = && e^{i\omega r_{\star}} \phi^{\text{out}}_{\text{hor}} + e^{-i\omega r_{\star}} \phi^{\text{in}}_{\text{hor}} \qquad \text{(horizon)}, \end{eqnarray}  
where $r_{\star} = \int dr/f$ is the usual tortoise coordinate, and the functions $\phi^{\text{in/out}}_{\text{hor/} \infty}$ are regular and non-zero at the relevant boundary. The boundary conditions appropriate for QNMs are such that the perturbation is ingoing at the BH horizon and outgoing at infinity. These conditions can be easily formulated in Eddington-Finkelstein coordinates, and they amount to setting $\phi_{\infty}^{\text{in}} = \phi_{\text{hor}}^{\text{out}} = 0$. 
It is useful to define a new function $ \widetilde{\phi}$ with this asymptotic behaviour stripped off, 
\be \phi = \Big(1-\frac{r_0}{r}\Big)^{-\frac{i\omega r_0}{d-3}} e^{i\omega r} \widetilde{\phi}, \ee
which is then regular and finite at either boundary when the above conditions are imposed. 

\begin{figure}[h!]
\centering
\includegraphics[width=1.0 \textwidth]{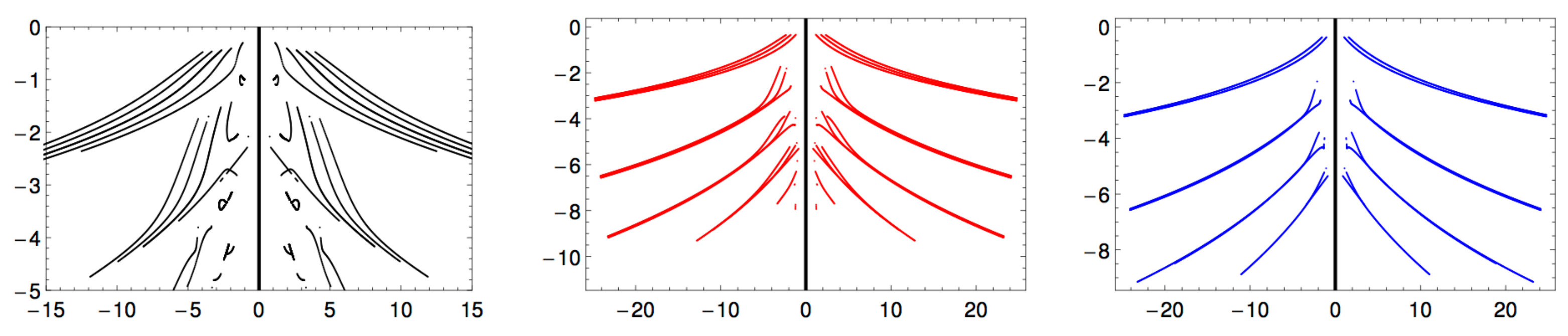}
\caption{\textbf{Schwarzschild}. The complex QNM frequencies for scalar ({\it left panel}), vector ({\it middle panel}), and tensor ({\it right panel}) perturbations. In these plots, the dimension ranges from $d=6$ to $d=100$. For the scalar plot, $\widetilde{\ell}_S = 2,3,4,5,6$ modes are displayed, while for vectors $\widetilde{\ell}_V = 2,3,4$, and for tensors $\widetilde{\ell}_T = 1,2$.}
\label{Fig:schw_all}
\end{figure}
\begin{figure}[h!]
\centering
\includegraphics[width=1.0 \textwidth]{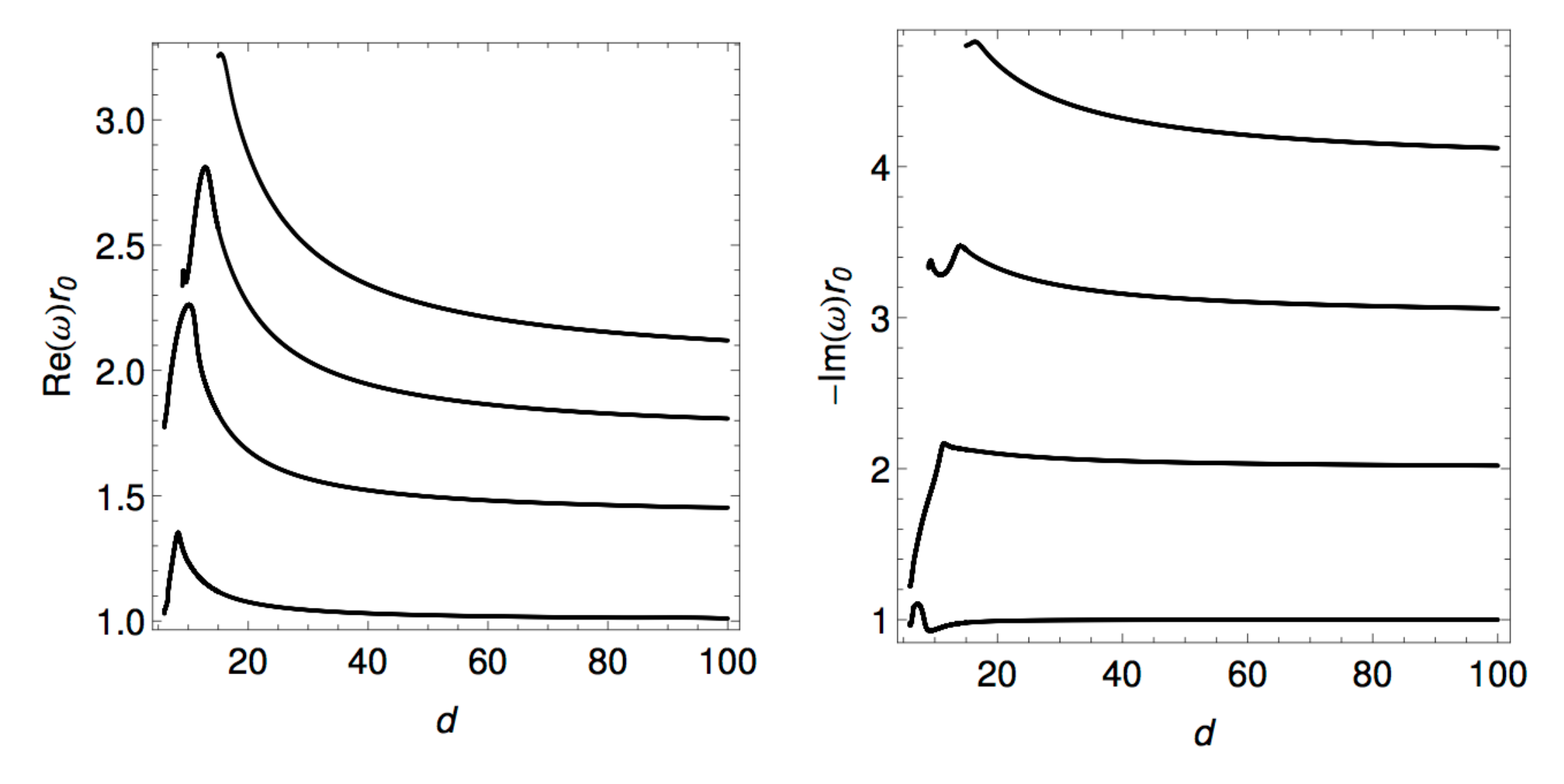}
\caption{\textbf{Schwarzschild}. The real and imaginary parts of the saturating scalar QNM's for $\widetilde{\ell}_S = 2,3,4,5$. Higher $\widetilde{\ell}_S$ curves lie above lower $\ell_S$ curves.}
\label{Fig:saturating_scalars}
\end{figure}

We then solved these equations numerically for the new $\widetilde{\phi}$ variables using the following scheme: introduce a compactified radial coordinate $r = r_0/z$, where $z \in [0,1]$, with $z=0$ corresponding to infinity, and $z=1$ corresponding to the horizon. This interval is then discretized using a Chebyshev grid. The QNMs were then solved for using one of two methods. The first relies on converting the equations into an eigenvalue problem for the frequencies $\omega$, which can then be solved using \emph{Mathematica}'s built-in routine \emph{Eigensystem}. More details of this method and the discretization scheme can be found in \cite{Dias:2010eu}. The second method is based on an application of the Newton-Raphson root-finding algorithm, and is detailed in \cite{Cardoso:2013pza}. The strength of the first method is that it gives many QNMs simultaneously, allowing for easy determination of the spectra. The second method can only be used to compute a single mode at a time, and only when a seed is known that is sufficiently close to the true answer. However, this method is much quicker as both the size of the grid and numerical precision increases, and can be used to push the numerics to extreme regions of the parameter space. For example, using the Newton-Raphson method, dimensions as large as $d=100$ were attainable, which is quite remarkable considering the steepness of the warp function $f(r) = 1 - (r_0/r)^{d-3}$ near the horizon.

\begin{figure}[th!]
\centering
\includegraphics[width=0.9 \textwidth]{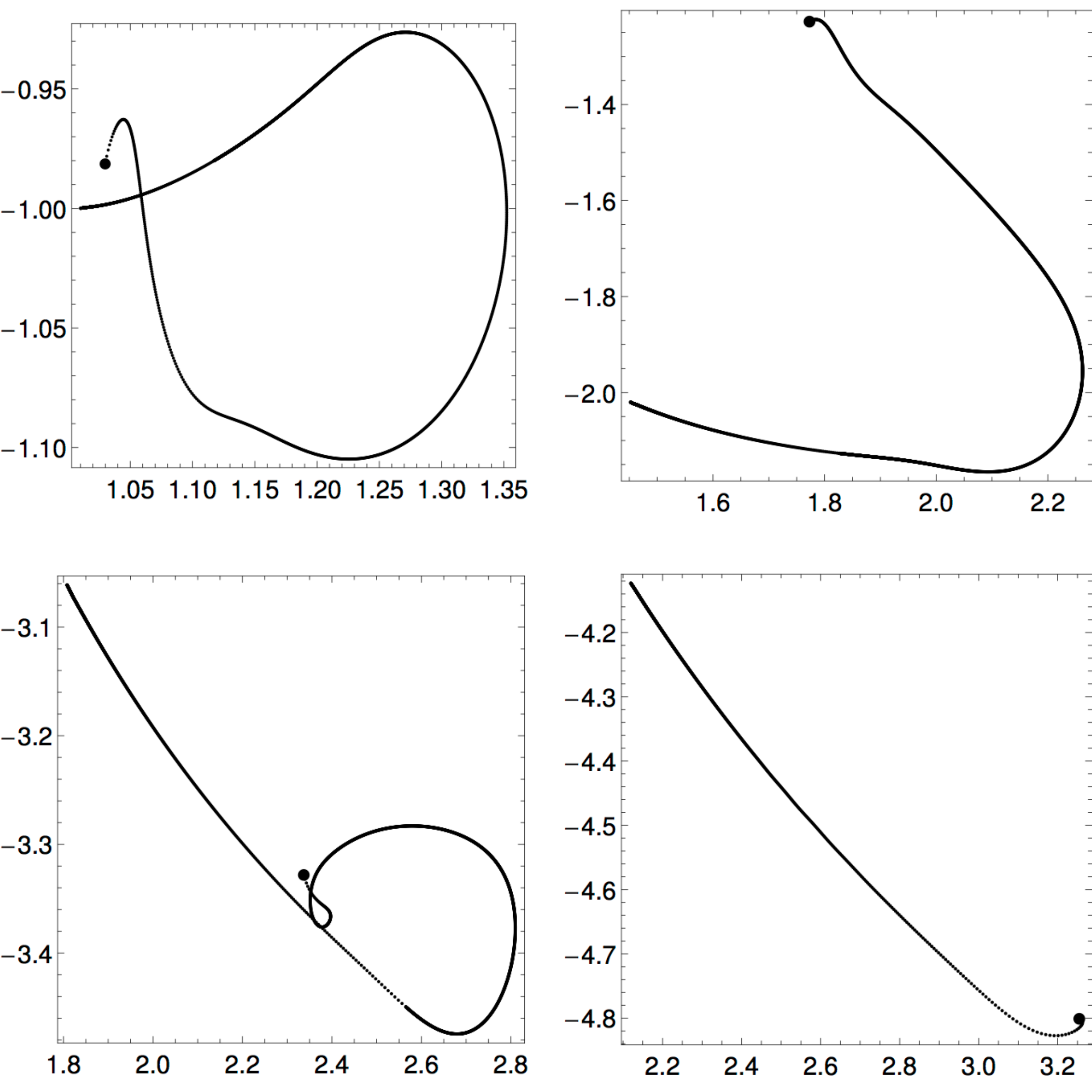}
\caption{\textbf{Schwarzschild}. The saturating scalar QNM's in the complex $\omega$ plane for $\widetilde{\ell}_S = 2$ (top left), $\widetilde{\ell}_S = 3$ (top right), $\widetilde{\ell}_S = 4$ (bottom left), $\widetilde{\ell}_S = 5$ (bottom right). The curves begin at $d=6,6,9,15$ (large dots) for $\widetilde{\ell}_S = 2,3,4,5$, respectively, and $d$ increases along the curve, reaching $d=100$ at the other endpoint.}
\label{Fig:saturating_scalars_complex}
\end{figure}

We are now ready to present our results. In Fig.~\ref{Fig:schw_all} we display the QNM  frequencies for scalar ({\it left panel}), vector ({\it middle panel}), and tensor ({\it right panel}) perturbations. We will analyze these results in more detail below, but two interesting features are immediately obvious. The first is that many of the modes scale with $d$, so that both their decay rate Im$(\omega)$ and oscillation frequency Re$(\omega$) increase in magnitude as the dimension is increased. The second is that many of the curves seem to lie on top one another. The curves that scale together have different angular quantum numbers $\widetilde{\ell}$. Evidently in the large-$d$ limit, the difference between these QNMs with $\widetilde{\ell} \sim \mathcal{O}(d^0)$ goes to zero.

A perhaps less obvious feature is arguably the most interesting: in the scalar plot there are QNM  curves that do not scale with $d$, and in fact seem to not stray too far from their low-$d$ values. We dub these modes {\it saturating modes}, and will study them in detail below. Let us also mention that saturating modes also exist for vectors, but they cannot be seen from the above plots because they are purely imaginary. We found no saturating modes for tensors. 

\begin{figure}[h!]
\centering
\includegraphics[width=1.0 \textwidth]{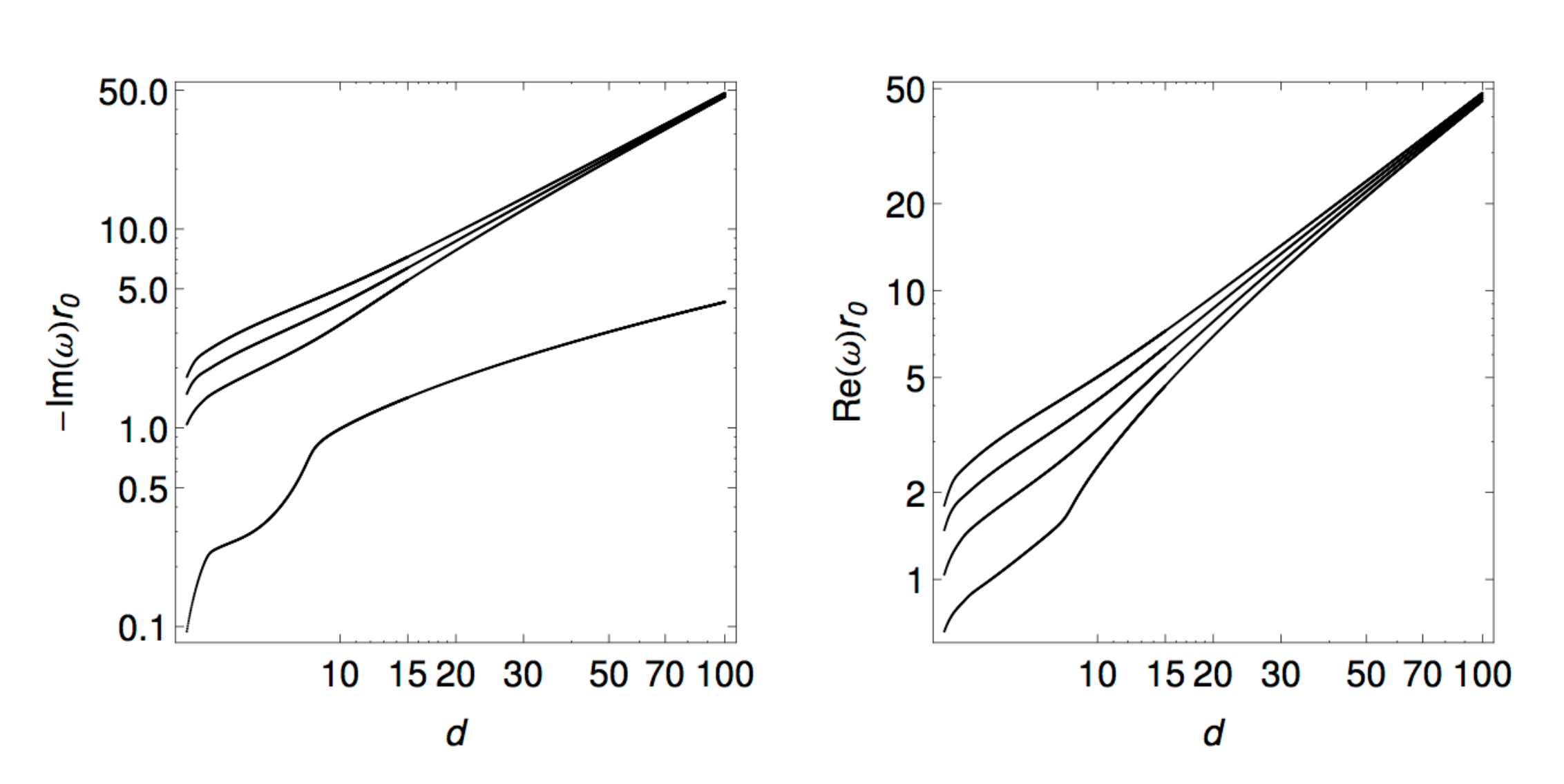}
\caption{\textbf{Schwarzschild}. Plot of non-saturating scalar modes for $\widetilde{\ell}_S = 2,3,4,5$. For both plots the curves appear in terms of increasing $\widetilde{\ell}_S$, from bottom to top.}
\label{Fig:scaling_scalars}
\end{figure}
%
%%%%%%%%%%%%%%%%%%%%%%%%%%%%%%
\subsubsection{Scalar modes} \noindent

Amongst the three sectors, the scalar potential is the most complicated, and it is presumably this structure which allows for the interesting behaviour observed as $d$ is varied. We start by presenting our scalar results for the saturating QNMs. In Fig.~\ref{Fig:saturating_scalars}, we display the real and imaginary part of the saturating modes for $\widetilde{\ell}_S = 2,3,4,5$. These modes are clearly saturating to finite values as $d\rightarrow \infty$. In Fig.~\ref{Fig:saturating_scalars_complex} we plot these saturating QNM's in the complex plane. In all cases the curves begin at the large dot and execute interesting trajectories as $d$ increases. 
The values of the saturating modes for $d=100$ are likely close to their limiting values, and are:
\be(\widetilde{\ell}_S, \omega r_0) \simeq (2, 1.01 - 1.00 i), \quad (3, 1.45-2.02i), \quad (4, 1.81 - 3.06i), \quad (5, 2.12 - 4.12i) . \ee
One would expect a simple analytic formula to describe these results. It is not obvious how the real part is changing as a function of $\widetilde{\ell}_S$, but the imaginary part seems to obey $\lim_{d\rightarrow \infty} \text{Im}(\omega) = -(\widetilde{\ell_S}-1)$. The error of the numerical data from these values is consistent with the corrections being $\mathcal{O}(d^{-1})$.

In addition to the saturating modes, there are also modes which scale with $d$ and that we might call {\it non-saturating} or {\it scaling modes}. Indeed, as can be seen from Fig.~\ref{Fig:schw_all}, many QNMs scale the same way in the large-$d$ limit. In Fig.~\ref{Fig:scaling_scalars} we plot one such group of scaling QNMs. For these curves we can extract their dependence on $d$. It is power law with roughly the dependence
\begin{equation}\label{scaling}
 {\rm Im}(\omega\, r_0) \sim d^{1/2},\qquad  {\rm Re}(\omega\, r_0) \sim d. 
\end{equation} 
We stress that these results pertain just to the group of modes plotted, and it could well be the case that there are many different scalings. It is harder to extract the power laws of the other groups of scaling modes as they enter the scaling regime at larger $d$'s than the group displayed.

%%%%%%%%%%%%%%%%%%%%%%%%%%%%%%
\subsubsection{Vector modes} 
\begin{figure}[h!]
\centering
\includegraphics[width=0.5 \textwidth]{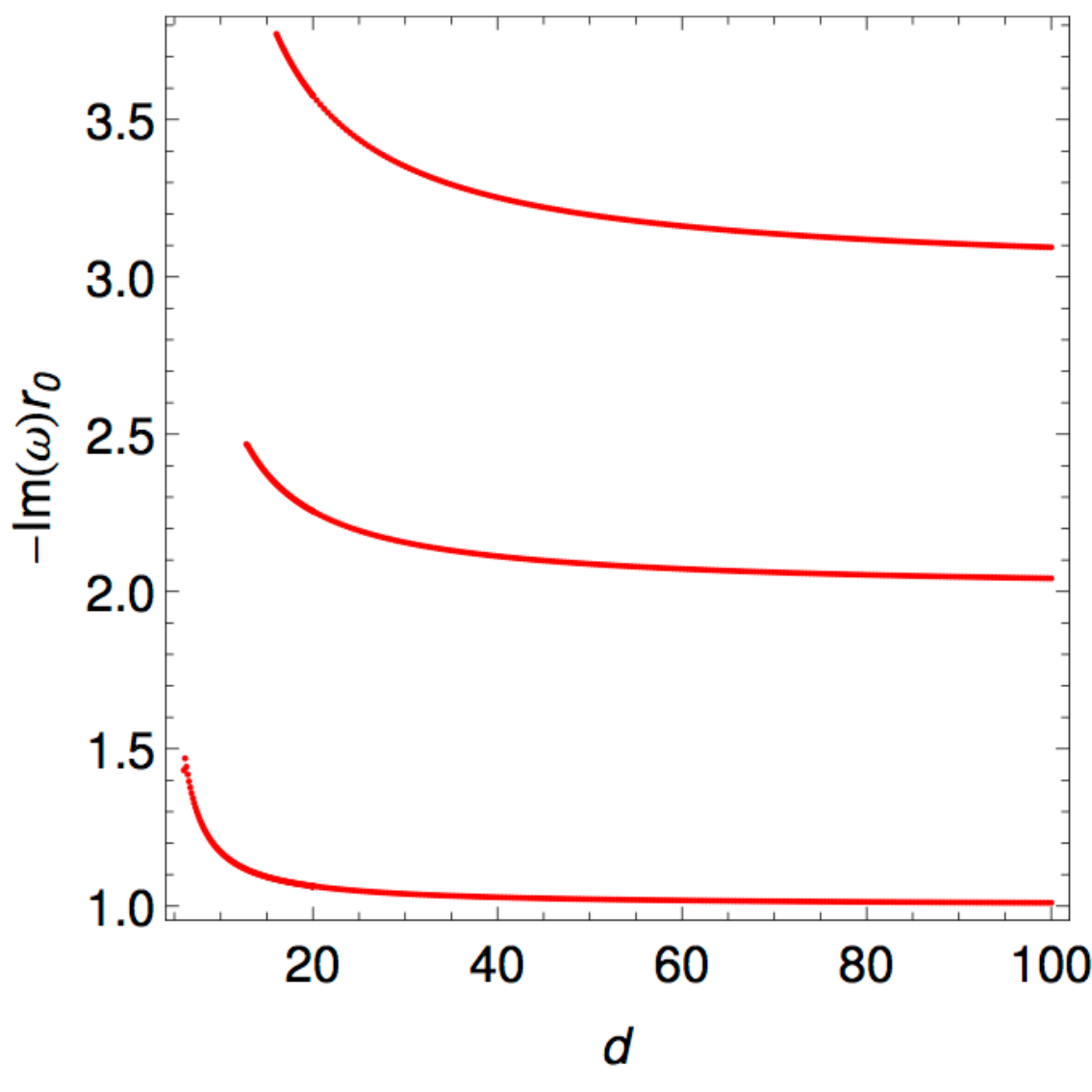}
\caption{\textbf{Schwarzschild}. The imaginary part of the saturating vector QNM's for $\widetilde{\ell}_V = 2,3,4$. Higher $\widetilde{\ell}_V$ curves lie above lower $\widetilde{\ell}_V$ curves.}
\label{Fig:saturating_vectors}
\end{figure}

\begin{figure}[h!]
\centering
\includegraphics[width=1.0 \textwidth]{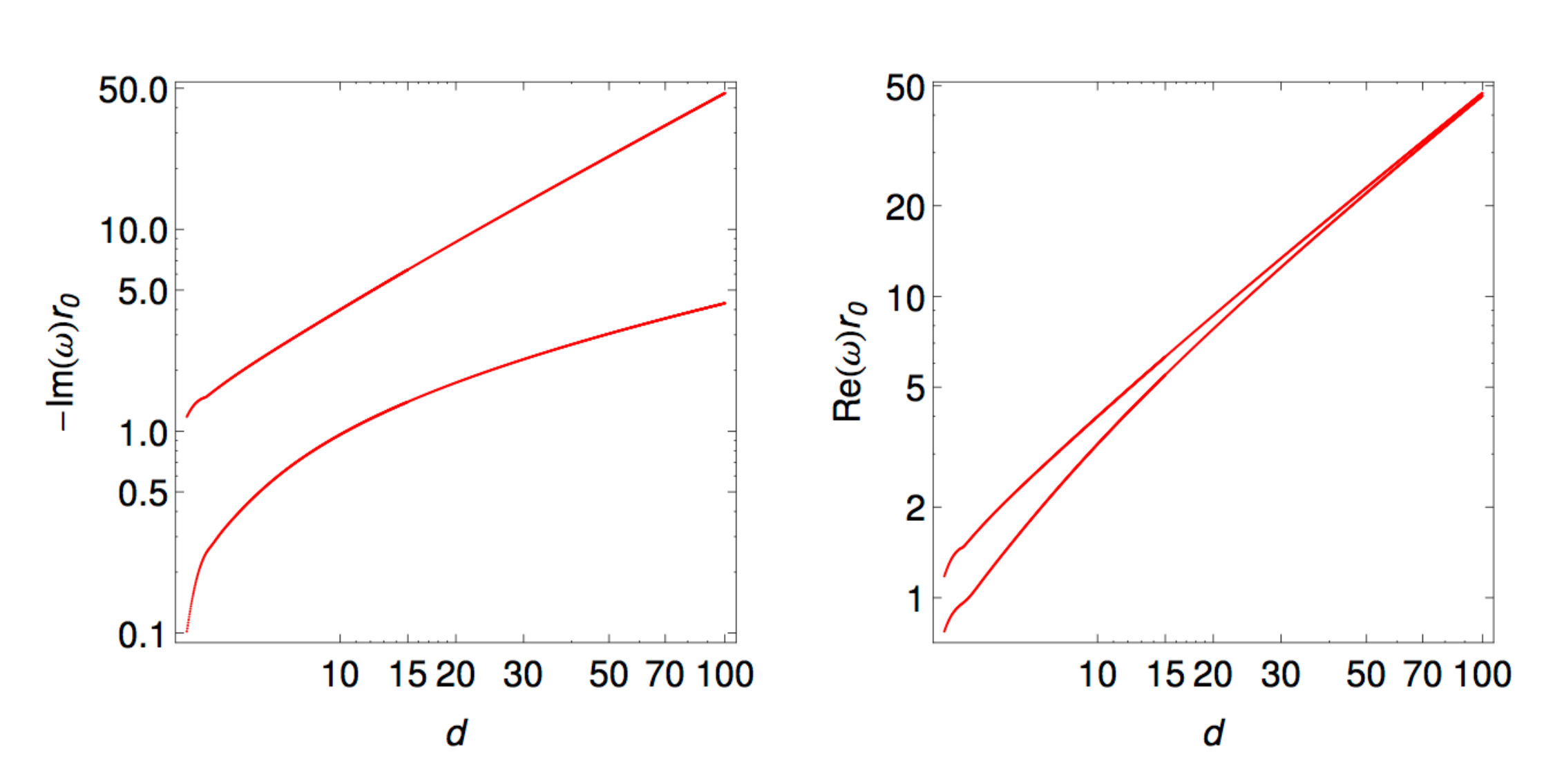}
\caption{\textbf{Schwarzschild}. Plot of non-saturating vector modes for $\widetilde{\ell}_V = 2,3$. For both plots the curves appear in terms of increasing $\widetilde{\ell}_V$, from bottom to top.}
\label{Fig:scaling_vectors}
\end{figure}

\noindent
We now turn to discuss the vector modes. We again find both saturating and scaling modes. Interestingly, the saturating vector modes are purely imaginary. These are plotted in Fig.~\ref{Fig:saturating_vectors}. 
The values of the saturating vector modes for $d=100$ are: 
\be(\widetilde{\ell}_V, \omega r_0) \simeq (2, -1.01i ), \quad (3, -2.04i), \quad (4, -3.09i) . \ee
Again the data suggests the relation $\lim_{d\rightarrow \infty} \text{Im}(\omega) = -(\widetilde{\ell}_V-1)$. Turning to the modes that scale with $d$, we can make the same plot as in the scalar case, plotting QNM's of different $\widetilde{\ell}_V$ that seem to scale the same way. The results are plotted in Fig.~\ref{Fig:scaling_vectors}. These modes also have the same scaling \eqref{scaling} as in the scalar case, namely Im$(\omega r_0) \sim d^{1/2}$, and Re$(\omega r_0) \sim d$.
%

%%%%%%%%%%%%%%%%%%%%%%%%%%%%%%
\subsubsection{Tensor modes} 
\begin{figure}[h!]
\centering
\includegraphics[width=1.0 \textwidth]{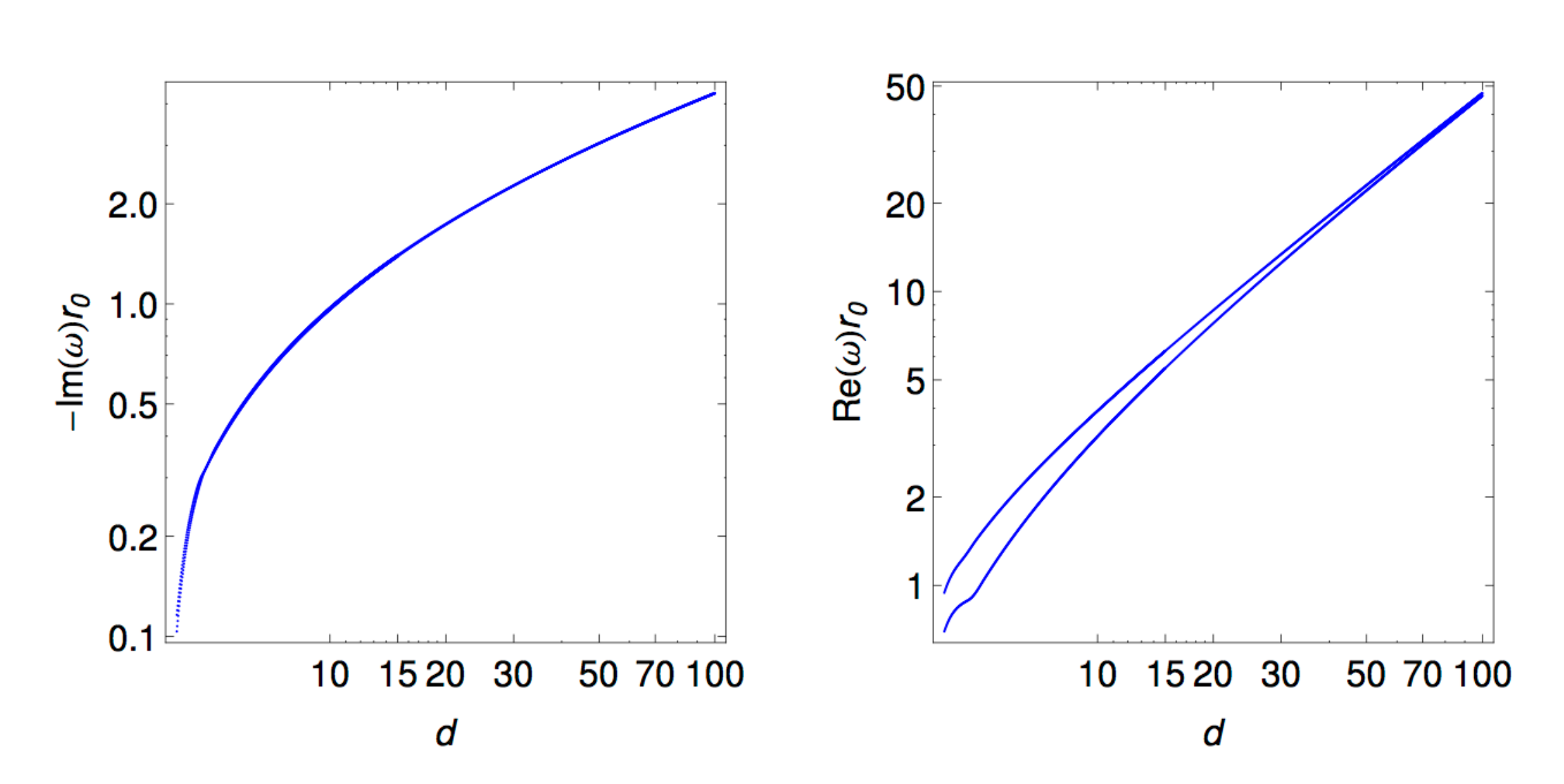}
\caption{\textbf{Schwarzschild}. Plot of non-saturating tensor modes for $\widetilde{\ell}_T = 1,2$. For both plots the curves appear in terms of increasing $\widetilde{\ell}_T$, from bottom to top. For the imaginary plot, the two curves are so close as to be indistinguishable.}
\label{Fig:scaling_tensors}
\end{figure}

\noindent
For the tensor modes, we observe no saturating modes. A few of the lowest-lying scaling modes are plotted in Fig.~\ref{Fig:scaling_tensors}. Once again, these modes have the same scaling  \eqref{scaling}  as in the scalar and vector cases, i.e.  Im$(\omega r_0) \sim d^{1/2}$, and Re$(\omega r_0) \sim d$.
%
%%%%%%%%%%%%%%%%%%%%%%%%%%%%%%
\subsubsection{Discussion of results} \noindent
The physics of the saturating modes is very interesting. In terms of the Schwarzschild time $t$, the modes that scale with $d$ decay increasingly rapidly as $d$ increases, whilst the saturating modes have a finite decay rate even in the infinite-$d$ limit. Thus, for phenomena for which $t$ is the relevant time, the QNM mode spectra consists only of the saturating modes and all the others become irrelevant. This is a particularly sharp characterization of the way in which the large-$d$ limit simplifies the physics.

\begin{figure}[h!]
\centering
\includegraphics[width=1.0 \textwidth]{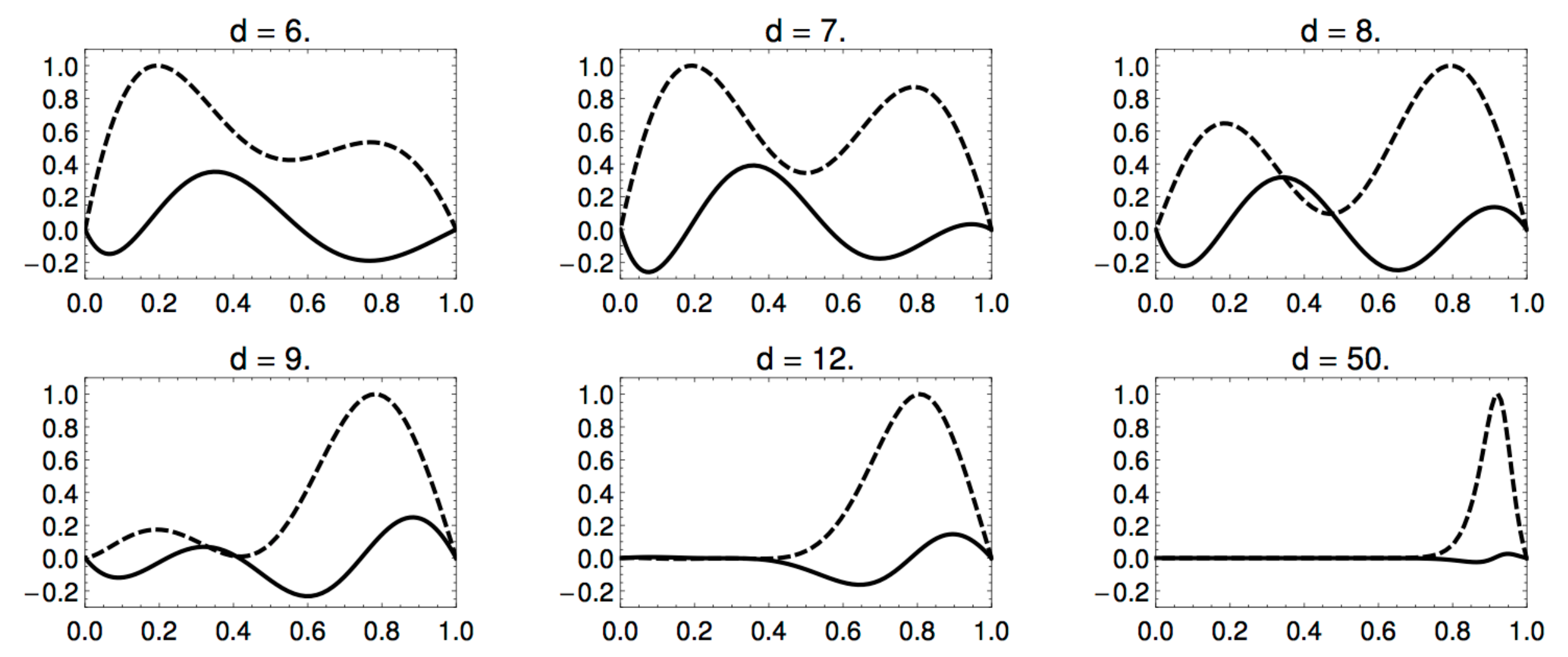}
\caption{\textbf{Schwarzschild}. The real (dashed) and imaginary (dotted) parts of the scalar gauge invariant wavefunctions $\widetilde{\phi}_S$ for $\widetilde{\ell}_S = 2$. As the dimension is increased, the wavefunctions become localized near the horizon.}
\label{Fig:saturating_scalars_wavefunc}
\end{figure}

Another important feature of the saturating modes is that they are localized near the horizon. In Fig.~\ref{Fig:saturating_scalars_wavefunc} we plot the real and imaginary parts of the scalar gauge invariant wavefunction $\widetilde{\phi}_S$ for $\widetilde{\ell}_S =2$ and various $d$. It is clear that as $d$ is increased, the wavefunction becomes increasingly localized around $z=1$, which corresponds to the horizon. The near-horizon geometry was shown to take the form of the direct product of a 2d string BH and a sphere \cite{Emparan:2013xia},
\be ds^2 = \frac{4r_0^2}{\tilde{n}^2} \Big(-\tanh^2 \rho d \hat{t}^2 + d\rho^2 \Big) + r_0^2 d\Omega_{\tilde{n}+1}^2, \ee
where $\tilde{n}=d-3$ and the time coordinate of the near-horizon geometry is related to the usual time via $\hat{t} = \tilde{n} t/(2r_0)$. In terms of this time coordinate, the saturating modes do not decay or oscillate in the infinite-$d$ limit, %
\be \exp(-i\omega t) = \exp(-2i \omega r_0 \hat{t}/\tilde{n}) \sim 1, \ee 
whereas scaling modes may decay or oscillate depending on how $\omega$ scales with $d$. Ref.~\cite{Emparan:2013xia} argued that modes with $\omega \sim \mathcal{O}(d^0)$ could be said to decouple from the asymptotic region, and the localization of the wavefunctions of these modes confirms this.

The presence of the saturating modes connects nicely to recent observations concerning unstable perturbations of rotating BHs in higher dimensions, to which we now turn. 

%

%%%%%%%%%%%%%%

%%%%%%%%%%%%%%%%%%%%%%%%%%%%%%
%%%%%%%%%%%%%%%%%%%%%%%%%%%%%%
\section{Cohomogeneity-1 Myers-Perry black holes
\label{sec:MPEAM}} \noindent
As discussed in Sec.~\ref{subsec:Intro}, when all the angular momenta are equal and non-zero for an odd dimensional Myers-Perry BH, there is a dramatic enhancement of the symmetry and the equations of motion of linearized gravitational perturbations can be reduced to systems of ODE's. As in the Schwarzschild case, the gravitational perturbations can be classified into three sectors, scalar, vector, and tensor, but this time depending on how the metric perturbation transforms under the isometries of the $\mathbb{CP}^N$ base space. The tensor sector is the simplest, where just as in the Schwarzschild case the linearized Einstein equations reduce to a single Schr\"{o}dinger equation. We shall not consider tensor perturbations here, as they were studied in  \cite{Kunduri:2006qa}, where no instabilities were found. The study of scalar modes was first initiated in \cite{Dias:2010eu}, where axisymmetric perturbations were studied, and continued in \cite{Hartnett:2013fba}, where non-axisymmetric perturbations were considered. Instabilities were found for both types of perturbations, in agreement with the predictions of Emparan and Myers \cite{Emparan:2003sy}. The axisymmetric instabilities are particularly interesting because they indicate the existence of new families of BHs with a single rotational symmetry \cite{Dias:2010eu}, and the non-axisymmetric $m$-bar mode instabilities are important because they occur for much slower rotations than the axisymmetric instabilities, and are hence the most dominant.

In this section we study scalar and vector perturbations of these BHs. We begin by reviewing the harmonic tensors needed for the separation of variables, and then discuss our numerical results.
%%%%
%%%%
\subsection{Charged harmonic tensors on $\mathbb{CP}^N$} \noindent
We now review charged scalar and vector harmonic tensors on $\mathbb{CP}^N$ which allow the separation of variables for linearised gravitational perturbations. Charged scalar harmonics were studied in \cite{Hoxha:2000jf}, and vector harmonics in \cite{Durkee:2010ea}. By charged harmonics we mean those tensors which are eigentensors with respect to the derivative operator
\be \hat{D}_a \equiv \hat{\nabla}_a-i m A_a, \ee
where $\hat{\nabla}$ is the covariant derivative on $\mathbb{CP}^N$, and $A_a$ is again related to the K\"{a}hler form via $J = dA/2$. This is the natural derivative operator given the appearance of the Hopf fibration in the BH metric. The charge of a given harmonic is $m$, which we take to be an integer. 
\\ \\
\textbf{Scalar harmonics} \\
Charged scalar harmonics are  functions of the $\mathbb{CP}^N$ coordinates which satisfy
\be (\hat{D}^2 + \lambda_S) \mathbb{S} = 0. \ee
Here the eigenvalue is a function of two quantized parameters, $(\kappa,m)$:
\be \lambda_S = l(l+2N)-m^2, \quad l = 2\kappa + |m|, \ee
where $\kappa = 0,1,2...$, and $m \in \mathbb{Z}$. Charged scalar-derived vectors can be obtained by differentiating,
\be \mathbb{S}_a = - \frac{1}{\sqrt{\lambda_S}} 
\hat{D}_a \mathbb{S}. \ee
These can be further decomposed into  Hermitian and anti-Hermitian parts
\be J_a^{\hspace{3pt} b} \mathbb{S}_b^{\pm} = \mp  i \mathbb{S}^{\pm}_a. \ee
Lastly, the scalar-derived tensors are given by
\be
\mathbb{S}_{ab}^{++} = \hat{D}^+_{(a}\mathbb{S}^{+}_{b)}, \qquad \mathbb{S}_{ab}^{--} = \hat{D}^-_{(a}\mathbb{S}^{-}_{b)}, \qquad \mathbb{S}^{+-}_{ab} = \hat{D}^+_{(a}\mathbb{S}^-_{b)} + \hat{D}^-_{(a}\mathbb{S}^+_{b)} - \frac{1}{2N}\hat{g}_{ab} \hat{D} \cdot \mathbb{S} . \ee
\\ \\
\textbf{Vector harmonics} \\
Next we consider the vector harmonics which only exist for $N\ge 2$. These are also eigenfunctions of $\hat{D}^2$ which transform as vectors in $\mathbb{CP}^N$ and which are also transverse with respect to $\hat{D}^a$:
\be (\hat{D}^2 + \lambda_V)\mathbb{V}_a = 0, \qquad \hat{D}^a \mathbb{V}_a = 0. \ee
Just as the scalar-derived vectors, these may be further characterized according to their eigenvalue under the complex structure,
\be J_a^{\hspace{3pt} b} \mathbb{V}_b = -i \epsilon \mathbb{V}_a, \qquad \epsilon = \pm 1. \ee
The vector-derived tensors are given by
\be \mathbb{V}_{ab}^{\pm} = - \frac{1}{\sqrt{\lambda_V}} \hat{D}_{(a}^{\pm} \mathbb{V}_{b)}. \ee
The eigenvalues were first computed in \cite{Durkee:2010ea} for the uncharged case, $m=0$,
\be 
\lambda_V^{(m=0)} = 4\kappa(\kappa+2)+2(N+1)(2\kappa+3). 
\ee
In Appendix \ref{app:vectors} we extend this result for non-zero $m$ in the $N=2$ case:
\be \lambda_V^{(N=2)} = 4\kappa \left( \kappa + 3 + |2+m\epsilon| \right) + 6|2+m\epsilon|+6+2m\epsilon. 
\label{eq:cpnvector} \ee

%%%%
\subsection{Perturbation decomposition and equations} \noindent

In order to implement the harmonic decomposition of the perturbation, it will be useful to introduce the 1-forms, $e^A$, where $A \in (0,1,2)$:
\be e^0 = \frac{p(\hat{r})^{1/2}}{h(\hat{r})^{1/2}} dt, \quad e^1 = p(\hat{r})^{-1/2} d\hat{r}, \quad e^2 = \hat{r} h(\hat{r}) (d\psi + A_a dx^a - u(\hat{r})dt). \ee
The scalar sector of metric perturbations is then
\begin{eqnarray}
h_{AB} &=& f_{AB} \mathbb{S}, \\
h_{Aa} &=& \hat{r}(f_A^+ \mathbb{S}^+_a + f_A^- \mathbb{S}^-_a ), \\
h_{ab} &=& -\frac{\hat{r}^2}{\lambda^{1/2}}(H^{++} \mathbb{S}^{++}_{ab} + H^{--} \mathbb{S}^{--}_{ab} + H^{+-} \mathbb{S}^{+-}_{ab}) + \hat{r}^2 H_L \hat{g}_{ab}\mathbb{S},
\end{eqnarray}
and the vector sector is
\be h_{AB}=0,\qquad h_{Aa} = \hat{r} f^{(V)}_A \mathbb{V}_a, \qquad  h_{ab} = -\frac{\hat{r}^2}{\lambda^{1/2}} (H^{+} \mathbb{V}^{+}_{ab} + H^{-} \mathbb{V}^{-}_{ab}). \ee
The $(t,\psi)$-dependence can be separated out if all of the above coefficient functions are proportional to $e^{-i(\omega t - m\psi)}$, which we can assume to be the case since $\partial_t$ and $\partial_\psi$ are Killing vectors of the background. The value of $m$ provides an important characterization of the perturbation. Those with $m=0$ are axisymmetric, while those with $m \neq 0 $ are non-axisymmetric.
\\ \\
This decomposition in terms of $\mathbb{CP}^N$ harmonic tensors parallels the one based on $S^{2N+1}$ tensors used in the KI  formalism. Here, however, it is not possible to construct a gauge invariant master variable, and therefore we will need to impose a gauge and solve coupled ODE's. The gauge we will work in is the traceless transverse gauge,
\be h = g^{\mu\nu}h_{\mu\nu} = 0, \quad \nabla^{\mu} h_{\mu\nu} = 0. \ee
In this gauge the Einstein equations take the simple form:
\be \nabla^2 h_{\mu\nu} + 2 R_{\mu\rho\nu\sigma}h^{\rho\sigma} = 0. \ee

%%%%
%%%%
\subsection{Numerical results} \noindent
Here we present our results for the QNMs of equal angular momenta Myers-Perry (MP) BHs in odd dimensions $d \ge 5$. 

The numerical methods used are the same as in Sec.~\ref{sec:Schw}, but adjusted to allow for coupled ODE's rather than a single ODE. Once again a compactified radial coordinate $\hat{r}=r_+/z$ is used, and the grid is Chebyshev, as before. Here we omit a detailed description of the boundary conditions. They are still determined by the physical condition of being ingoing at the horizon and outgoing at infinity, but the exact form they take depends on the perturbation sector, the dimension, and the $\mathbb{CP}^N$ quantum numbers. The reason for the perhaps unexpected dependence on these last two quantities is due to the fact that for certain dimensions and quantum numbers various $\mathbb{CP}^N$ harmonics vanish. For example, in $N=1$,  $\mathbb{S}^{+-} = 0$ and in all $N$, $\mathbb{S}^{++} = 0$ for $\kappa = 1$, $m>0$. We refer the interested reader to Ref.~\cite{Dias:2010eu} for a discussion of boundary conditions that can easily be adapted to specific cases.

%%%%%%%%%%%%%%%%%%%%
\subsubsection{Scalar modes} \noindent

The scalar sector of perturbations is again the most complicated, involving the largest number of perturbation functions. Recall that charged scalar harmonics on $\mathbb{CP}^N$ are characterized by two integers, $(\kappa,m)$. Axisymmetric modes ($m=0$) were first studied in \cite{Dias:2010eu} where it was found that the $(2,0)$ mode was ultraspinning unstable for odd $d \ge 7$ (see footnote \ref{foot:ultra}). Ref.~\cite{Hartnett:2013fba} studied scalar perturbations for $m\neq0$ where a bar-mode instability was found for the $(0,m)$ mode for $m \ge 2$. 

\begin{figure}[ht]
\centering
\includegraphics[width=1.0 \textwidth]{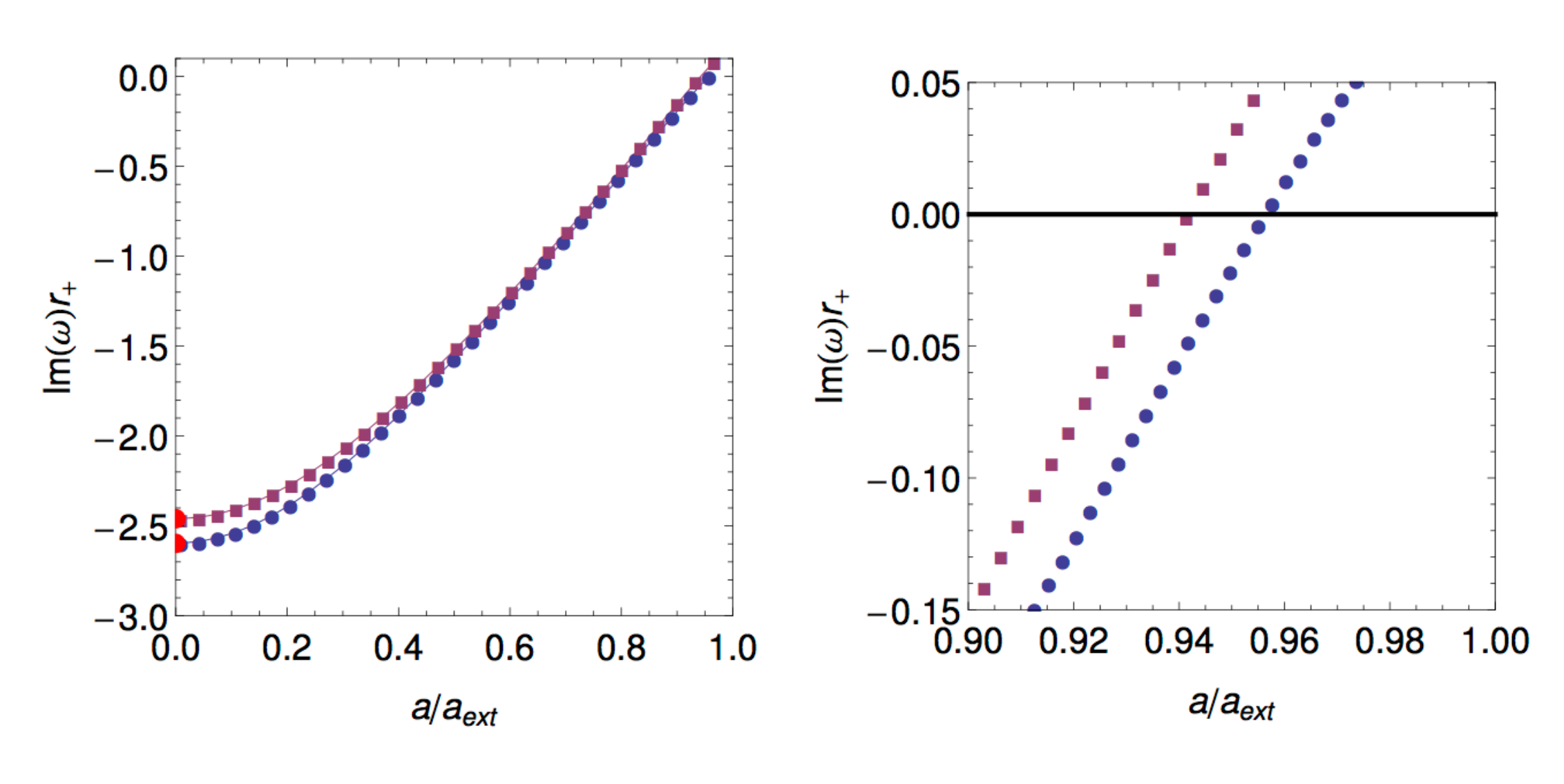}
\caption{\textbf{EAM MP.} \emph{Left Panel:} The purely imaginary $(2,0)$ frequency for $d=11$ (circles) and $d=13$ (squares). The large dots at $a=0$ correspond to a Schwarzschild vector modes with $\widetilde{\ell}_V = 3$. \emph{Right Panel:} A zoomed in plot showing the modes crossing the instability threshold.}
\label{Fig:mp_scalars_k2m0}
\end{figure}

Ref.~\cite{Dias:2010eu} first found the axisymmetric, ultraspinning instability by studying the related problem of the Gregory-Laflamme instability for these rotating BHs. It was expected that the instability would produce a zero mode at the threshold of instability, i.e. $\omega = 0$, and that for larger rotations it would take the form $\omega = i K(a)$, with $K(a)$ a positive an increasing function of the rotation $a$. The approach of \cite{Dias:2010eu} only allowed for the determination of $\omega$ at and above the threshold of instability. However, our methods allow us to follow this mode all the way down to zero rotation and find its Schwarzschild limit. This is shown in Fig.~\ref{Fig:mp_scalars_k2m0} in which  the QNM frequency associated with the $(2,0)$ ultraspinning instability is plotted for $d=11$ and $d=13$ \footnote{We could of course study other dimensions quite easily, but oddly it becomes numerically difficult to isolate this mode for smaller values of $d$, although it is undoubtedly present.}. Interestingly, we find that $\omega$ is always purely imaginary and, at zero rotation, this mode connects to a Schwarzschild vector mode with $\widetilde{\ell}_V = 3$.

\begin{figure}[ht]
\centering
\includegraphics[width=1.0 \textwidth]{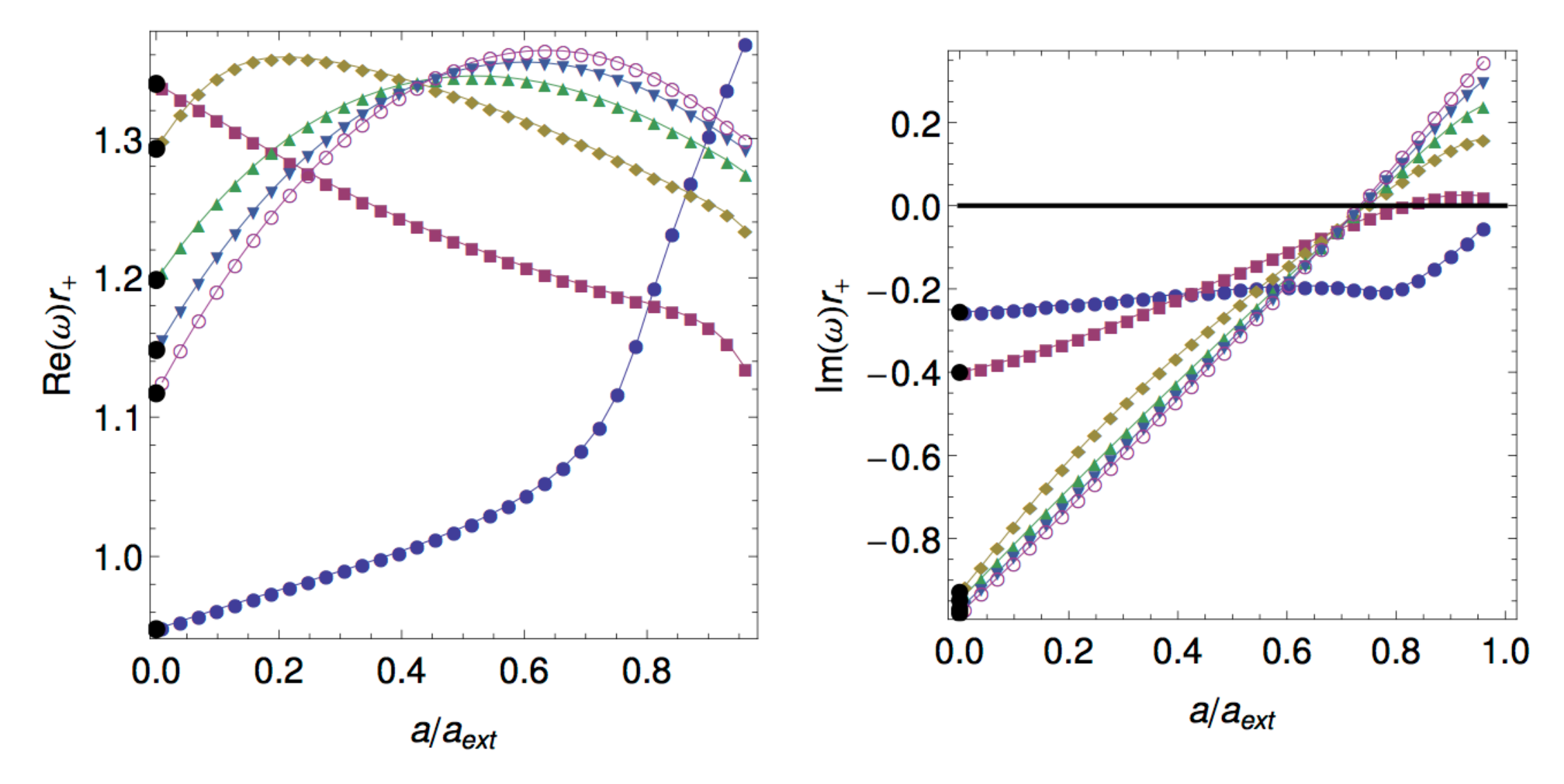}
\caption{\textbf{EAM MP.} Real (\emph{left panel}) and imaginary (\emph{right panel}) parts of the QNM frequency for the $(0,2)$ scalar mode for $d = 5$ (filled-in circles), $d=7$ (filled-in squares), $d=9$ (filled-in diamonds), $d=11$ (filled-in triangles), $d=13$ (filled-in upside-down triangles), and $d=15$ (open circles). For zero rotation the frequencies reduce to the $\widetilde{\ell}_S= 2$ scalar modes of Schwarzschild, which are depicted by large black dots. These were calculated using a separate code based on the KI  master equation.}
\label{Fig:mp_scalars_k0m2}
\end{figure}

\begin{figure}[ht]
\centering
\includegraphics[width=1.0 \textwidth]{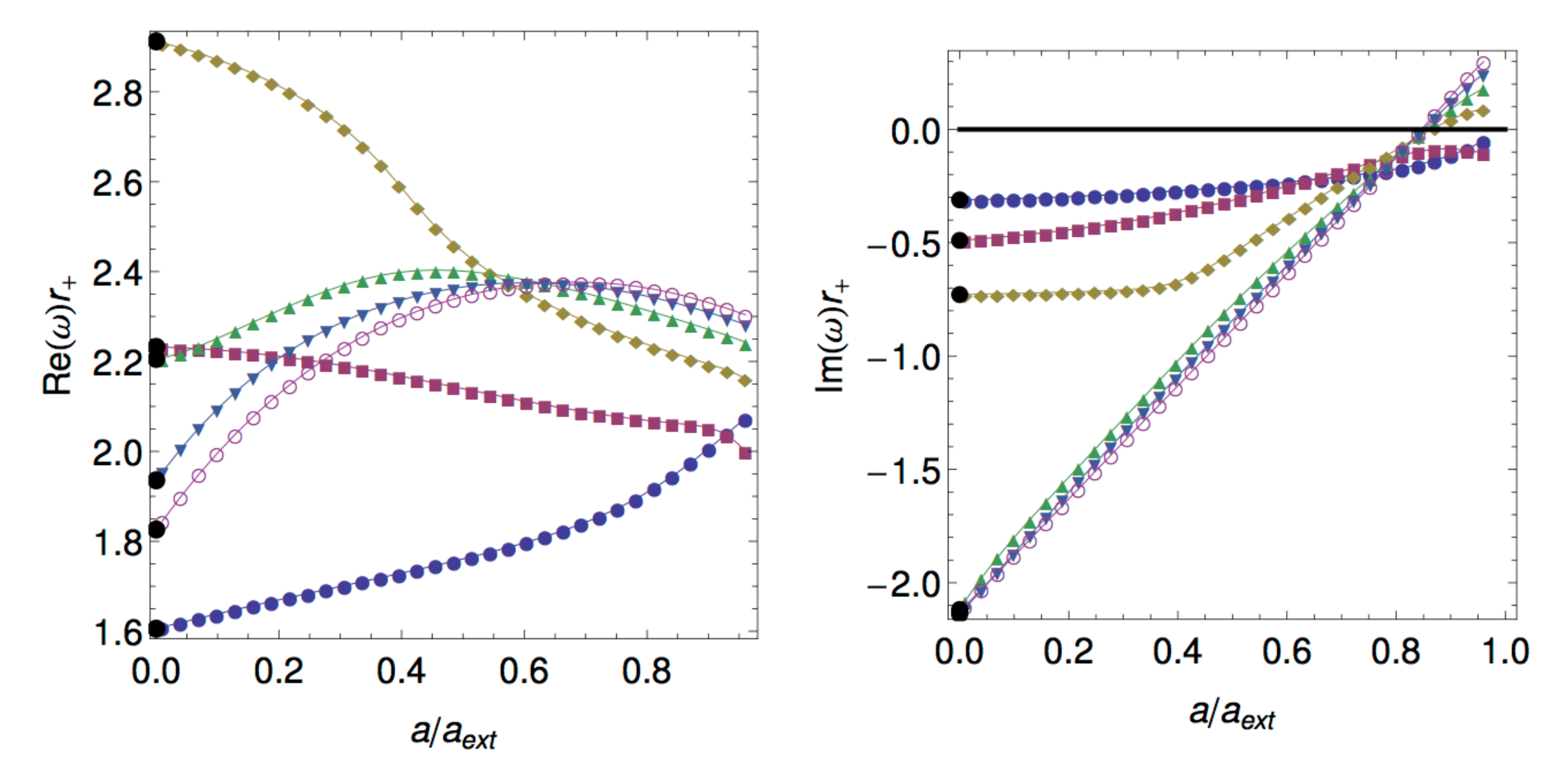}
\caption{\textbf{EAM MP.} Real (\emph{left panel}) and imaginary (\emph{right panel}) parts of the QNM frequency for the $(0,3)$ scalar mode for $d = 5$ (filled-in circles), $d=7$ (filled-in squares), $d=9$ (filled-in diamonds), $d=11$ (filled-in triangles), $d=13$ (filled-in upside-down triangles), and $d=15$ (open circles). For zero rotation the frequencies reduce to  the $\widetilde{\ell}_S=3$ scalar modes of Schwarzschild, which are depicted by large black dots. These were calculated using a separate code based on the KI  master equation.}
\label{Fig:mp_scalars_k0m3}
\end{figure}

\begin{figure}[ht]
\centering
\includegraphics[width=1.0 \textwidth]{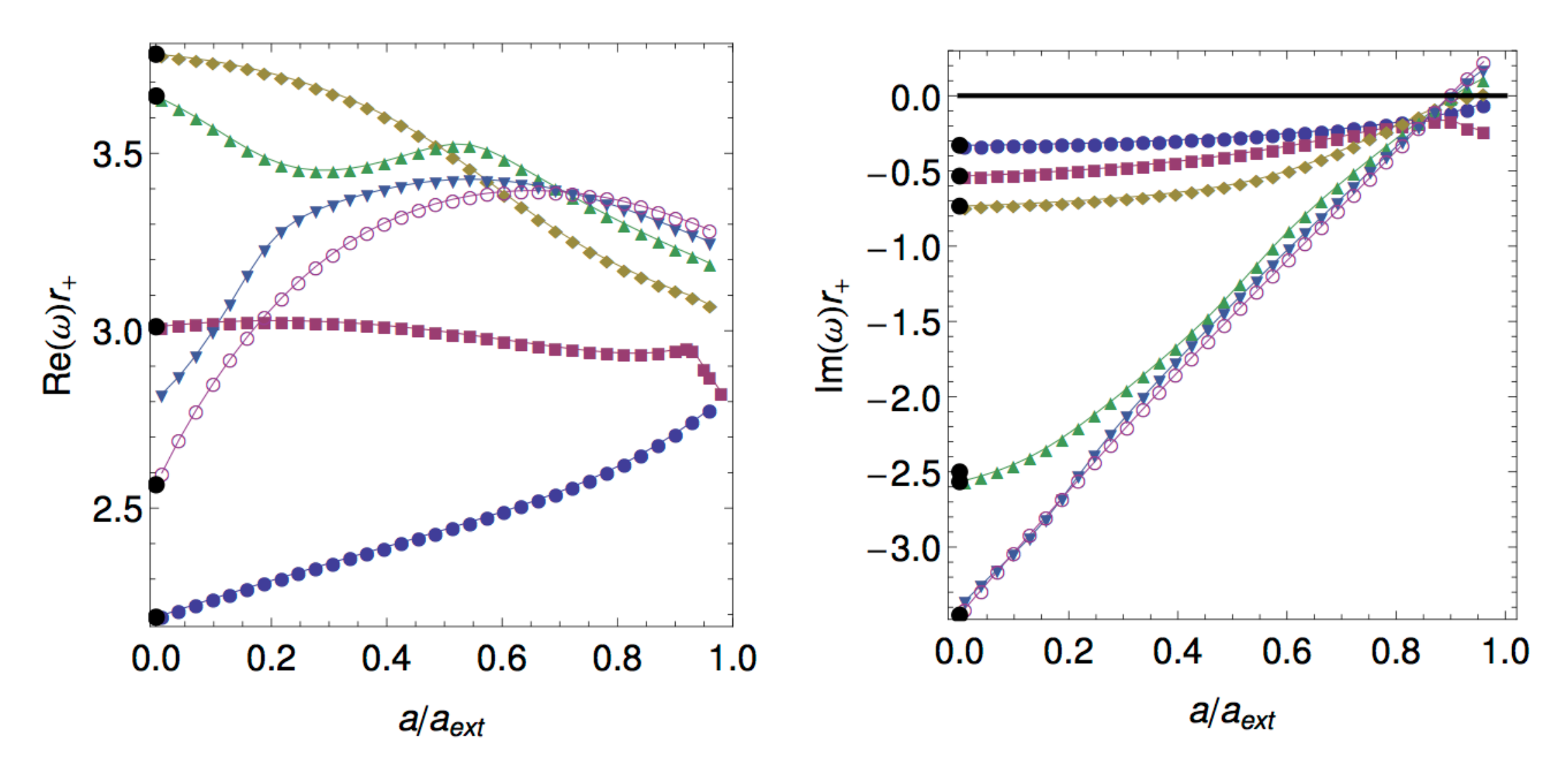}
\caption{\textbf{EAM MP.} Real (\emph{left panel}) and imaginary (\emph{right panel}) parts of the QNM frequency for the $(0,4)$ scalar mode for $d = 5$ (filled-in circles), $d=7$ (filled-in squares), $d=9$ (filled-in diamonds), $d=11$ (filled-in triangles), $d=13$ (filled-in upside-down triangles), and $d=15$ (open circles). For zero rotation the frequencies reduce to  the $\widetilde{\ell}_S= 4$ scalar modes of Schwarzschild, which are depicted by large black dots. These were calculated using a separate code based on the KI  master equation.}
\label{Fig:mp_scalars_k0m4}
\end{figure}

Next we consider the bar-mode instability. In Fig.'s~\ref{Fig:mp_scalars_k0m2}, \ref{Fig:mp_scalars_k0m3}, and \ref{Fig:mp_scalars_k0m4} we plot the real and imaginary parts of the dominantly unstable bar-mode\footnote{By dominantly unstable we mean the frequency with the largest value of Im$(\omega)$, considered as function of the rotation. Here we stress that as the rotation is tuned, the QNM frequencies can cross, and we are plotting the modes which attain the largest Im$(\omega)$ for all $a$.} for the $(0,m)$ mode for $m=2,3,4$. The $(0,2)$ mode is unstable for $d \ge 7$, while the $(0,3)$ and $(0,4)$ modes are unstable for $d \ge 9$. Of all the bar-mode instabilities found, the $(0,2)$ mode is most unstable, i.e. it has the largest growth rate, Im$(\omega)$. These bar-modes connect to Schwarzschild scalar modes with $\widetilde{\ell}_S = 2$ at zero rotation.

\begin{figure}[ht]
\centering
\includegraphics[width=1.0 \textwidth]{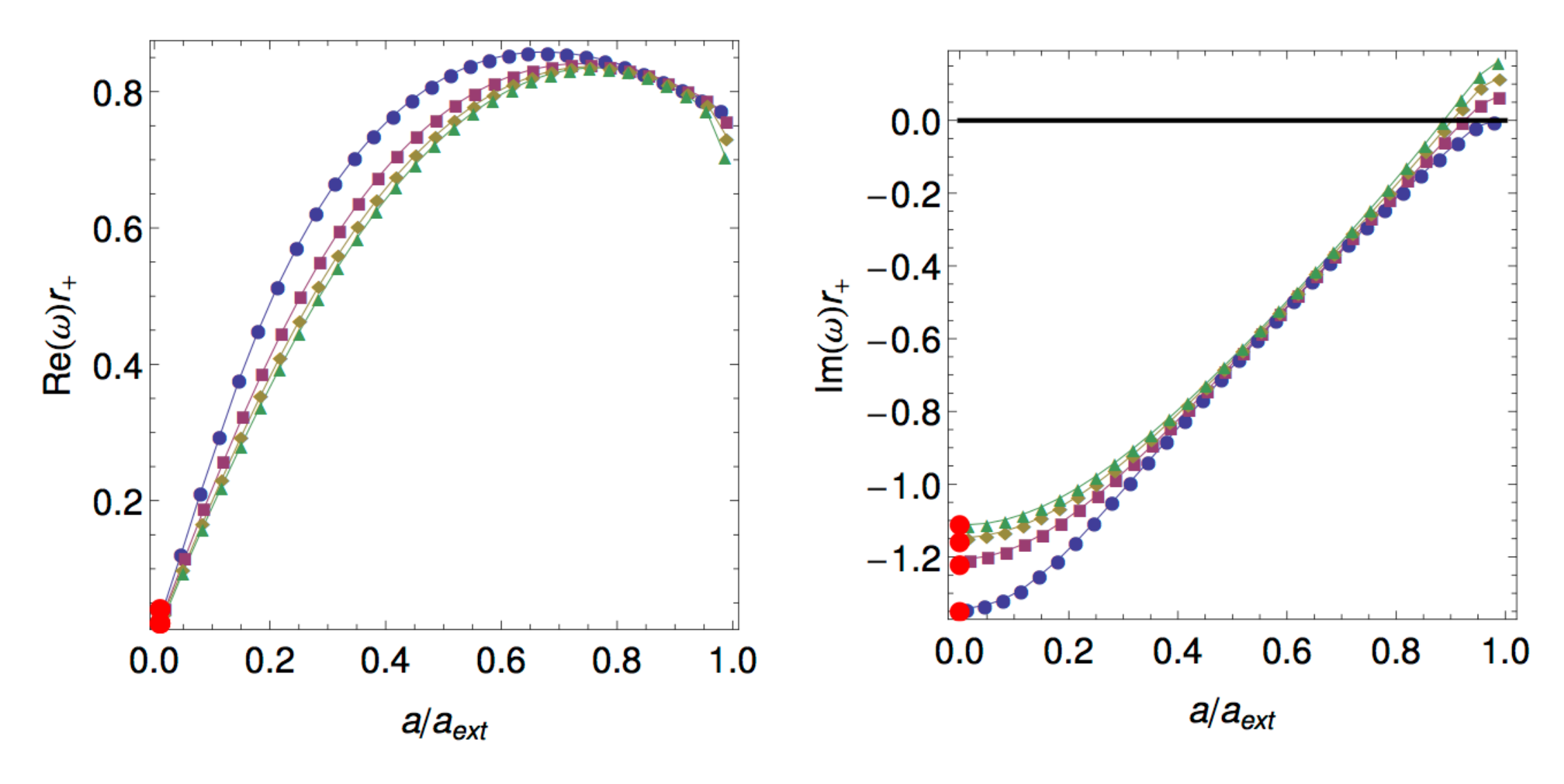}
\caption{\textbf{EAM MP.} Real (\emph{left panel}) and imaginary (\emph{right panel}) parts of the QNM frequency for the $(1,1)$ scalar mode for $d=7$ (circles), $d=9$ (squares), $d=11$ (diamonds), and $d=13$ (triangles). For zero rotation the frequencies reduce to $\widetilde{\ell}_V = 5$ vector modes of Schwarzschild, which are depicted by red dots. These were calculated using a separate code based on the KI  master equation.}
\label{Fig:mp_scalars_k1m1}
\end{figure}

To conclude our investigation of the scalar sector, in Fig.~\ref{Fig:mp_scalars_k1m1} we plot the dominant QNM for the $(\kappa, m)=(1,1)$ sector. Here we find instabilities for $d \ge 9$. These modes connect to the Schwarzschild vector mode with $\widetilde{\ell}_V = 5$ at zero rotation. Of course there are an infinite number of modes we have omitted, but it's quite reasonable that the physical importance of these will be subdominant to the modes studied here. In Table \ref{Table:EAMtable} we list the critical rotations for the instabilities studied.


A very important quantity for non-axisymmetric modes is the so-called superradiant factor, 
$\varpi = \text{Re}(\omega) - m \Omega_H $. 
The energy flux through the horizon is proportional to this factor. When it is negative, the energy flux across the horizon is negative and the perturbation is said to be {\it superradiant}. The first law, applied to a process that extracts the energy $\delta E=-\text{Re}(\omega)$ and the angular momentum $\delta J=-m$  to the BH, yields that the change in horizon area is controlled by $\varpi$:  $\delta A_H \propto [m \Omega_H - \text{Re}(\omega)]  = -\varpi $. Therefore, the second law, $\delta A_H \ge 0$, requires that any unstable mode (whose growth rate is sourced by energy and momenta extracted from the BH) must satisfy $\varpi \leq 0$. We have explicitly checked that  $\varpi <0$ for the non-axisymmetric bar-mode instabilities we found numerically. Note that these modes have Im$(\omega) > 0$ and $m \neq 0$, so in a non-linear time evolution the system will have to radiate since the associated linear mode breaks axisymmetry.

\begin{table}[ht]
\begin{eqnarray}
\nonumber
\begin{array}{||c|c|c|c|c|c||}
\hline\hline d & (2,0) & (0,2) & (0,3) & (0,4) & (1,1) \\
\hline 7 & 0.99998 & 0.8109 & \text{stable} & \text{stable} & \text{stable} \\
\hline 9 & 0.9979 & 0.7463 & 0.8644 & 0.9291 & 0.9252 \\
\hline 11 & 0.9921 & 0.7413 & 0.8547 & 0.9052 & 0.9024 \\
\hline 13 & 0.9850 & 0.7369 & 0.8504 & 0.9011 & 0.8873 \\
\hline 15 & 0.9777 & 0.7331 & 0.8464 & 0.8973 & - \\
\hline\hline
\end{array}
\end{eqnarray}
\caption{Critical rotation $a_c/a_{\text{ext}}$ at which the  instabilities set in for $(\kappa,m)$ scalar sector of perturbations. $d=5$ is linearly stable. Dashes indicate modes for which no data exists. The $(2,0)$ data is taken from \cite{Dias:2010eu}, \cite{Durkee:2010ea}. In the dimensions where we independently have data for the $(2,0)$ mode we agree with \cite{Dias:2010eu}, \cite{Durkee:2010ea} to within a few percent.} \label{Table:EAMtable}
\end{table}
%
%%%%%%%%%%%%%%%%%%%%%%%%
\subsubsection{Vector modes} \noindent
Since $\mathbb{CP}^N$ charged vector harmonics only exist for $N\ge 2$, vector perturbations only exist for $d \ge 7$. In order to study these perturbations, the spectrum of charged vector harmonics is needed. For $N=2$ we derived this in Appendix \ref{app:vectors}. For all other $N$, the result is only known for uncharged $(m=0)$ harmonics. Therefore, we are able to exhaustively study vector perturbations only in $d=7$. We also studied axisymmetric perturbations for $d=9,11,13,15$. In no cases were instabilities found.

\begin{figure}[ht]
\centering
\includegraphics[width=1.0 \textwidth]{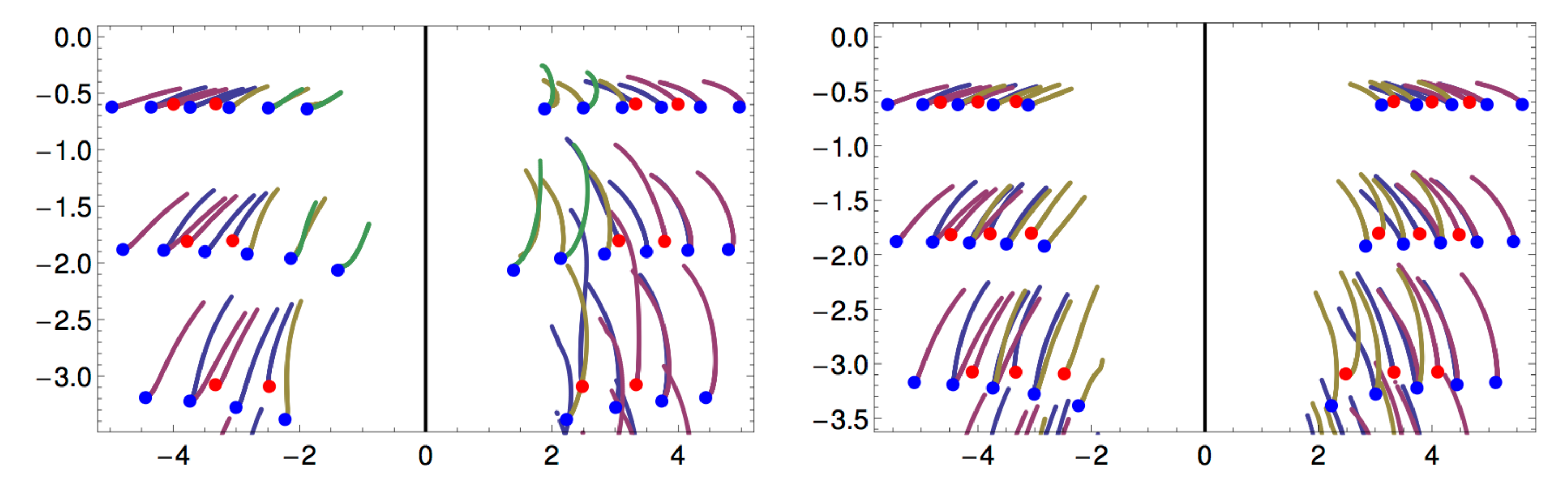}
\caption{\textbf{EAM MP.} \emph{Left Panel:} $\kappa =0$ vector QNM frequencies for $(m,\epsilon) = (1,1)$ (magenta), $(2,1)$ (blue), $(1,-1)$ (yellow), $(2,-1)$ (green). \emph{Right Panel:} $\kappa = 1$ vector QNM frequencies for $m = 0$ (magenta), and $(m,\epsilon) = (1,1)$ (blue), $(1,-1)$ (yellow). At zero rotation the curves connect to Schwarzschild vector modes (red dots) and tensor modes (blue dots) of various $\tilde{\ell}$ values.}
\label{Fig:mp_vectors}
\end{figure}

We now present our results for $d=7$. Since there are no special modes to single out, i.e. ones that become unstable, in Fig.~\ref{Fig:mp_vectors} we plot the complex QNM frequencies for some of the $d=7$ vector modes. Here we need to be careful. Using our explicit construction of the charged vector harmonics, we can verify that for $\kappa = 0$, one or both vector-derived tensors vanish (recall this also happened for the scalar perturbations). To discuss this, it is useful to introduce the parameter $\alpha = 2+m\epsilon$. For $\alpha = 0$, both vector-derived tensors vanish. For $\alpha > 0, \epsilon = \pm 1$, $V^{\pm} = 0$, and for $\alpha < 0, \epsilon = \pm 1$, $V^{\mp} = 0$.

%

%%%%%%%%%%%%%%%%%%%%%%%%
\subsection{Unstable black holes in the large-$d$ Limit} 
\begin{figure}[ht]
\centering
\includegraphics[width=0.7 \textwidth]{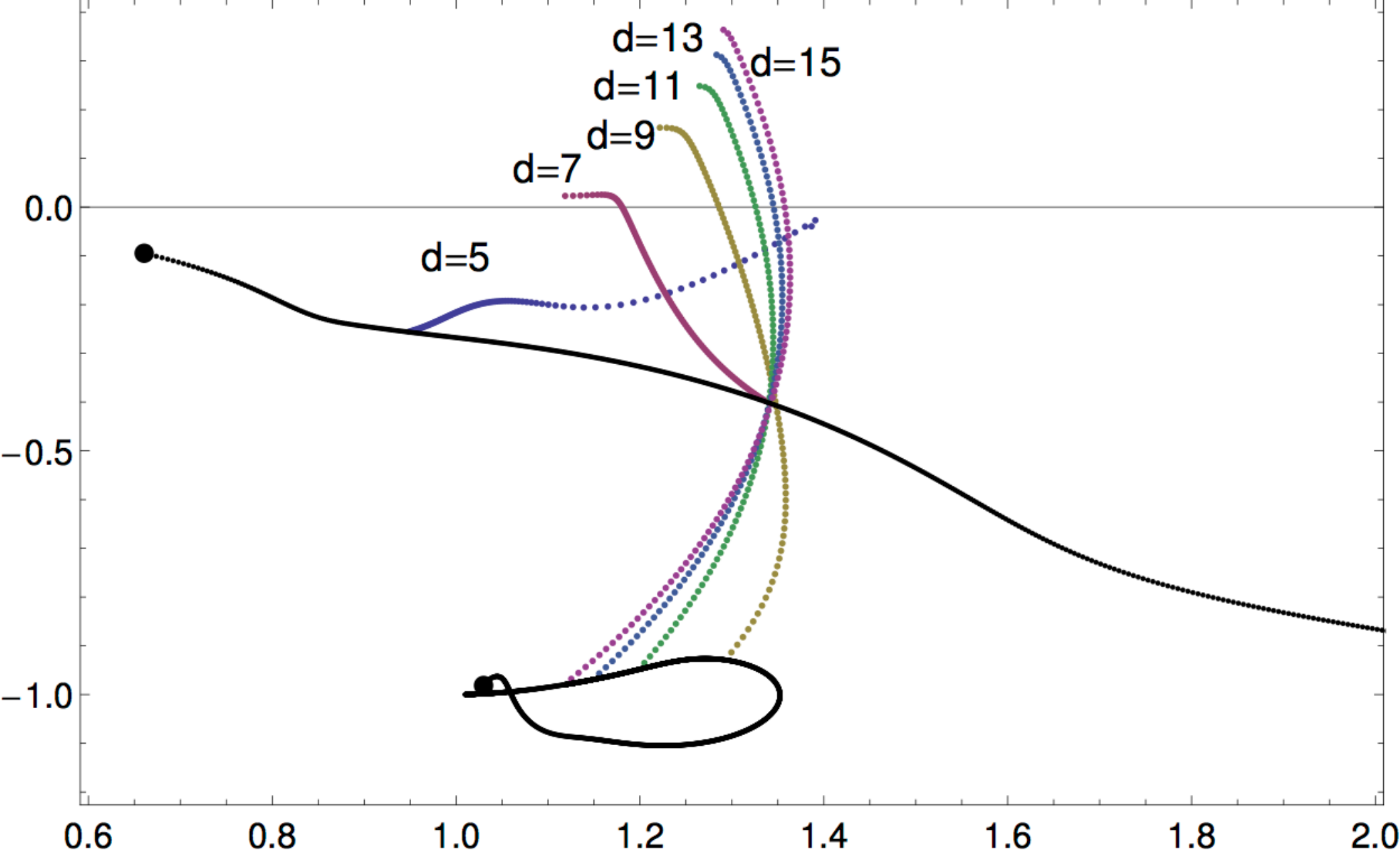}
\caption{\textbf{EAM MP.} The QNM frequency $\omega$ in the complex plane for different BHs. The curves in black represent the $\widetilde{\ell}_S = 2$ Schwarzschild frequencies (the smallest $d$ is indicated with a large dot. The dimension then increases along the curves). The top and bottom curves correspond to the scaling and saturating modes discussed in Sec.~\ref{sec:Schw}, respectively. The coloured curves correspond to the $(0,2)$ scalar frequencies for equal angular momenta MP BHs. At zero rotation the MP frequencies agree with the Schwarzschild modes, and as the rotation increases the frequencies move upwards in the complex plane, becoming unstable for $d\ge 7$. For $d=5,7$, the MP curves connect to the scaling Schwarzschild mode, while for $d\ge 9$ they connect to the saturating mode.}
\label{Fig:barmode}
\end{figure}

\noindent
The large range of dimensions studied in this section allows for some interesting observations concerning the large-$d$ limit. As the dimension grows, it seems like some of the QNM frequencies are approaching limiting values. In particular, the critical rotation seems to approach a limiting value. This motivates a connection with the saturating Schwarzschild modes, discussed in Sec.~\ref{sec:Schw}. In Fig.~\ref{Fig:barmode} we plot in black both the low-lying $\widetilde{\ell}_S = 2$ saturating and scaling QNM curves of the Schwarzschild solution in the complex $\omega$ plane. On top of this we also show, in color, the complex frequencies associated with the $(\kappa,m)=(0,2)$ bar-mode instability for equal angular momenta MP BHs. For $d=5,7$, the MP curves connect to the scaling Schwarzschild mode but, for $d=9,11,13,15$, they connect to the saturating Schwarzschild mode. Similar results hold for the other instabilities found. In particular, the $(0,m)$ bar-modes connect to the saturating Schwarzschild scalars with $\widetilde{\ell}_S = m$, and, for large enough dimension, the $(\kappa,m)=(2,0)$ ultraspinning mode connects to the saturating Schwarzschild vector mode with $\widetilde{\ell}_V = 3$. We expect the $(\kappa,m)=(1,1)$ instability to connect to a saturating vector for $\widetilde{\ell}_V = 5$, but we have not verified this, since as $\widetilde{\ell}$ increases it takes larger $d$ values to see the QNM begin to saturate. To summarize, it seems that in all cases we can study, the instabilities of these MP BHs, for sufficiently large $d$, can be connected to saturating Schwarzschild modes in the zero rotation limit. 

Thus, we see that the existence of saturating modes in the Schwarzschild geometry seems to be essential to allow for instabilities of rotating BHs for arbitrarily large-$d$. Indeed, if there were no such saturating modes, then an unstable mode of a large-$d$ MP BH would necessarily start off far from the origin in the complex plane, and would need to move a large distance in order to become unstable for a finite value of the rotation. This seems particularly unlikely for the equal angular momenta case as the rotation cannot be taken arbitrarily large. Therefore, any instability of equal angular momenta MP BHs that persists for arbitrarily large-$d$ is very likely connected to a saturating Schwarzschild mode. One of the motivations for studying the equal angular momenta case was the expectation that it might be representative of generic MP BHs which have no vanishing angular momenta. These BHs also have an upper bound on the rotations, and are likely to suffer from instabilities similar to the equal angular momenta BHs. \emph{Therefore, we conjecture that, for sufficiently large-$d$, all unstable modes of MP BHs with no vanishing angular momenta are connected to saturating Schwarzschild modes in the zero rotation limit.}

\begin{figure}[ht]
\centering
\includegraphics[width=0.5 \textwidth]{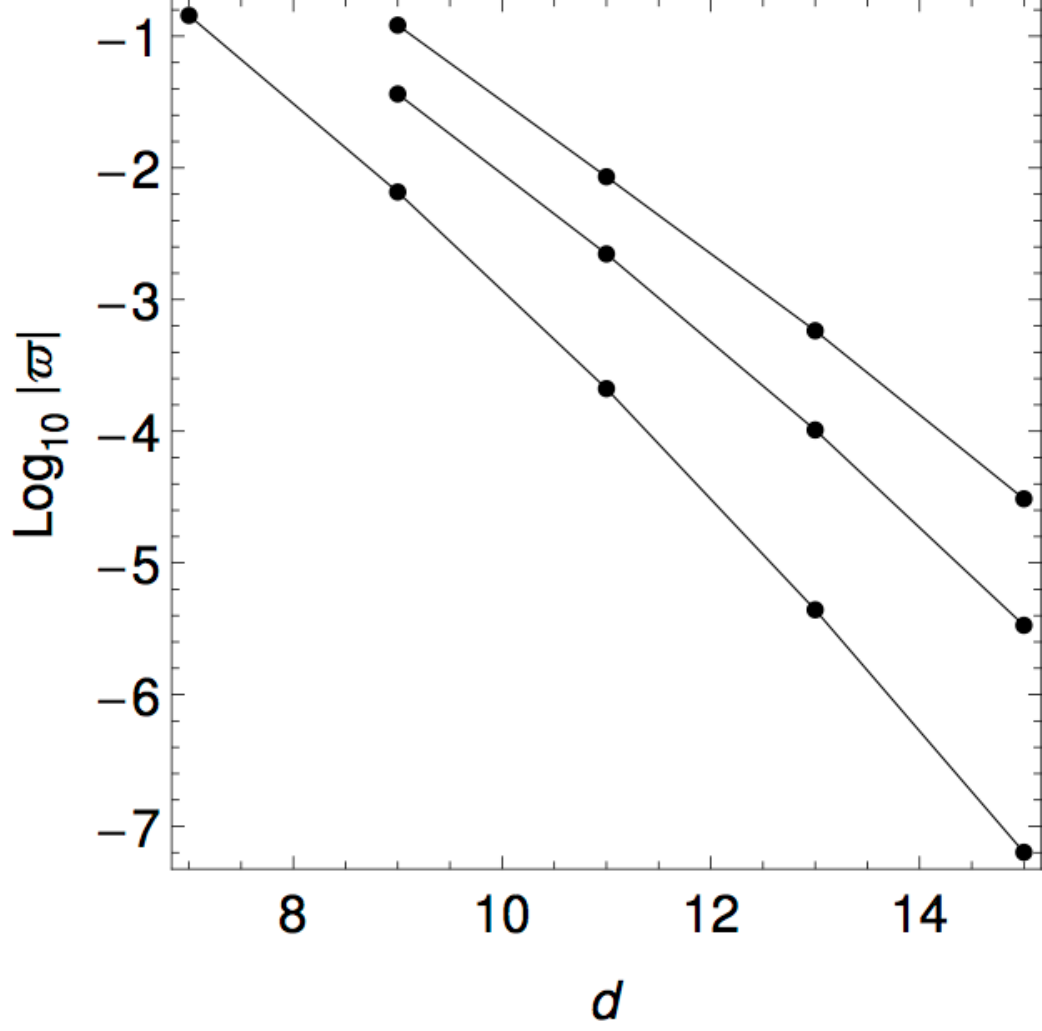}
\caption{\textbf{EAM MP.} The logarithm of the absolute value of the superradiant factor for the scalar (0,2) mode (bottom curve), (0,3) mode (middle curve), and (0,4) mode (top curve), plotted as a function of $d$.}
\label{Fig:mp_scalars_superradiant}
\end{figure}

Next, we discuss a second interesting feature of large-$d$ unstable BHs. Above we remarked that any unstable modes must satisfy $\varpi \leq 0$. Suppose that both $\varpi$ and Im$(\omega)$ cross zero at the same critical rotation for some bar-mode instability (so with $m\neq 0$). This is a particularly interesting possibility because then we exactly have $\omega = m \Omega_H$ at the threshold of the instability, and although the perturbation breaks the $t,\psi$-translational symmetries, it is invariant under the horizon-generating Killing field, $K= \partial_t + \Omega_H \partial_{\psi}$. This is not an academic assumption since it actually happens for superradiant instabilities in the Kerr-AdS BH, and it signals the existence of a new family of BHs that are neither axisymmetric nor time independent, but are invariant under the linear combination $K$  (see \cite{Kunduri:2006qa,Cardoso:2013pza} and references therein). It is thus interesting to investigate whether or not this is also the case for the asymptotically flat bar-mode instabilities, i.e. if  $\omega = m \Omega_H$ when Im$(\omega)=0$ in the MP BHs. In Fig.~\ref{Fig:mp_scalars_superradiant} we plot $\log_{10} (-\varpi)$ at the threshold of the instability (where Im$(\omega)=0$) for the $(0,m)$ bar-modes studied \footnote{The negative of the superradiant factor is considered because by the area law one must have $\varpi \le 0$ at the threshold.}. Interestingly, it appears that as the dimension increases, this factor quickly goes to zero \footnote{We do not observe a similar trend for the other bar-mode instability studied, the $(1,1)$ mode. It could be that such an effect is not present for this mode, or it could be that it is, and we simply don't have data for large enough dimensions to observe it.}. This agrees with analytic results for the large-$d$ limit of the $m-$bar instability \cite{Emparan:unpub}. Thus, in analogy with the single Killing field BHs that have been conjectured to exist in AdS \cite{Cardoso:2013pza}, it seems like large-$d$ asymptotically flat BHs only approximately allow for such solutions, with the approximation becoming better as $d$ increases.

In Sec.~\ref{sec:Schw} saturating QNMs of Schwarzschild were studied, and the existence of these modes supported the idea that gravity in the large-$d$ limit still retains some very interesting features. The results of this section further support this interpretation, as we have seen that some of the most interesting physics of rotating BHs, namely linearly instabilities, survives the large-$d$ limit.

%
%%%%%%%%%%%%%%%%

%%%%%%%%%%%%%%%%%%%%%%%%%%%%%%
%%%%%%%%%%%%%%%%%%%%%%%%%%%%%%
\section{Singly Spinning MP black holes\label{sec:IntrossMP}}
%%%%%%%%%%%%%%%%
The singly spinning MP BH  can be written as direct sum of the metrics of a 4-dimensional orbit space $g_{AB} \left( x^C\right) $, with $A,B,C=\{ t,r,\widetilde{x},\phi \}$, and  spherical fibres:
\begin{equation}\label{SMP:sumSpaces}
ds^2=g_{\mu\nu} dx^\mu dx^\nu=g_{AB}  \left( x^C\right) dx^A dx^B +R^2\left( x^C\right) d\Omega_{d-4}^2,
\end{equation} 
where $R=r \widetilde{x}$ and $d\Omega_{d-4}^2=\gamma_{i,j}d\hat{x}^i d\hat{x}^j $  is the line element of a unit-radius $(d-4)$-sphere (we will use small latin indices $i,j$ to describe the coordinates on the sphere). 
We can decompose perturbations on this background according to how they transform under diffeomorphisms of the sphere $S^{d-4}$. More concretely, an arbitrary metric perturbation $h_{\mu\nu}$ can be decomposed into perturbations of scalar, transverse vector, and transverse traceless tensor types on $S^{d-4}$.\footnote{\label{foot:harmonics} Note that the KI formalism for the Schwarzschild BH of Section \ref{sec:Schw}  uses the harmonic decomposition of perturbations on a $S^{d-2}$ sphere \cite{Kodama:2003jz}, while in the singly spinning MP background of this Section we will use a harmonic decomposition with respect to $S^{d-4}$. To distinguish the different dimensionality of these two families of harmonics we will use a different notation, i.e. ${\bf S},{\bf V}_a, {\bf T}_{ab}$ and  $\ell_S, \ell_V, \ell_T$  for the decomposition on the $S^{d-4}$ instead of the notation $S, V_a, T_{ab}$ and  $\widetilde{\ell}_S, \widetilde{\ell}_V, \widetilde{\ell}_T$ employed in the $S^{d-2}$ decomposition.} 
Ref. \cite{Kodama:2009bf} studied in detail the tensor sector of perturbations. They found no instability in this sector and computed its QNM frequencies. We will have nothing to add to this tensor sector analysis.
Our aim in this subsection is to study the spectrum of QNMs and instabilities in the scalar and vector sectors since this  study is missing. We are particularly motivated by the fact that two known instabilities of the MP BH, namely the bar-mode \cite{Emparan:2003sy,Shibata:2009ad,Shibata:2010wz} and ultraspinning \cite{Emparan:2003sy,Dias:2009iu,Dias:2010maa} instabilities  are precisely in the scalar sector of perturbations.   

We briefly summarize the novel results that emerge from the study we do next. We will find the (most relevant, low-lying) scalar and vector QNMs of the singly spinning MP BH and will compute the timescale of the bar-mode instability that is present in $d\geq 6$. As a major result we find that the $d=5$ singly spinning MP BH is linearly stable in the sense that we find no linear instability; in particular, we find that it does not have a linear bar-mode instability (a bar-mode instability was reported to be present also in the $d=5$ BH in  the time evolution study of \cite{Shibata:2009ad,Shibata:2010wz}).

Technically, we  find convenient to introduce the dimensioneless rotational parameter $\alpha$ and new coordinates $\{ T,y,x, x^i \}$  related to the standard coordinates  $\{ t,r,\widetilde{x}, x^i \}$  as 
\begin{equation}\label{SMP:coord}
\alpha=a/r_+\,; \qquad \quad t=r_+ T\,, \qquad  r=\frac{r_+}{1-y^2}\,, \qquad \widetilde{x}=x\sqrt{2-x^2},
\end{equation} 
where $y$ is a compact radial coordinate, $0 \leq y \leq 1$, with horizon at $y=0$ and asymptotic spatial infinity at $y=1$, and the new polar coordinate $x$ ranges between $0 \leq x \leq 1$.
With these new coordinates $g_{tt}$ and $g_{\phi\phi}$ vanish quadratically as $y^2$ and $(1-x^2)^2$ at the horizon and pole, respectively, as they should at a bolt. Moreover, in this coordinate frame the boundary conditions will be much simpler.
In these coordinates the geometry of the singly spinning MP BH reads
\begin{eqnarray}\label{SMP:ds2}
&&ds^2=\frac{\Delta_y \Sigma _y}{\rho_y}\,y^2 dT^2 + \frac{ 4 \Sigma_y}{\left(1-y^2\right)^4 \Delta _y}\,dy^2
 +\frac{4 \Sigma _y}{\left(2-x^2\right) \left(1-y^2\right)^2}\,dx^2  +\frac{x^2 \left(2-x^2\right)}{\left(1-y^2\right)^2}\,d\Omega_{d-4}^2   \nonumber\\
&&\hspace{1cm}+\frac{\left(1-x^2\right)^2 \rho _y}{\left(1-y^2\right)^2 \Sigma_y} \left(d\phi-\frac{\alpha  \left(1-y^2\right)^2 }{\rho _y} \left(\alpha ^2 \left(1-y^2\right)^2+1-y^2 \Delta _y\right) dT \right)^2
\end{eqnarray} 
where 
\begin{eqnarray}\label{SMP:ds2func}
&&\Delta_y=   \frac{\left(1-y^2\right)^2}{y^2} \frac{\Delta (y)}{r_+^2}\,, \qquad   
\Sigma_y=\alpha^2 x^2 \left(2-x^2\right) \left(1-y^2\right)^2+1\,, \nonumber\\
&&
\rho _y=\left[\alpha ^2 \left(1-y^2\right)^2+1\right]^2-\alpha ^2 \left(1-x^2\right)^2 \left(1-y^2\right)^2 y^2 \Delta _y\,.
\end{eqnarray} 

%%%%%%%%%%%%%%%%
\subsection{Scalar QNMs ($d\geq 5$). Bar-mode and ultraspinning instabilities \label{sec:singleMPscalar}}
%%%%%%%%%%%%%%

Scalar perturbations of the singly spinning MP background can be expanded in terms of a basis of scalar harmonic  $ {\bf S}$ on the unit sphere $S^{d-4}$ that solve the eigenvalue equation (see footnote \ref{foot:harmonics})
  \begin{equation}\label{SMP:scalarH} 
\left(\Box_{S^{d-4}}+\lambda_S \right) {\bf S} =0\,,
\end{equation} 
where $\lambda_S$ is the eigenvalue, and $\Box=D^iD_i$ with $D$ being the derivative defined by the metric $\gamma_{ij}$ of the base space $S^{d-4}$.
Regularity of the scalar harmonics requires 
 \begin{equation}\label{SMP:scalarH2} 
\lambda_S=\ell_S \left(\ell_S+d-5 \right)\,,\qquad \hbox{with} \quad \ell_S =0,1,2,\cdots.
\end{equation} 
Note that the angular base space is exactly a sphere. Therefore the perturbation equations and solutions  are independent of the azimuthal quantum number of the $S^{(d-4)}$. They only depend on the quantum number $\ell_S$ that measures the number of nodes along the polar direction of the $S^{(d-4)}$.
Out of this scalar harmonic we can construct a  scalar-type  vector harmonic  $ {\bf S}_i$ and a  traceless scalar-type tensor harmonic  $ {\bf S}_{ij}$ as 
\begin{equation}
   {\bf S}_i = -\frac{1}{\sqrt{\lambda_S}}  D_i  {\bf S} \,, \qquad 
   {\bf S}_{ij} = \frac{1}{\lambda_S}  D_i D_j  {\bf S}
                 + \frac{1}{d-4}\gamma_{ij}  {\bf S} \,.      
\end{equation}
Scalar perturbations are then given by 
\begin{eqnarray}\label{pert:scalar}
 h_{ab}=  f_{ab} e^{-i \omega t}e^{i m \phi }  {\bf S}, \quad  
 h_{ai}= f_a  e^{-i \omega t}e^{i m \phi }   {\bf S}_i  , \quad 
 h_{ij}= e^{-i \omega t}e^{i m \phi }\left( H_L\gamma_{ij}  {\bf S} + H_T  {\bf S}_{ij} \right) \,,
\end{eqnarray}
with $f_{ab},f_a,H_T,H_L$ functions of $\{r,\widetilde{x} \}$, and we used the fact that $\partial_t$ and $\partial_\phi$ are Killing vector fields of the background to do a Fourier decomposition along these directions.

We will restrict our analysis to $s$-wave modes, i.e. modes with $\ell_S=0$, which effectively means that we set $f_a=0=H_T$. This considerably reduces the computational cost of our task since we ``just" have to solve a  coupled PDE system of ten equations for ten variables $f_{ab},H_L$; a task that is itself already hard even numerically. Moreover, the most interesting properties of the scalar perturbations, namely the bar-mode and ultraspinning instabilities, are precisely in this $s$-wave sector.  

In the sequel we describe the procedure we find most tractable to solve the technical problem at hand. We find convenient to introduce the tetrad basis 
\begin{eqnarray}\label{SMP:tetrad}
&& e^{(1)}=dt - \alpha  \left(1-x^2\right)^2 d\phi\,,\qquad e^{(2)}=dy\,,\qquad  e^{(3)}=dx\,,\nonumber\\
&& e^{(4)}=  - \alpha  \left(1-y^2\right)^2 dt +\left[1+ \alpha^2  \left(1-y^2\right)^2 \right] d\phi \,,\qquad 
e^{(i)}=\hat{e}^{i}\,,
\end{eqnarray} 
where $\hat{e}^i$ are a vielbein for the metric $d\Omega_{(d-4)}^2$ of the unit-radius $(d - 4)$-sphere.
This basis has a diagonal metric (with elements that are not equal to unity) and considerably simplifies the computations.
The most general scalar perturbation has eleven non-vanishing tetrad components namely, $e^{-i \omega t}e^{i m  \phi}h_{(a)(b)}$, with $a,b=1,2,3,4$ and $h_{(i)(i)}=e^{-i \omega t}e^{i m \phi}h_{(\Omega)(\Omega)}$ for $i=5,\cdots d$. Since $h$ is a symmetric tensor this gives a total of 11 unknown functions.

We choose to work in the traceless-transverse (TT) gauge,
\begin{equation}\label{SMP:TT}
h^{(a)}_{\phantom{(a)}(a)}=0\,,\qquad \nabla^{(a)}h_{(a)(b)}=0\,.
\end{equation}
In this gauge the linearised Einstein equations read
\begin{equation} (\Delta_L h)_{(a)(b)} \equiv - \nabla_{(c)} \nabla^{(c)} h_{(a)(b)}
-2\, R_{(a)\phantom{(a)}{(b)}}^{\phantom{(a)}(c)\phantom{(b)}(d)}
h_{(c)(d)}=0 \,, \label{SMP:LichnerowiczEq}
\end{equation}
where  $\Delta_L$ is the Lichnerowicz operator and $R$ the Riemann tensor.

The traceless condition can immediately be used to eliminate $h_{(\Omega)(\Omega)}$ since it  can be written as an algebraic relation as a function of $\{ h_{(1)(1)}, h_{(2)(2)},h_{(3)(3)},h_{(4)(4)}\}$.
The transverse conditions give algebraic relations that could be used to eliminate further variables but we find that this yields complicated equations of motion that increase the numerical error in our computations.
Instead we identify the following system of PDEs
\begin{equation}\label{SMP:EOM}
\begin{aligned}
&(\Delta_L h)_{(1)(3)} =0\,,\qquad (\Delta_L h)_{(2)(2)} =0\,,\qquad (\Delta_L h)_{(2)(3)} =0\,,\\
&(\Delta_L h)_{(2)(4)} =0\,,\qquad(\Delta_L h)_{(3)(3)} =0\,,\qquad(\Delta_L h)_{(3)(4)} =0\,,\\
&\nabla^{(a)}h_{(a)(1)}=0\,,\qquad \nabla^{(a)}h_{(a)(2)}=0\,,\qquad \nabla^{(a)}h_{(a)(3)}=0\,,\quad \nabla^{(a)}h_{(a)(4)}=0\,,
\end{aligned}
\end{equation}
which give a system of 10 independent equations to solve for the 10 independent variables $h_{(a)(b)}$, with $a,b=1,2,3,4$. Note that we have explicitly checked that this system of 10 equations closes the Lichnerowicz system, i.e. that the equations \eqref{SMP:EOM} imply
that the remaining equations in \eqref{SMP:LichnerowiczEq} are also obeyed.

%%%%%%%%%%%%%%%%
\subsubsection{Boundary conditions \label{sec:singleMPscalar1}}
%%%%%%%%%%%%%%

To discuss the boundary conditions (BCs) on the future event horizon ${\cal H}^+$ we introduce the ingoing Eddington-Finkelstein (EF) coordinates $\{v,\widetilde{\phi}\}$ that are regular at ${\cal H}^+$
\begin{equation}\label{SMP:EF}
dT=dv-\frac{2 \left(1+\alpha ^2 \left(1-y^2\right)^2\right)}{y \left(1-y^2\right)^2 \Delta_y}\,dy\,,\qquad  d\phi= d\widetilde{\phi}-\frac{2 \alpha }{y \Delta_y}\,dy\,.
\end{equation} 
The BC at the future horizon requires that the metric components $h_{(a)(b)}$ are smooth functions of the ingoing EF coordinates $\{v,y,x,\widetilde{\phi},x^i\}$ at $y=0$. This requires the BCs
\begin{eqnarray}\label{SMP:confBCy0}
&& 
h_{(1)(1)}{\bigl |}_{y=0} \sim y^{-2\,i \,\frac{\omega -m \Omega_H}{4 \pi T_H}} H_{11}(x)\,,\qquad
h_{(1)(2)}{\bigl |}_{y=0} \sim y^{-2\,i \,\frac{\omega -m \Omega_H}{4 \pi T_H}-1} H_{12}(x)\,,
\nonumber\\
&& 
h_{(1)(3)}{\bigl |}_{y=0} \sim y^{-2\,i \,\frac{\omega -m \Omega_H}{4 \pi T_H}}H_{13}(x)\,,\qquad
h_{(1)(4)}{\bigl |}_{y=0}\sim y^{-2\,i \,\frac{\omega -m \Omega_H}{4 \pi T_H}} H_{14}(x)\,,
\nonumber\\
&& 
h_{(2)(2)}{\bigl |}_{y=0} \sim y^{-2\,i \,\frac{\omega -m \Omega_H}{4 \pi T_H}-2}H_{22}(x)\,,\qquad
h_{(2)(3)}{\bigl |}_{y=0} \sim  y^{-2\,i \,\frac{\omega -m \Omega_H}{4 \pi T_H}-1}H_{23}(x)\,,
\nonumber\\
&& 
h_{(2)(4)}{\bigl |}_{y=0} \sim  y^{-2\,i \,\frac{\omega -m \Omega_H}{4 \pi T_H}-1} H_{24}(x)\,,\qquad
h_{(3)(3)}{\bigl |}_{y=0} \sim y^{-2\,i \,\frac{\omega -m \Omega_H}{4 \pi T_H}} H_{33}(x)\,,
\nonumber\\
&& 
h_{(3)(4)}{\bigl |}_{y=0}  \sim y^{-2\,i \,\frac{\omega -m \Omega_H}{4 \pi T_H}-1} H_{34}(x)\,,\qquad
h_{(4)(4)}{\bigl |}_{y=0}  \sim y^{-2\,i \,\frac{\omega -m \Omega_H}{4 \pi T_H}} H_{44}(x)\,,
 \end{eqnarray} 
 where $H_{ab}(x)$ are smooth functions of $x$.

At spatial infinity, $y\to 1$, i.e.  $r\to\infty$, we demand that the perturbations preserve the asymptotic flatness of the spacetime. This means that they must decay strictly faster than the asymptotic Minkowski background asymptotic. In this asymptotic region, the  Lichnerowicz equation \eqref{SMP:LichnerowiczEq} reduces to $\Box h_{(a)(b)} =0$. We would like to solve this system to find the exact decay of the asymptotic solutions. {\it \`A priori} this is a hard task since, in the spherical coordinate system we work, the non-vanishing connections in the differential operator means that we have a coupled system of 10 differential equations to solve for. However, we can make our life considerably easier and get the desired result by solving a single ODE. The procedure is the following. Consider Minkowski spacetime and the perturbation components  $h_{ab}^\prime$ in Cartesian coordinates.  In these conditions the affine connection vanishes and $\Box h_{ab} ^\prime=0$ reduces simply to $\partial_c \partial^c h_{ab} ^\prime=0$. A Fourier decomposition  in the time direction, $h_{ab}^\prime=e^{-i \omega t}h_{ab}^\prime$ allows to write it as $\nabla^2 h_{ab}^\prime=-\omega^2 h_{ab}^\prime$. Using the spherical harmonic decomposition of the perturbation $h_{ab}^\prime=h_{ab}^\prime(r) Y_\ell(x)$ and writing the spatial operator in spherical coordinates  this equation reads 
\begin{equation}\label{SMP:waveInf}
\frac{1}{r^{d-2}} \,\partial_r \left(r^{d-2} \partial_r  h_{ab}^\prime\right)+\frac{\ell(\ell+d-3)}{r^{d-2}}\,h_{ab}^\prime=-\omega ^2 h_{ab}^\prime .
\end{equation} 
Near spatial infinity, $h_{ab}^\prime$ behaves as $h_{ab}^\prime{\bigl |}_{r\to\infty} \sim C_{in}\,  r^{-\frac{d-2}{2}} \, e^{-i \omega r}+C_{out} \, r^{-\frac{d-2}{2}} \, e^{i \omega r}$ where the amplitudes $\{C_{in} ,C_{out}\}$ are a function of $\{ A,B\}$. Asymptotically we want outgoing waves so we impose the BC $C_{in} =0$. We conclude that all the Cartesian components of the perturbation must decay asymptotically as $h_{ab}^\prime \sim r^{-\frac{d-2}{2}} e^{i \omega r}$ in order to have outgoing BCs that preserve the asymptotic Minkowski structure of the spacetime. We can now apply a coordinate transformation from Cartesian to our spherical coordinates $\{t,y,x,\phi,x_{S^d-4}\}$ to find the decays of $h_{(a)(b)}$ in the spherical frame. In this process we keep in mind that we are interested only in scalar perturbations so we do not consider the vector contributions here. We find that scalar perturbations with asymptotically Minkowski outgoing BCs behave as:

\begin{eqnarray}\label{SMP:confBCy1}
&& 
h_{(1)(1)}{\bigl |}_{y=1} \sim \left(1-y^2\right)^{\frac{d-2}{2}} e^{\frac{i \omega }{1-y^2}} H_{11}(x)\,,\qquad
h_{(1)(2)}{\bigl |}_{y=1} \sim \left(1-y^2\right)^{\frac{d-2}{2}-2} e^{\frac{i \omega }{1-y^2}} H_{12}(x)\,,
\nonumber\\
&& 
h_{(1)(3)}{\bigl |}_{y=1} \sim \left(1-y^2\right)^{\frac{d-2}{2}-1} e^{\frac{i \omega }{1-y^2}} H_{13}(x)\,,\qquad
h_{(1)(4)}{\bigl |}_{y=1} \sim  \left(1-y^2\right)^{\frac{d-2}{2}-1} e^{\frac{i \omega }{1-y^2}}H_{14}(x)\,,
\nonumber\\
&& 
h_{(2)(2)}{\bigl |}_{y=1} \sim \left(1-y^2\right)^{\frac{d-2}{2}-4} e^{\frac{i \omega }{1-y^2}}  H_{22}(x)\,,\qquad
h_{(2)(3)}{\bigl |}_{y=1} \sim  \left(1-y^2\right)^{\frac{d-2}{2}-3} e^{\frac{i \omega }{1-y^2}} H_{23}(x)\,,
\nonumber\\
&& 
h_{(2)(4)}{\bigl |}_{y=1} \sim  \left(1-y^2\right)^{\frac{d-2}{2}-3} e^{\frac{i \omega }{1-y^2}} H_{24}(x)\,,\qquad
h_{(3)(3)}{\bigl |}_{y=1} \sim \left(1-y^2\right)^{\frac{d-2}{2}-2} e^{\frac{i \omega }{1-y^2}}H_{33}(x)\,,
\nonumber\\
&& 
h_{(3)(4)}{\bigl |}_{y=1}  \sim \left(1-y^2\right)^{\frac{d-2}{2}-2} e^{\frac{i \omega }{1-y^2}}H_{34}(x)\,,\qquad
h_{(4)(4)}{\bigl |}_{y=1}  \sim \left(1-y^2\right)^{\frac{d-2}{2}-2} e^{\frac{i \omega }{1-y^2}}H_{44}(x)\,,\nonumber\\
&& 
\end{eqnarray} 
 where $H_{ab}(x)$ are smooth functions of $x$.
 
To find the BCs at the equator, $x=0$ where $g_{\Omega\Omega}\to 0$, we require  that the metric perturbation $h_{ab}dx^a dx^b$ is a regular symmetric 2-tensor when expressed in coordinates where the background metric is regular. A procedure similar to the one described in detail in Section 3.3 of \cite{Dias:2010maa} yields that smooth BCs at $x=0$ require that
\begin{eqnarray}\label{SMP:confBCx0}
&& 
h_{(a)(b)}{\bigl |}_{x=0} \sim x H_{ab}(y)\,,\quad \hbox{for}\quad  (a)(b)=\{ (1)(3),(2)(3),(3)(4)\};\nonumber\\
&& 
h_{(a)(b)}{\bigl |}_{x=0} \sim H_{ab}(y),\quad \hbox{otherwise}
 \end{eqnarray} 
 where $H_{ab}(y)$ are smooth functions of $y$.

Finally we discuss the BCs that the metric perturbations must satisfy at the axis of rotation, $x=1$, where $\partial_\phi$ vanishes. Near $x=1$, a generic component of the metric behaves as $h_{(a)(b)}=(1-x)^{\beta_j}h_{(a)(b)}(y) $ for some constant $\beta_j$ that generically is a function of the azimuthal quantum number $m$ associated with the Killing field $\partial_\phi$.  Our task is to find the ten $\beta_j$'s that yield smooth perturbations at $x=1$. These can be determining introducing the Cartesian coordinates $\{X,Y\}$ as $\rho\equiv 1-x=\sqrt{X^2+Y^2}$ and $\phi=\hbox{ArcTan}\left(Y/X\right)$ and then requiring the absence of non-analytical or divergent contributions (e.g. of the type $\sqrt{X^2+Y^2}$ or $\left( X\pm i\,Y\right)^{-1}$) on each component of the metric perturbation  in this frame. This requires the BCs at $x=1$:
\begin{eqnarray}\label{SMP:confBCx1}
&& 
h_{(1)(1)}{\bigl |}_{x=1} \sim \left(1-x\right)^{m}  H_{11}(y)\,,\qquad
h_{(1)(2)}{\bigl |}_{x=1} \sim   \left(1-x\right)^{m}  H_{12}(y)\,, \nonumber\\
&& 
h_{(1)(3)}{\bigl |}_{x=1} \sim   \left(1-x\right)^{m-1} H_{13}(y)\,,\qquad
h_{(1)(4)}{\bigl |}_{x=1} \sim \left(1-x\right)^{m} H_{14}(y)\,,\nonumber\\
&& 
h_{(2)(2)}{\bigl |}_{x=1} \sim \left(1-x\right)^{m}  H_{22}(y)\,,\qquad
h_{(2)(3)}{\bigl |}_{x=1} \sim \left(1-x\right)^{m-1} H_{23}(y)\,,
\nonumber\\
&& 
h_{(2)(4)}{\bigl |}_{x=1} \sim \left(1-x\right)^{m}  H_{24}(y)\,,\qquad
h_{(3)(3)}{\bigl |}_{x=1} \sim \left(1-x\right)^{\beta_1}  H_{33}(y)\,,
\nonumber\\
&& 
h_{(3)(4)}{\bigl |}_{x=1}  \sim \left(1-x\right)^{\beta_2}  H_{34}(y)\,,\qquad
h_{(4)(4)}{\bigl |}_{x=1}  \sim \left(1-x\right)^{\beta_3}  H_{44}(y)\,,
\end{eqnarray} 
 where $H_{ab}(y)$ are smooth functions of $y$ and the exponents $\beta_{1,2,3}$ depende on $m$ and are given by
\begin{equation}\label{SMP:confBCx1aux} 
\left\{
\begin{array}{ll}
\{\beta_1,\beta_3,\beta_3\}= \{1,2,3\} \,, & \qquad \hbox{if} \quad m=1\,, \\
\{\beta_1,\beta_3,\beta_3\}= \{m-2,m-1,m \} \,, & \qquad \hbox{if} \quad m\geq 2\,, 
\end{array}
\right.
\end{equation} 
For reasons that will be explained in Section \ref{sec:Conc}, we will not present results for $m=0$ scalar modes.
%%%%%%%%%%%%%%%%
\subsubsection{Numerical procedure  \label{sec:singleMPscalar2}}
%%%%%%%%%%%%%%

To solve numerically the equations of motion it is a good idea to factor out the singularities and/or leading behaviour identified in \eqref{SMP:confBCy0}, \eqref{SMP:confBCy1}, \eqref{SMP:confBCx0} \eqref{SMP:confBCx1}  which allows to work with manifestly analytic functions. Hence we introduce the new independent variables $q_1,\cdots,q_{10}$ defined as
\begin{eqnarray}\label{SMP:qs}
&& 
h_{(1)(1)}  = \left(1-x\right)^{m}  y^{-2\,i \,\frac{\omega -m \Omega_H}{4 \pi T_H}} \left(1-y^2\right)^{\frac{d-2}{2}} e^{\frac{i \omega }{1-y^2}}  q_1\,,\nonumber\\
&&
h_{(1)(2)} =  \left(1-x\right)^{m} y^{-2\,i \,\frac{\omega -m \Omega_H}{4 \pi T_H}-1} \left(1-y^2\right)^{\frac{d-2}{2}-2} e^{\frac{i \omega }{1-y^2}}  q_2,,
\nonumber\\
&& 
h_{(1)(3)} = x \left(1-x\right)^{m-1}  y^{-2\,i \,\frac{\omega -m \Omega_H}{4 \pi T_H}} \left(1-y^2\right)^{\frac{d-2}{2}-1} e^{\frac{i \omega }{1-y^2}} q_3  \,,\nonumber\\
&&
h_{(1)(4)} = \left(1-x\right)^{m}  y^{-2\,i \,\frac{\omega -m \Omega_H}{4 \pi T_H}} \left(1-y^2\right)^{\frac{d-2}{2}-1} e^{\frac{i \omega }{1-y^2}} q_4 \,,
\nonumber\\
&& 
h_{(2)(2)} =  \left(1-x\right)^{m}  y^{-2\,i \,\frac{\omega -m \Omega_H}{4 \pi T_H}-2} \left(1-y^2\right)^{\frac{d-2}{2}-4} e^{\frac{i \omega }{1-y^2}} q_5 \,,\nonumber\\
&&
h_{(2)(3)} = x \left(1-x\right)^{m-1} y^{-2\,i \,\frac{\omega -m \Omega_H}{4 \pi T_H}-1} \left(1-y^2\right)^{\frac{d-2}{2}-3} e^{\frac{i \omega }{1-y^2}} q_6\,,
\nonumber\\
&& 
h_{(2)(4)} = \left(1-x\right)^{m} y^{-2\,i \,\frac{\omega -m \Omega_H}{4 \pi T_H}-1}  \left(1-y^2\right)^{\frac{d-2}{2}-3} e^{\frac{i \omega }{1-y^2}} q_7 \,,\nonumber\\
&&
h_{(3)(3)} = \left(1-x\right)^{\beta_1} y^{-2\,i \,\frac{\omega -m \Omega_H}{4 \pi T_H}} \left(1-y^2\right)^{\frac{d-2}{2}-2} e^{\frac{i \omega }{1-y^2}} q_8 \,,
\nonumber\\
&& 
h_{(3)(4)} = x \left(1-x\right)^{\beta_2} y^{-2\,i \,\frac{\omega -m \Omega_H}{4 \pi T_H}-1} \left(1-y^2\right)^{\frac{d-2}{2}-2} e^{\frac{i \omega }{1-y^2}} q_9 \,,\nonumber\\
&&
h_{(4)(4)} = \left(1-x\right)^{\beta_3} y^{-2\,i \,\frac{\omega -m \Omega_H}{4 \pi T_H}}  \left(1-y^2\right)^{\frac{d-2}{2}-2} e^{\frac{i \omega }{1-y^2}} q_{10}\,.
 \end{eqnarray} 
 where $\beta_{1,2,3}$ are defined in \eqref{SMP:confBCx1aux}.
 
Introducing these new definitions in the equations of motion \eqref{SMP:EOM}, and solving these equations using a standard Taylor expansion around each of the four boundaries, it is now straightforward to find the final BCs we need  to impose in each of the variables $q_j(x,y)$. Namely, we find simple Robin, Neumann or Dirichlet BCs for all $q_j$'s at the equator $x=0$, axis of rotation $x=1$ and horizon $y=0$. At the  asymptotic boundary, $y=1$ some of the $q_j$'s obey Dirichlet BCs and the others are subject to less simple Robin BCs. 

The equations of motion \eqref{SMP:EOM} constitute a coupled system of ten partial differential equations that forms a quadratic eigenvalue problem in the frequency $\omega$, for a given mode $m$. To solve this eigenvalue problem, 
we use a pseudospectral collocation procedure to descretize our PDE system. We use a collocation grid, in the $x$ and $y$ directions, on Gauss-Chebyshev-Lobbato points. 
Alternatively, to check results and especially when we want to increase the accuracy of our results at lower computational cost, we use the novel numerical procedure introduced and described in \cite{Cardoso:2013pza} and already used in previous sections. This numerical method is based on the Newton-Raphson root-finding algorithm that searches for specific QNMs, once a seed solution is given.

As described in detail in Section \ref{sec:Schw}, we have also written an independent code
 to search directly for the QNMs of the Schwarzschild BH using the Kodama-Ishibashi (KI) decomposition on a $S^{d-2}$\cite{Kodama:2003jz}.
These independent results are very useful to check the numerical results we get with the codes for the single spin MP BH when the rotation vanishes. Recall that the KI formalism for the Schwarzschild BH uses the harmonic decomposition of perturbations on a $S^{d-2}$ sphere \cite{Kodama:2003jz} (while in the our spinning case we use a harmonic decomposition with respect to $S^{d-4}$). There are scalar, vector and tensor KI modes specified by quantum numbers that we will denote as $\widetilde{\ell}_S,\widetilde{\ell}_V,\widetilde{\ell}_T$, respectively. Non-trivial KI perturbations are described by integer $\widetilde{\ell}_S \geq 2,\widetilde{\ell}_V\geq 2,\widetilde{\ell}_T\geq 1$. The results from the KI code will be useful also to establish the relation between the harmonic decomposition on $S^{d-4}$, and associated quantum numbers $(\ell_S,m)$, that we use in our analysis of the single spin MP BH and the KI harmonic decomposition on $S^{d-2}$ and its quantum numbers  $\widetilde{\ell}_S,\widetilde{\ell}_V,\widetilde{\ell}_T$. In Section \ref{sec:Schw} we were mainly interested in the large-$d$ limit of Schwarzschild QNMs, while in the present section we will be interested in the results for $d=5,6,7$.

%%%%%%%%%%%%%%%%
\subsubsection{Results \label{sec:singleMPscalar3}}
%%%%%%%%%%%%%%

To discuss the results, first note that when the rotation vanishes, the Schwarzschild background has the symmetry $t \to -t $; consequently the associated QNM frequencies always come in trivial pairs of $\{\omega,-\omega^* \}$.  The $t \to -t $ symmetry is broken when the rotation is turned-on and for each pair of angular quantum numbers $\{\ell_S,m\}$ we have a pair of modes that is no longer trivially related.  However the singly spinning MP BHs have the $t -\phi$ symmetry and thus we can focus our attention only on modes with $m\geq 0$, say. Indeed there are two family of modes for each $m$, one with Re$(\omega)>0$ and the other with Re$(\omega)<0$. It follows from the  $t -\phi$ symmetry (i.e. the symmetry under the transformation $\{t,\phi \} \rightarrow \{-t,-\phi \}$)  that modes with negative azimuthal number $-m$ just trade the sign of the frequencies of the $m>0$ families. 

A second observation is that the QNM spectrum of scalar modes of the single spin MP BH is populated with many QNM, even when we reduce our search to a specific pair of $(\ell_S,m)$. The reason being that each pair of angular quantum numbers still has an infinite family of modes that have different radial overtone (i.e wavefunctions with different number of radial zeros). Typically, the modes with lowest radial overtone are the ones with lower frequency and in particular with smaller absolute value for the imaginary part of the frequency. Therefore they are the most important ones in a time evolution and we will typically focus our analysis in the modes with lower overtone.

Typically, we will express our results using the dimensionless frequency $\omega r_+$ and dimensionless rotation $a/r_+$. However, sometimes we will also find convenient to express our results in terms of the dimensionless quantities  $\omega r_0$ and dimensionless rotation $a/r_0$ where $r_0$ is the mass radius related to the horizon radius $r_+$ by   
\begin{equation}\label{SMP:r0rp}
r_0=r_+ \left(\frac{a^2}{r_+^2}+1\right)^{\frac{1}{d-3}}.
\end{equation} 
Finally note that we choose to present results only for $d=5,6,7$ dimensions since we expect higher dimensions to have a behavior that is similar to the cases $d=6,7$.

\begin{figure}[ht]
\centering
\includegraphics[width=.47\textwidth]{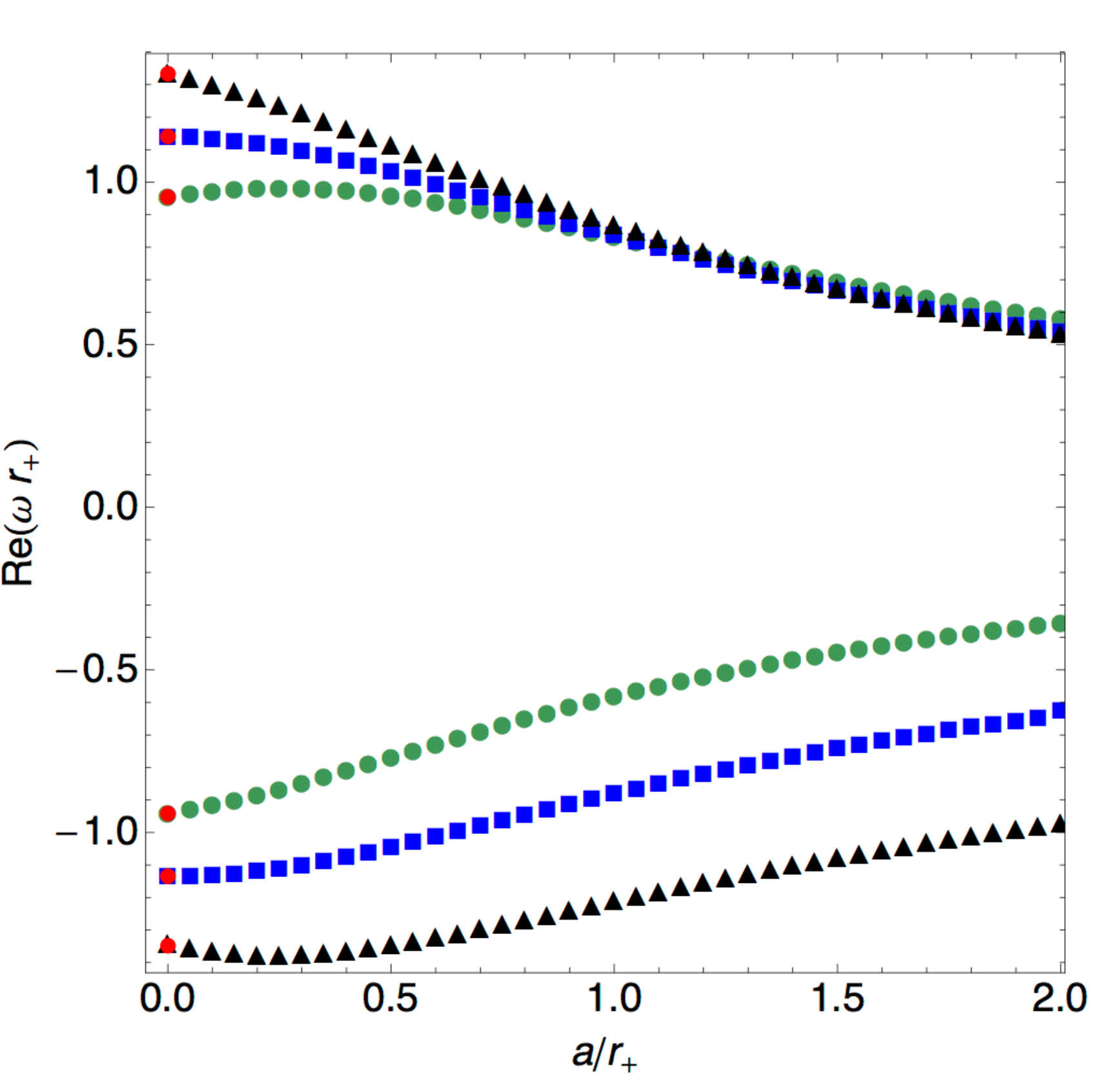}
\hspace{0.5cm}
\includegraphics[width=.47\textwidth]{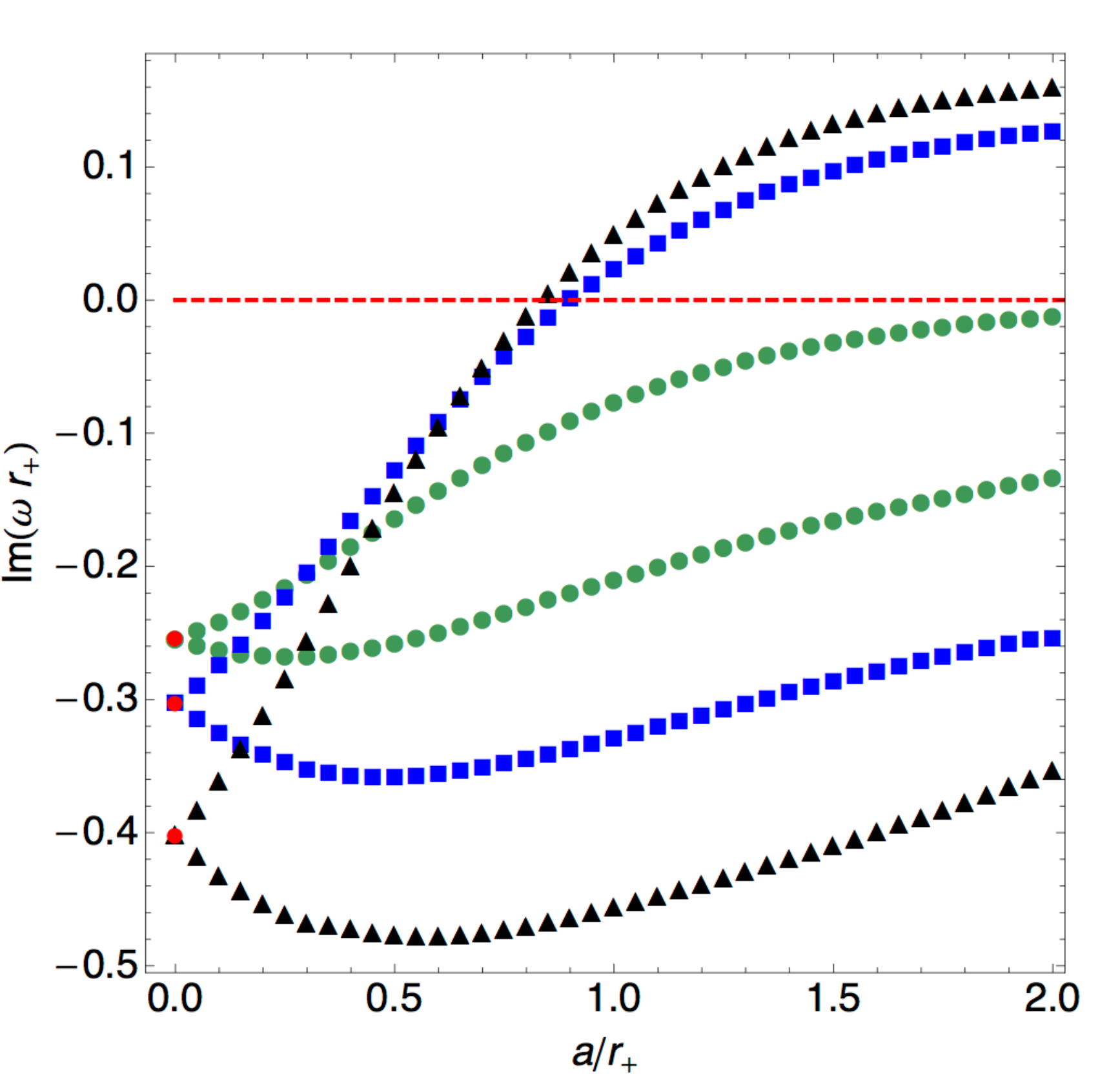}
\caption{{\bf Single MP.} Scalar modes with $(\ell_S,m)=(0,2)$ for $d=5$ (green disks), $d=6$ (blue squares) and $d=7$ (black triangles). These modes reduce to the  $\widetilde{\ell}_S=2$ KI scalar QNMs when $a/r_+\to 0$ (pinpointed as red dots; see Table \ref{Table:SchwQNMs}). 
}\label{Fig:scalarm2alld}
\end{figure}

\begin{figure}[ht]
\centering
\includegraphics[width=.47\textwidth]{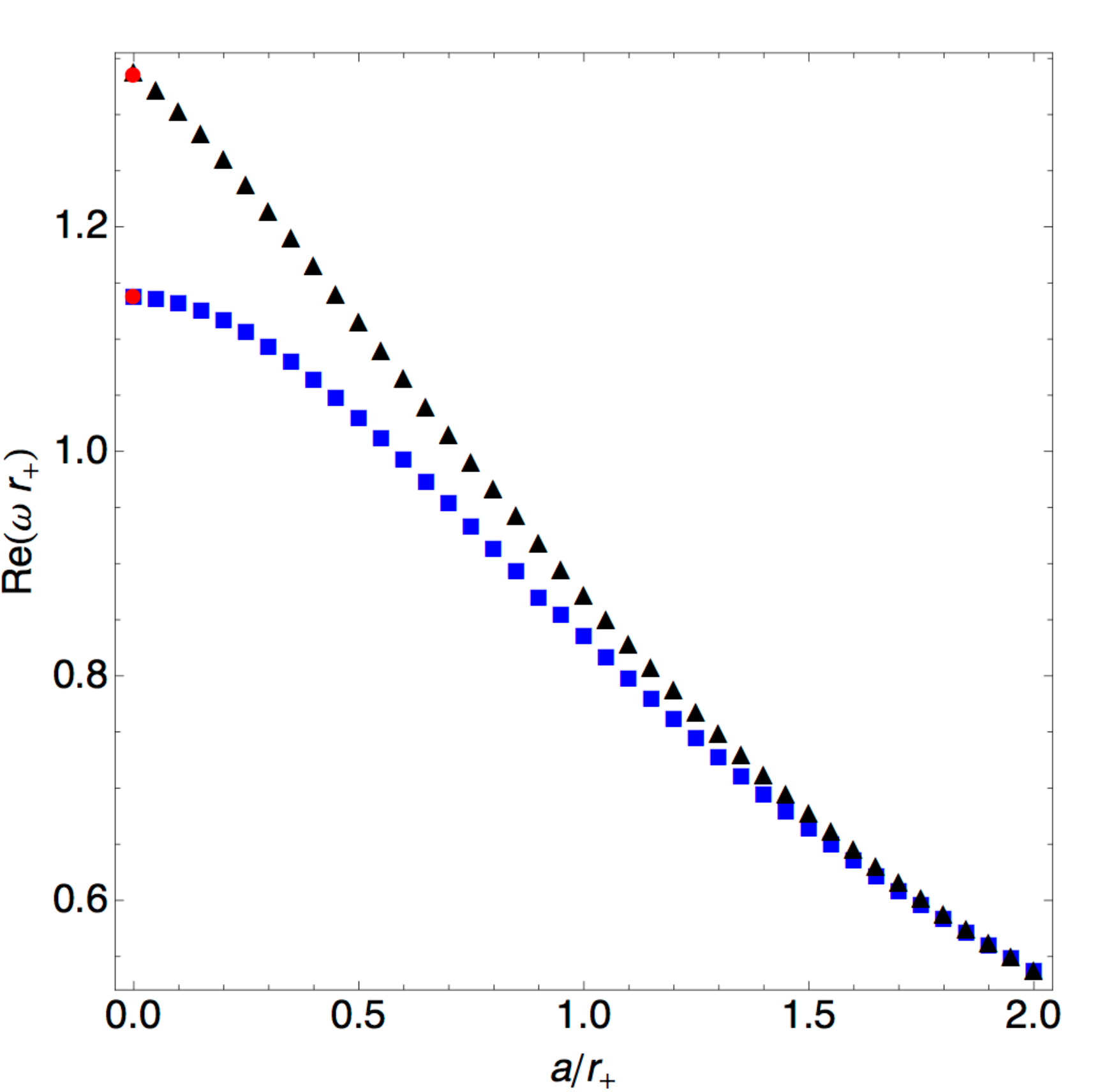}
\hspace{0.5cm}
\includegraphics[width=.485\textwidth]{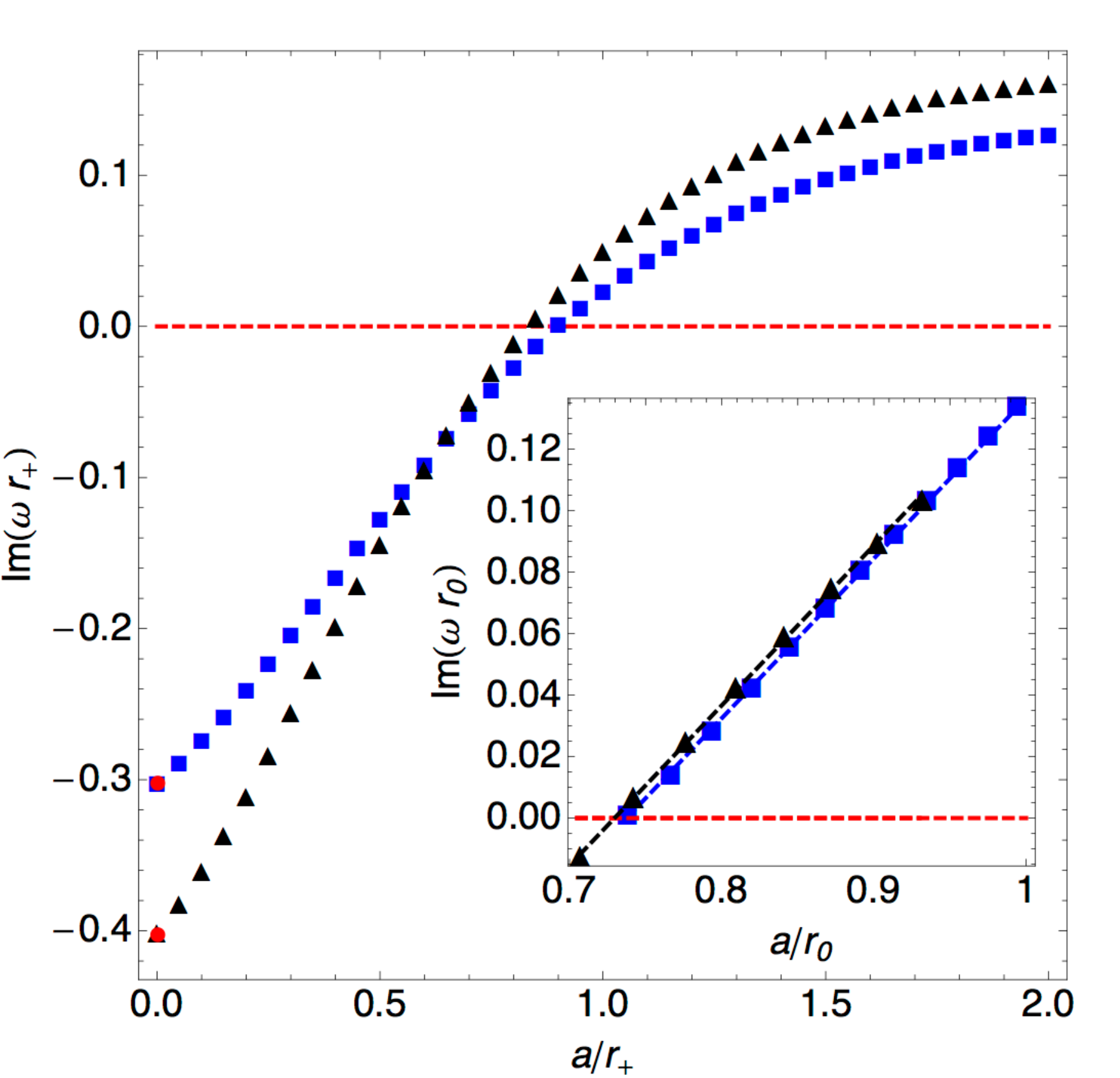}
\caption{{\bf Single MP.}  bar-mode instability in $d=6$ (blue squares) and $d=7$ (black triangles). {\it Left Panel:} Real part of the frequency ${\rm Re}(\omega r_+)$ as a function of the dimensionless rotation $a/r_+$. {\it Right Panel:} Imaginary part of the frequency ${\rm Im}(\omega r_+)$ as a function of $a/r_+$. 
The inset plot in the {\it Right Panel} details the bar-mode instability in $d=6$ and $d=7$ near the onset (this time in mass radius $r_0$ units) and is further discussed in the text.}\label{Fig:scalarm2barmode}
\end{figure}

\begin{figure}[ht]
\centering
\includegraphics[width=.47\textwidth]{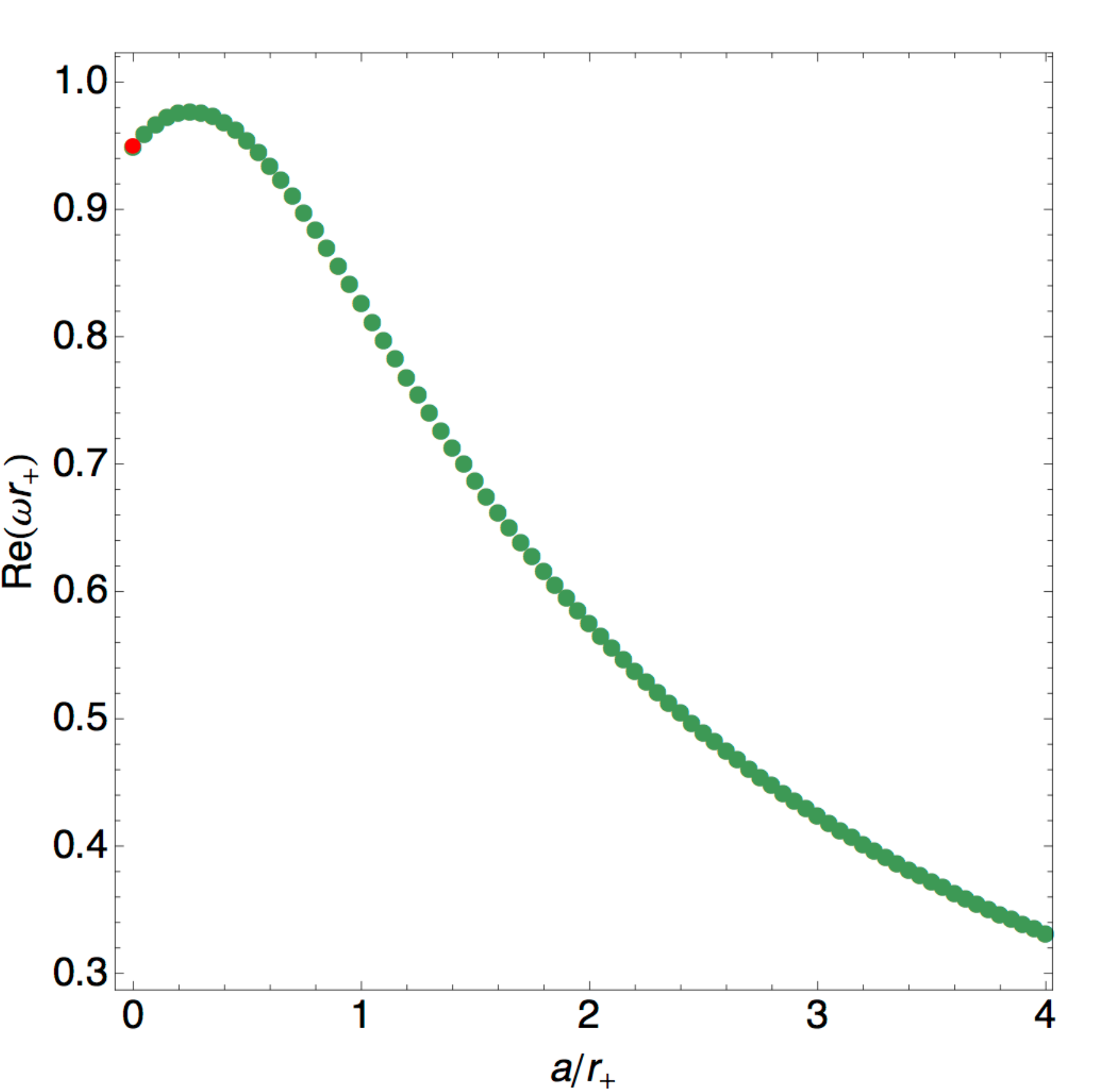}
\hspace{0.5cm}
\includegraphics[width=.485\textwidth]{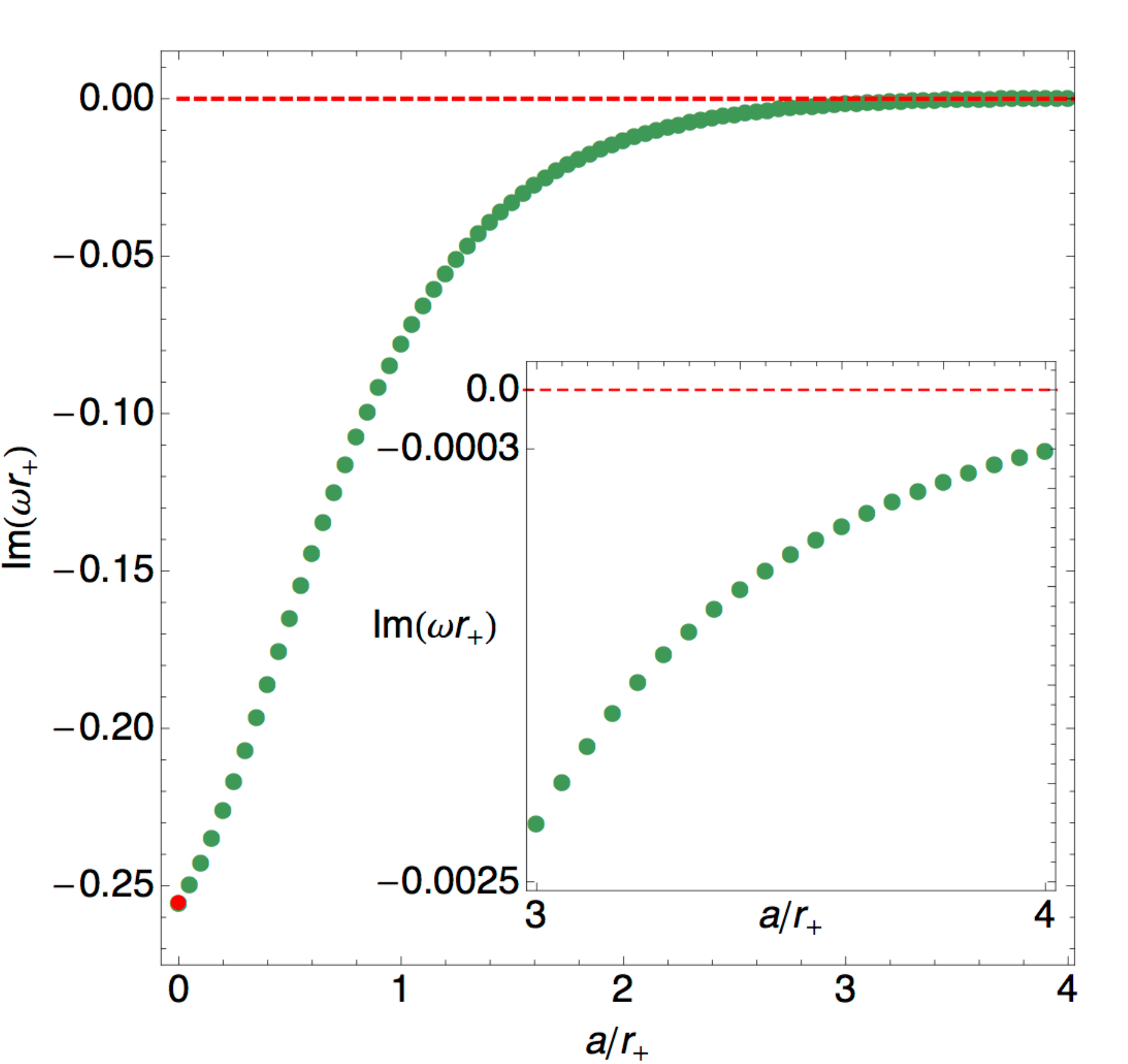}
\caption{{\bf Single MP.} Scalar modes with $(\ell_S,m)=(0,2)$ for $d=5$. There is no linear bar-mode instability.}\label{Fig:scalarm2d5}
\end{figure}

\begin{figure}[ht]
\centering
\includegraphics[width=.47\textwidth]{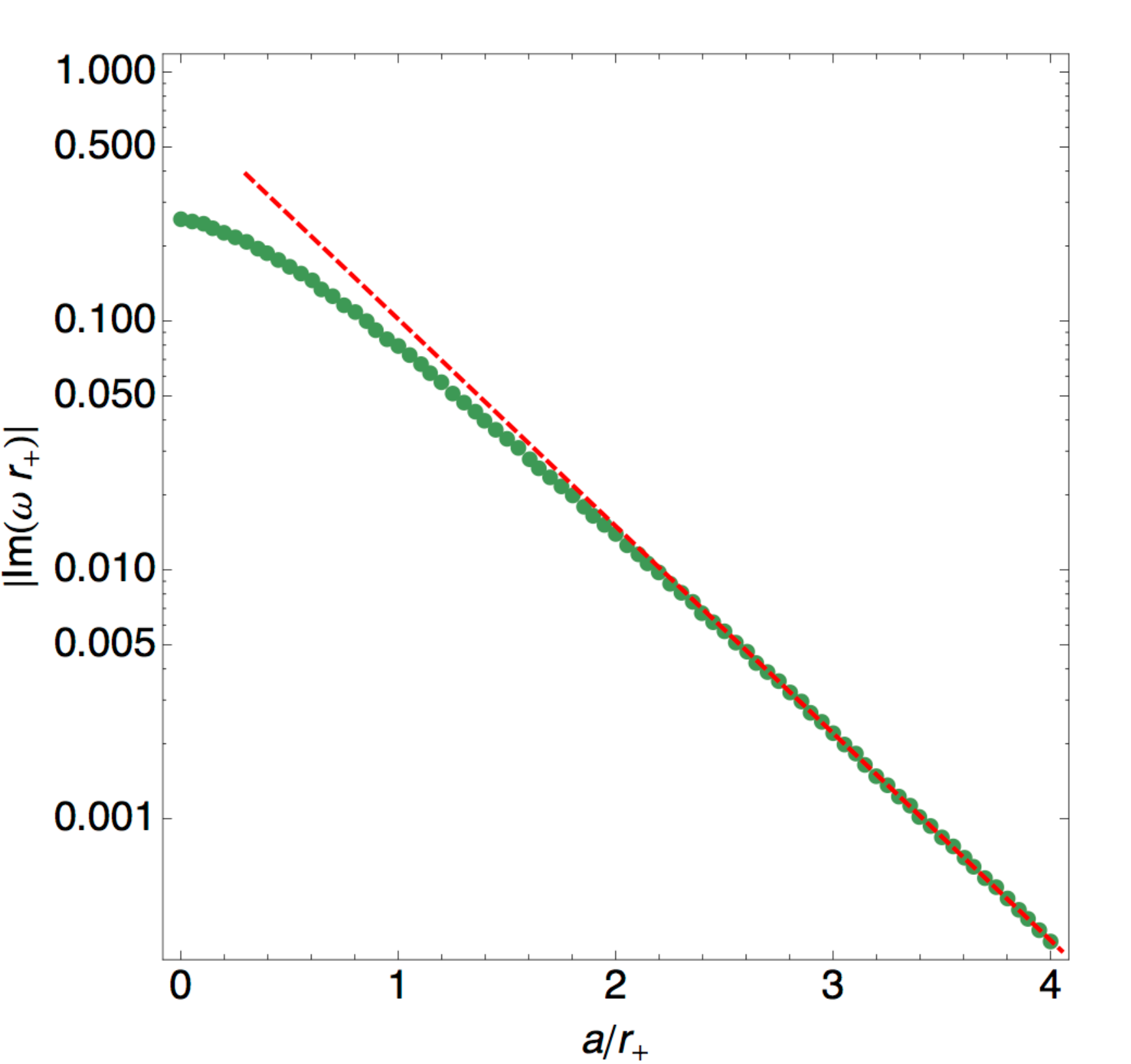}
\hspace{0.5cm}
\includegraphics[width=.47\textwidth]{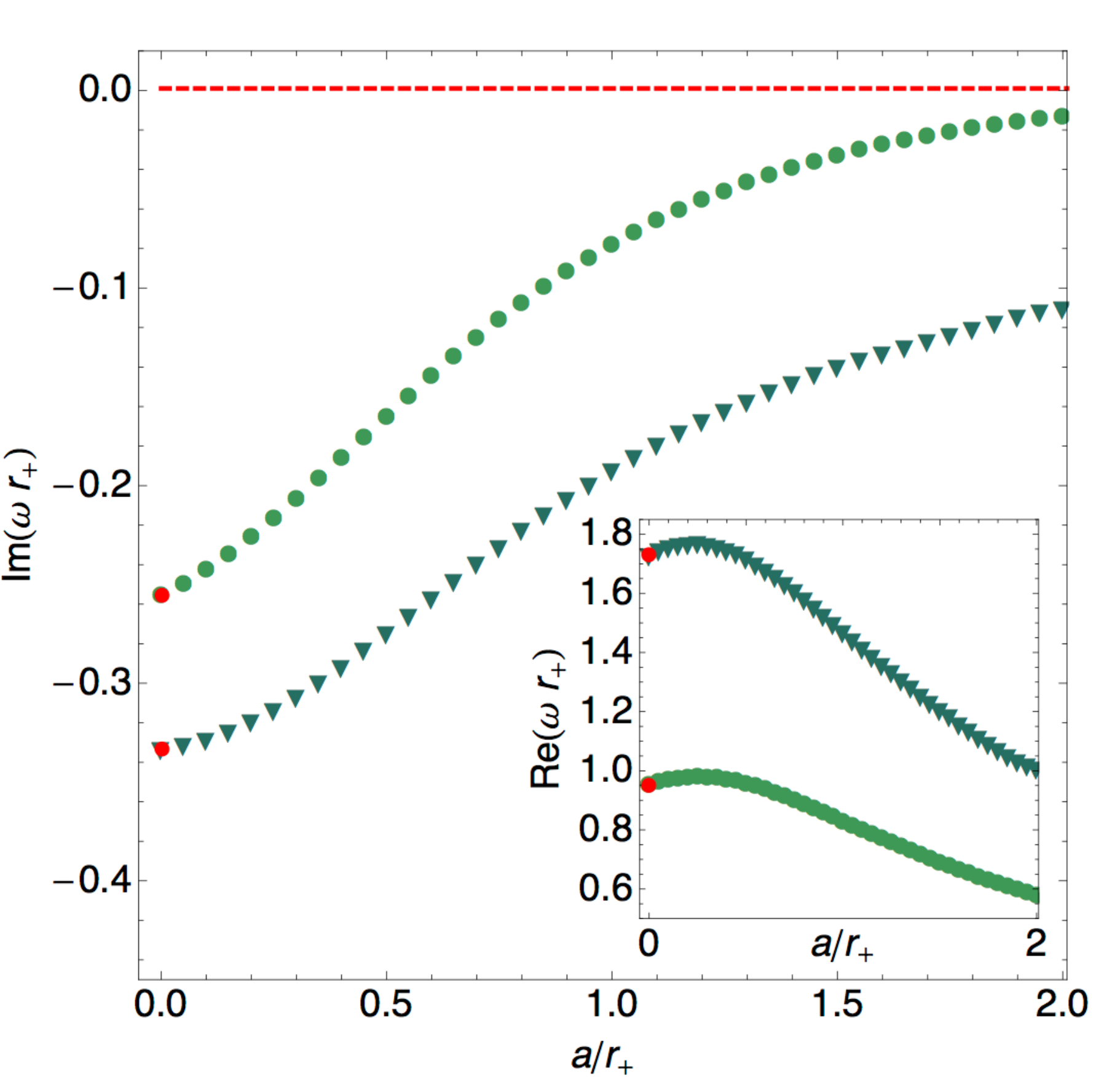}
\caption{{\bf Single MP.}  {\it Left Panel:}  Log plot for the scalar modes with $(\ell_S,m)=(0,2)$ in $d=5$. We find that 
${\rm Im}\left( \omega r_+\right)=A_0 e^{-\gamma a/r_+}$ with $\gamma\sim 1.9$ and $A_0\sim 0.69$. 
{\it  Right Panel:} Scalar modes with $(\ell_S,m)=(0,2)$ in $d=5$. The green disk curve is the one already shown in Fig.~\ref{Fig:scalarm2d5} and in the \emph{left panel}  of this Fig. that connects to the KI scalar mode with $\widetilde{\ell}_S=2$. The dark-green triangle curve is also a $(\ell_S,m)=(0,2)$ but with higher overtone that connects to the KI vector mode with $\widetilde{\ell}_V=3$ when $a/r_+\to 0$ (see  Table \ref{Table:SchwQNMs}).}
\label{Fig:scalarm2d5log}
\end{figure}

\begin{figure}[ht]
\centering
\includegraphics[width=.46\textwidth]{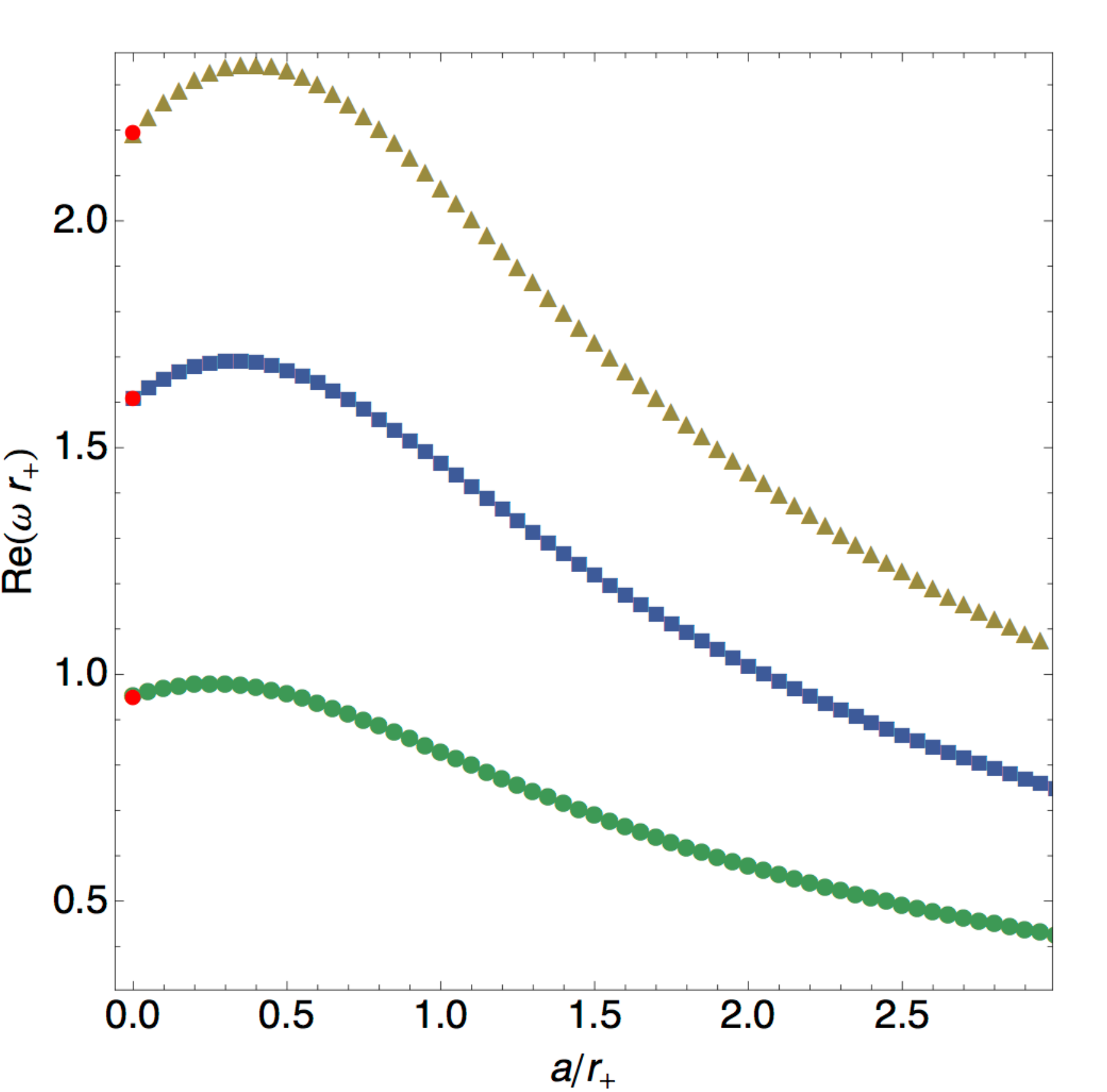}
\hspace{0.5cm}
\includegraphics[width=.485\textwidth]{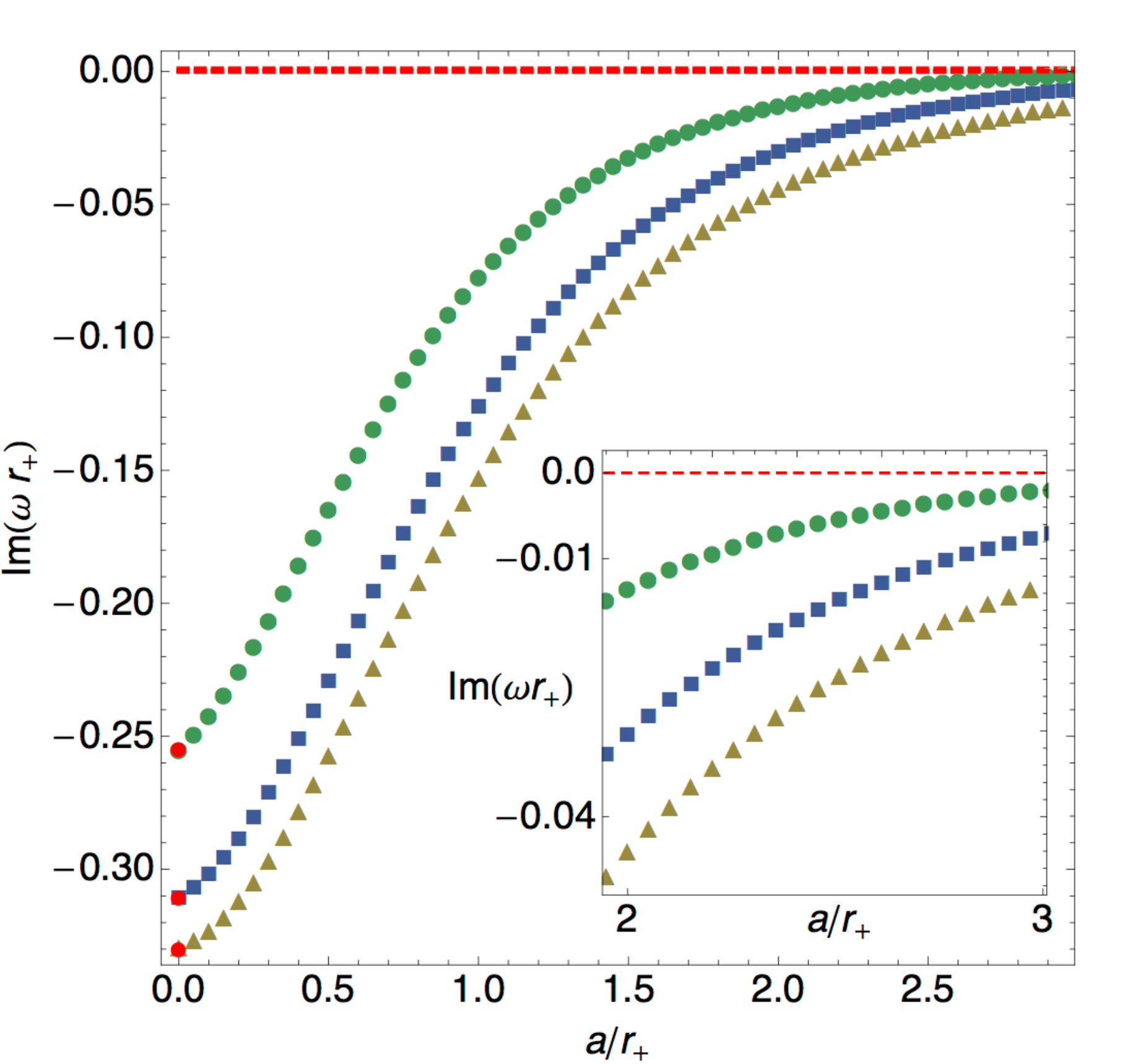}
\caption{{\bf Single MP.}  Scalar modes with $(\ell,m)=(0,2)$, $(\ell,m)=(0,3)$ and $(\ell,m)=(0,4)$  for $d=5$. 
The $(\ell,m)=(0,2)$ mode is the green disk curve already shown in previous plots while the $(\ell,m)=(0,3)$ and $(\ell,m)=(0,4)$ modes are, respectively, the blue square line and the brown triangle line which reduce to the $\widetilde{\ell}_S = 3$ and $\widetilde{\ell}_S = 4$ KI scalar modes when $a/r_+\to 0$.  The $m=3$ and $m=4$ scalar modes show no instability: one finds that the frequency approaches exponentially zero as the rotation grows large.}\label{Fig:scalarm3d5}
\end{figure}

\begin{figure}[ht]
\centering
\includegraphics[width=.46\textwidth]{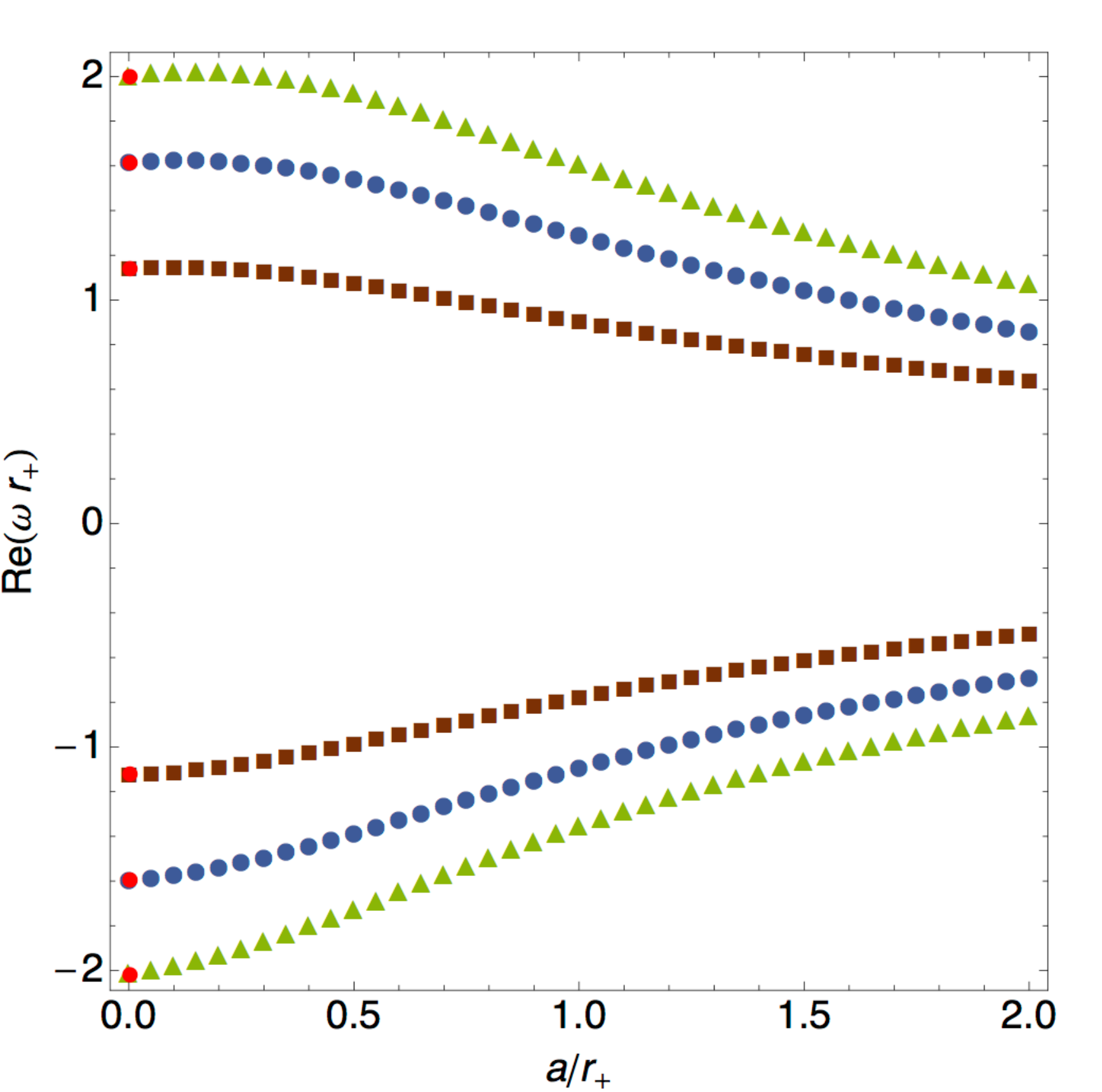}
\hspace{0.5cm}
\includegraphics[width=.485\textwidth]{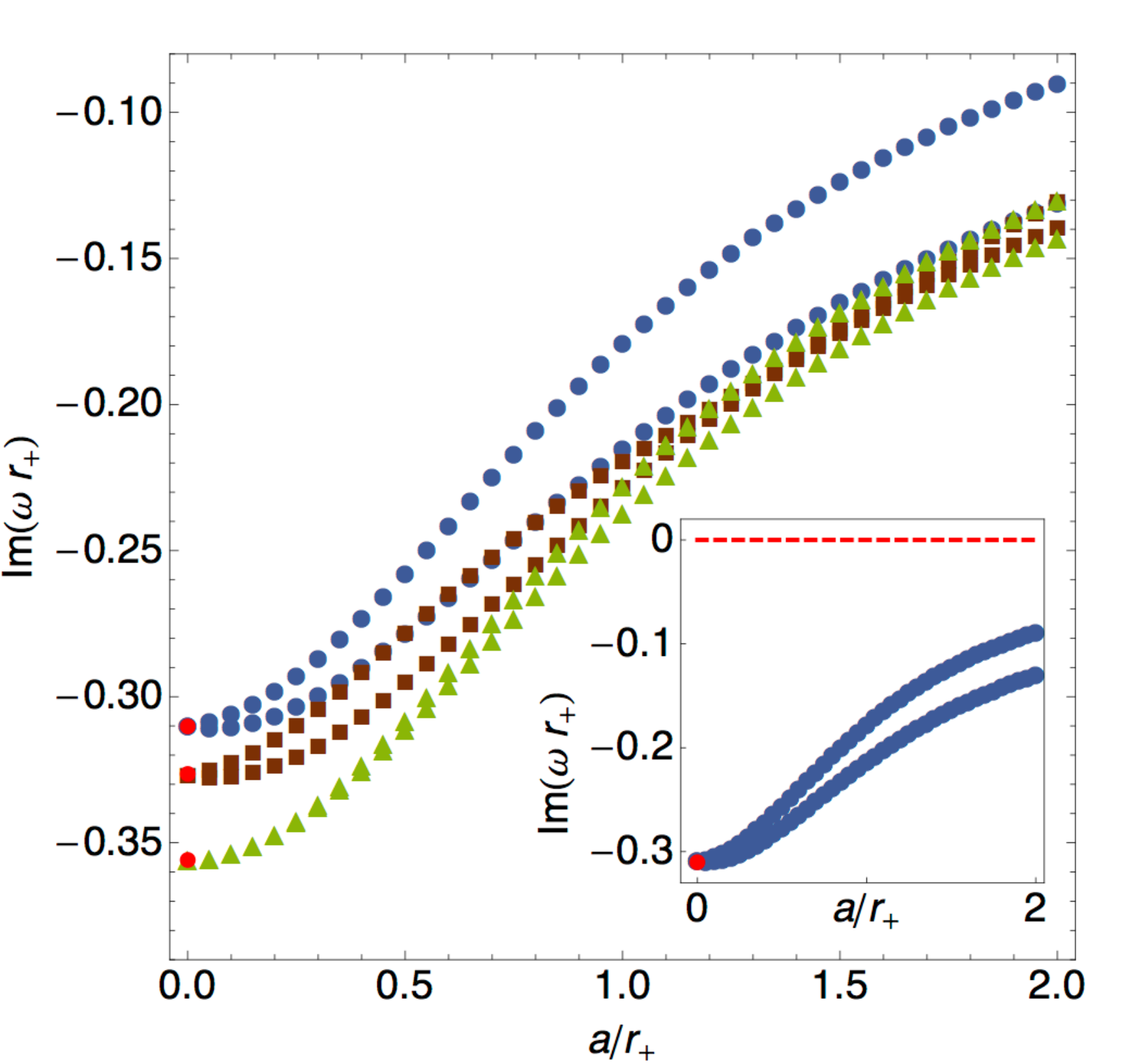}
\caption{{\bf Single MP.}  Scalar modes with $(\ell,m)=(0,1)$  for $d=5$. The blue disk line reduces to the  $\widetilde{\ell}_S=3$ KI scalar mode when $a/r_+\to 0$, while the  brown square (green triangle) line connects to the $\widetilde{\ell}_V=2$ KI vector mode ($\widetilde{\ell}_T=3$ KI tensor mode) when $a/r_+\to 0$. These Schwarzschild KI  QNMs are pinpointed as red dots. For $d\geq 6$ the behaviour of the $m=1$ QNMs has a similar behaviour. 
As highlighted in the inset plot of the {\it Right Panel} these modes are always stable, i.e. they have 
${\rm Im}(\omega r_+)<0$.}\label{Fig:scalarm1d5}
\end{figure}

Consider first the results when the rotation vanishes. The KI scalar, vector and tensor modes of the Schwarzschild BH (with lower overtone) in $d=5,6,7$ are listed in Table \ref{Table:SchwQNMs} (see footnote \ref{foot:L01}; these agree with the $d=5$ QNMs of \cite{Cardoso:2003qd}). Some of these KI modes will be represented by red dots in the plots of Figs \ref{Fig:scalarm2alld}-\ref{Fig:scalarm1d5} and they agree with zero rotation results of the single MP analysis. 

\begin{table}[ht]
\begin{eqnarray}
\nonumber
\begin{array}{||c|||c|c|||c|c|||c|c|||}\hline\hline
 d=5 & \quad \widetilde{\ell}_S \quad & \omega & \quad \widetilde{\ell}_V \quad & \omega
 &\quad \widetilde{\ell}_T \quad& \omega \\
\hline \hline
   & 2 & \pm 0.94774\, -0.25609\, i   &  2 &   \pm 1.13400 -0.32752\, i & 1 &  \pm   0.53384 -0.383375 \, i \\
 \hline
   & 3 &  \pm 1.60560-0.31096\, i   &  3 &  \pm  1.72536-0.33384 \, i & 2 &  \pm   1.51057 -0.357537 \, i \\
    \hline
   & 4 &  \pm 2.19240-0.32934\, i  &  4 &   \pm 2.28049 -0.34002 \, i &  3 &   \pm  2.00789 -0.355802\, i \\
 \hline
   & 5 &  \pm  2.74725 -0.33776\, i   &  5 &  \pm  2.81722 -0.34389\, i & 4 &    \pm 2.50629 -0.354993\, i \\
\hline\hline
\hline\hline
 d=6  & \quad \widetilde{\ell}_S \quad & \omega & \quad \widetilde{\ell}_V \quad & \omega
 &\quad \widetilde{\ell}_T \quad& \omega \\
\hline \hline
   & 2 &   \pm 1.13690 -0.303576 \, i   &  2 &   \pm 1.52466 -0.474124 \, i & 1 &   \pm  1.44651 -0.509268  \, i \\
 \hline
   & 3 &  \pm  1.92305 -0.399946 \, i   &  3 &   \pm 2.18807 -0.467877 \, i & 2 &    \pm 2.01153 -0.501938 \, i \\
    \hline
   & 4 &   \pm 2.62307 -0.437897 \, i &  4 &   \pm 2.82371 -0.472522\, i & 3 &    \pm 2.57909 -0.498873 \, i \\
 \hline
   & 5 &   \pm 3.51772 -0.461165 \, i &  5 &   \pm 3.44095 -0.477187 \, i & 4 &   \pm  3.1478 -0.497328 \, i \\
   \hline
   & 6 &   \pm 3.91091 -0.467149 \, i &  6 &   \pm 4.04645 -0.480773 \, i & 5 &    \pm 3.71715 -0.49645  \, i \\
\hline\hline
\hline\hline
 d=7  & \quad \widetilde{\ell}_S \quad & \omega & \quad \widetilde{\ell}_V \quad & \omega
 &\quad \widetilde{\ell}_T \quad& \omega \\
\hline \hline
   & 2 &  \pm  1.33916 -0.400860 \, i    &  2 &  \pm  1.93446 -0.612313 \, i   & 1 &  \pm   1.8814 -0.641077 I \, i  \\
 \hline
   & 3 &  \pm  2.23241 -0.489720 \, i    &  3 &  \pm  2.6374 0-0.594633 \, i & 2 &  \pm  2.49678 -0.631881 \, i  \\
    \hline
   & 4 &   \pm 3.01118 -0.533841\, i  &  4 &  \pm  3.32488 -0.594512 \, i  & 3 &  \pm   3.11407 -0.627613 \, i  \\
 \hline
   & 5 &   \pm 3.73988 -0.559174 \, i  &  5 &  \pm  3.99683 -0.597894 \, i  & 4 &   \pm  3.73239 -0.625325  \, i  \\
    \hline
   & 6 &   \pm 4.43954 -0.574854 \, i   &  6 &  \pm  4.65747 -0.601461 \, i  & 5 &   \pm  4.35131 -0.623967 \, i \\
\hline\hline
\end{array}
\end{eqnarray}
\caption{QNMs of the $d=5,6,7$ Schwarzschild BH with lower overtone using the KI harmonic decomposition on the $S^{d-2}$.} \label{Table:SchwQNMs}
\end{table}

Switching-on the rotation, consider first the scalar modes of the singly spinning MP BH with $\ell_S=0$ and $m=2$. There is a pair of curves (one with positive and the other with negative real part) with $(\ell_S,m)=(0,2)$ in each spacetime dimension $d$. 
Fig.~\ref{Fig:scalarm2alld} shows these pairs of curves for $d=5$ (green disk curves), $d=6$ (blue square curves)  and $d=7$ (black triangle curves); these are the pairs of curves for the zero overtone solutions. The {\it left panel} ({\it right panel}) plots the real (imaginary) part of the dimensional frequency $\omega r_+$ as a function of the dimensional rotation parameter $a/r_+$.
When the rotation vanishes these curves coincide exactly with the KI scalar mode with $\widetilde{\ell}_S=2$ of the Schwarzschild BH (see Table \ref{Table:SchwQNMs}). It is interesting to point out that, although not shown in Fig. \ref{Fig:scalarm2alld}, there are other $(\ell_S,m)=(0,2)$  scalar modes of MP $-$ with higher overtone and consequently with higher absolute value of the imaginary part of the frequency $-$ that connect to  even $\widetilde{\ell}_S=4,6,\dots$ KI scalar QNMs, to  odd  $\widetilde{\ell}_V=3,5,\dots$ KI vector QNMs, and to  even $\widetilde{\ell}_T=2,4,6,\dots$ KI tensor QNMs  when $a/r_+\to 0$. We concentrate on the lowest-lying modes that are less damped and thus dominate the time evolution of a perturbed BH. 

In Fig.~\ref{Fig:scalarm2alld} we find that there are two modes that stand out, because as the rotation increases their imaginary part goes from negative to positive. Therefore they are QNMs at low rotation but they become unstable modes after a critical rotation. These two modes are plotted in more detail in  Fig. \ref{Fig:scalarm2barmode}. They are the low-lying scalar modes  with  $(\ell_S,m)=(0,2)$ for $d=6$ (blue squares) and $d=7$ (black triangles). We identify them as the linear modes that are responsible for the  bar-mode instability that was first conjectured to exist by Emparan-Myers \cite{Emparan:2003sy} and later confirmed to be indeed present in a full non-linear time evolution analysis of a perturbed singly spinning MP BH by Shibata-Yoshino \cite{Shibata:2010wz}. Indeed, the critical rotation at which ${\rm Im}(\omega r_+)=0$ agrees precisely with the critical rotation above which the time evolution code of  Shibata-Yoshino \cite{Shibata:2010wz} finds the bar-mode instability. The values we find for these onset critical rotations 
are given in Table \ref{Table:BarMode} (to be compared with Table I of   \cite{Shibata:2010wz}).
\begin{table}[ht]
\begin{eqnarray}
\nonumber
\begin{array}{||c||c|c||}\hline\hline
 Bar-mode & \quad a_c/r_+  \quad &  a_c/r_0   \\
\hline \hline
   d=5 & \quad  \hbox{Not present} \quad & \quad  \hbox{Not present}  \quad \\
 \hline
  d=6 & \quad 0.903\pm 0.002   \quad&  \quad  0.740\pm 0.001 \quad \\
    \hline
   d=7 &  \quad 0.833\pm 0.002  \quad & \quad  0.730\pm 0.001 \quad \\
\hline\hline
\end{array}
\end{eqnarray}
\caption{Critical dimensionless rotation at which the bar-mode (linear) instability onset occurs.} \label{Table:BarMode}
\end{table}

In the inset plot of the {\it Right Panel} of Fig.~\ref{Fig:scalarm2barmode}  we zoom the bar-mode instability in $d=6$ and $d=7$ near its onset using the frequency and the rotation in mass radius units, $\omega r_0$ and $a/r_0$. Both in $d=6$ and $d=7$  we find that near the onset the unstable mode scales as ${\rm Im} \left( \omega r_0 \right)\sim C_\tau (a-a_{c})/r_0$ with  $C_\tau \sim 0.521$ where $a_c$ is the critical rotation above which the instability is present. This behaviour agrees with the linear relation predicted by the time evolution analysis of Shibata-Yoshino (see discussion associated with equation (61) of \cite{Shibata:2010wz}).

Coming back to Fig. \ref{Fig:scalarm2alld} we find that the  $(\ell_S,m)=(0,2)$  scalar mode does not become linearly unstable in $d=5$, contrary to what happens in $d\geq 6$. 
To analyse this in more detail, we have followed this  $(\ell_S,m)=(0,2)$ scalar mode up to rotations $a/r_+=4$ using a numerical code with very high precision. The associated results are plotted in Fig.~\ref{Fig:scalarm2d5}. Clearly we find no linear bar-mode instability in $d=5$.
Actually, as shown in more detail in the log-log plot of the {\it left panel} of Fig.~\ref{Fig:scalarm2d5log}, a fit of our numerical data indicates that the imaginary part of the frequency of this modes approaches exponentially zero from below but never becomes positive.
As we pointed out previously,  there are other $(\ell_S,m)=(0,2)$  scalar modes of MP (with higher radial overtone) besides the one shown in Fig.~\ref{Fig:scalarm2d5}. These modes connect to the even $\widetilde{\ell}_S=4,6,\dots$ KI scalar QNMs, to the odd  $\widetilde{\ell}_V=3,5,\dots$ KI vector QNMs, and to the even $\widetilde{\ell}_T=2,4,6,\dots$ KI tensor QNMs when $a/r_+\to 0$, all of which have lower ${\rm Im}(\omega r_+)$ than the $\widetilde{\ell}_S=2$ KI scalar QNM pinpointed as a red point in Fig.~\ref{Fig:scalarm2d5}. Nevertheless, given the negative result we get for the possibility of a $d=5$ bar-mode instability in the mode of Fig.~\ref{Fig:scalarm2d5}, we have done an exhaustive study of these other $(\ell_S,m)=(0,2)$  scalar modes of MP. We find that these modes always have, for a given rotation, ${\rm Im}(\omega r_+)$ that is more negative than the value shown in  Fig.~\ref{Fig:scalarm2d5}. (A particular example illustrating this study is shown in the {\it right panel} of Fig.~\ref{Fig:scalarm2d5log} where we look to the scalar mode that connects to the KI vector mode $\widetilde{\ell}_V=3$ when $a/r_+\to 0$). This in particular means that their ${\rm Im}(\omega r_+)$ does not become positive above a certain critical rotation and thus they are not associated to a linear bar-mode instability. 
We have also considered the possibility that a linear bar-mode instability could be associated to a scalar mode with $m>2$. However we did not find such a $m\geq 3$ mode with an imaginary part of the frequency that is positive; in fact the imaginary part of the frequency of $m\geq 3$ scalar modes is  more negative than the one of the  $m=2$ scalar mode plotted in  Fig.~\ref{Fig:scalarm2d5}.
An example that illustrates this discussion is shown in Fig.~\ref{Fig:scalarm3d5}  where the blue square curve describes the low-lying scalar mode with $m=3$ and the brown triangles describe the $m=4$ mode; these curves (the imaginary part of the frequency) are below the $m=2$ scalar mode curve (green disks).
Finally, in our data analysis we also rule out the less conventional possibility of an unstable mode that is not connected to a QNM of the Schwarzschild BH.  We conclude that the $d=5$ singly spinning MP BH is not unstable to a linear bar-mode instability.

Concluding, we find absolutely no evidence of a linear bar-mode instability in $d=5$. Note however that the non-linear numerical time evolution analysis of  \cite{Shibata:2009ad,Shibata:2010wz} does find a bar-mode instability for rotations $a_c/r_+ > 1.76$ (i.e. $a_c/r_0 > 0.87)$. 
Our linear results indicate that the $d=5$ instability appearing in the non-linear analysis of \cite{Shibata:2009ad,Shibata:2010wz} has no linear origin. 

We now proceed to the discussion of scalar modes of the singly spinning MP BH with $\ell_S=0$ and $m=1$. Again, there is a pair of curves (one with positive and the other with negative real part) with $(\ell_S,m)=(0,1)$ in each dimension. Fig.~\ref{Fig:scalarm1d5} gives detailed data for the $(\ell_S,m)=(0,1)$  scalar modes of the $d=5$ case. The blue disk line reduces to the  $\widetilde{\ell}_S=3$ KI scalar mode when $a/r_+\to 0$, while the  brown square (green triangle) line connects to the $\widetilde{\ell}_V=2$ KI vector mode ($\widetilde{\ell}_T=3$ KI tensor mode) when $a/r_+\to 0$. These Schwarzschild KI  QNMs (see Table \ref{Table:SchwQNMs}) are pinpointed as red dots. For $d\geq 6$, $(\ell_S,m)=(0,1)$ scalar QNMs behave similarly to those in $d=5$. In particular, we checked that these modes are always stable for any dimension. Not shown in Fig.~\ref{Fig:scalarm1d5}, for $d=5$ (and $d\geq 6$) there are other $m=1$ scalar modes of MP with higher absolute value of the imaginary part of the frequency that connect to the odd $\widetilde{\ell}_S=5,7,\cdots$ KI scalar QNMs, to the even  $\widetilde{\ell}_V=4,6,\cdots$ KI vector QNMs, and to the odd $\widetilde{\ell}_T=5,7,\cdots$ KI tensor QNMs. 

%%%%%%%%%%%%%%%%
\subsection{Vector QNMs ($d\geq 6$) \label{sec:singleMPvector}}
%%%%%%%%%%%%%%

We now turn to the study of vector perturbations, which are out of vector harmonics ${\bf V}_i$ 
\begin{equation}\label{SMP:vectorhab} 
   h_{ab}=0 \,, \qquad 
  h_{ai}= e^{-i\omega t}e^{i m \phi} h_a {\bf V}_i ,\qquad 
  h_{ij} = -\frac{1}{2 \sqrt{\lambda_V }}\,e^{-i\omega t}e^{i m \phi}  h_T  D_{(i}{\bf V}_{j)} \,, 
\end{equation} 
where $\{h_a, h_T\}$ are functions of $(r,\widetilde{x})$, and $ {\bf V}_i$ denotes a transverse vector harmonic on $S^{d-4}$: 
\begin{equation}
D_i  {\bf V}^i=0, \qquad \left (D^2 + \lambda_V \right)  {\bf V}_i=0
\end{equation} 
Regularity of the vector harmonics requires 
 \begin{equation}\label{SMP:vectorH2} 
\lambda=\ell_V \left(\ell_V+d-5 \right)-1\,,\qquad \hbox{with} \quad \ell_V =1,2,3.
\end{equation} 
Harmonics with $ \ell_V =1$ are special since they are Killing vectors on $S^{d-4}$.

%%%%%%%%%%%%%%%%
\subsubsection{Boundary conditions and numerical approach \label{sec:singleMPvector1}}
%%%%%%%%%%%%%%

Like in the scalar sector, we work in the traceless-transverse (TT) gauge, and we insert the vector perturbations \eqref{SMP:vectorhab}   into the linearised Einstein equations, $(\Delta_L h)_{ab}=0$.
The  TT gauge conditions can be used to express $h_T$ as an algebraic function of $h_a $ and their first derivatives.
We have then 4 independent variables, $h_1, h_2, h_3$ and $h_4$. The following equations, 
\begin{equation}\label{SMP:EOMvector}
(\Delta_L h)_{1\Omega} =0\,,\qquad (\Delta_L h)_{2\Omega} =0\,,\qquad (\Delta_L h)_{3\Omega} =0\,,\qquad(\Delta_L h)_{4\Omega} =0\,,
\end{equation}
(where here the subscript $\Omega$ describes the azimuthal coordinate of the $S^{(d-4)}$) constitute a system of four independent equations to solve for the four independent variables $h_{a}$, 
and that closes the full Lichnerowicz system.

The singular or leading behaviour of the $h_a$'s at each of the four boundaries can be determined following similar  procedures to those described in the scalar sector case. Again, it is convenient to factor them out to work with manifestly analytical functions. Therefore we introduce the new independent functions $q_1,\cdots,q_{4}$:
\begin{eqnarray}\label{SMP:vectorqs}
&& 
h_{1}  = x^{\ell+1}\left(1-x\right)^{m}  y^{-2\,i \,\frac{\omega -m \Omega_H}{4 \pi T_H}} \left(1-y^2\right)^{\frac{d-2}{2}-1} e^{\frac{i \omega }{1-y^2}}  q_1\,,\nonumber\\
&&
h_{2} =  x^{\ell+1} \left(1-x\right)^{m} y^{-2\,i \,\frac{\omega -m \Omega_H}{4 \pi T_H}-1} \left(1-y^2\right)^{\frac{d-2}{2}-3} e^{\frac{i \omega }{1-y^2}}  q_2,,
\nonumber\\
&& 
h_{3} = x^{\ell} \left(1-x\right)^{\alpha_3}  y^{-2\,i \,\frac{\omega -m \Omega_H}{4 \pi T_H}} \left(1-y^2\right)^{\frac{d-2}{2}-2} e^{\frac{i \omega }{1-y^2}} q_3  \,,\nonumber\\
&&
h_{4} = x^{\ell+1} \left(1-x\right)^{\alpha_4}  y^{-2\,i \,\frac{\omega -m \Omega_H}{4 \pi T_H}} \left(1-y^2\right)^{\frac{d-2}{2}-2} e^{\frac{i \omega }{1-y^2}} q_4 \,,
 \end{eqnarray} 
 where
 \begin{equation} \label{SMP:confBCx1auxV} 
\left\{
\begin{array}{ll}
\{ \alpha_3, \alpha_4 \}=  \{1,2 \} \,, & \qquad \hbox{if} \quad m=0 \,, \\
\{ \alpha_3, \alpha_4 \}= \{m-1,m \} \,, & \qquad \hbox{if} \quad m \geq 1\,, 
\end{array}
\right.
\end{equation} 

Inserting \eqref{SMP:vectorqs} into the equations of motion \eqref{SMP:EOMvector} and Taylor expanding to solve these equations  around each of the four boundaries it is straightforward to find the BCs we need  to impose on each of the  $q_j(x,y)$. Namely, the horizon has simple Dirichlet or Neumann BCs (depending on the $q_j$'s)  and  we also find very simple Robin, Neumann or Dirichlet BCs for all $q_j$'s at the equator $x=0$ and axis of rotation $x=1$ and horizon $y=0$. At the the asymptotic boundary, $y=1$, $q_2$ has a Dirichlet BC while the other $q_j$'s are subject to Robin BCs (the expressions for which are rather lengthy and so we omit their presentation here). 

We solve numerically our system using the same two numerical methods that were described in the scalar sector of perturbations.

%%%%%%%%%%%%%%%%
\subsubsection{Results \label{sec:singleMPvector2}}
%%%%%%%%%%%%%%

\begin{figure}[ht]
\centering
\includegraphics[width=.47\textwidth]{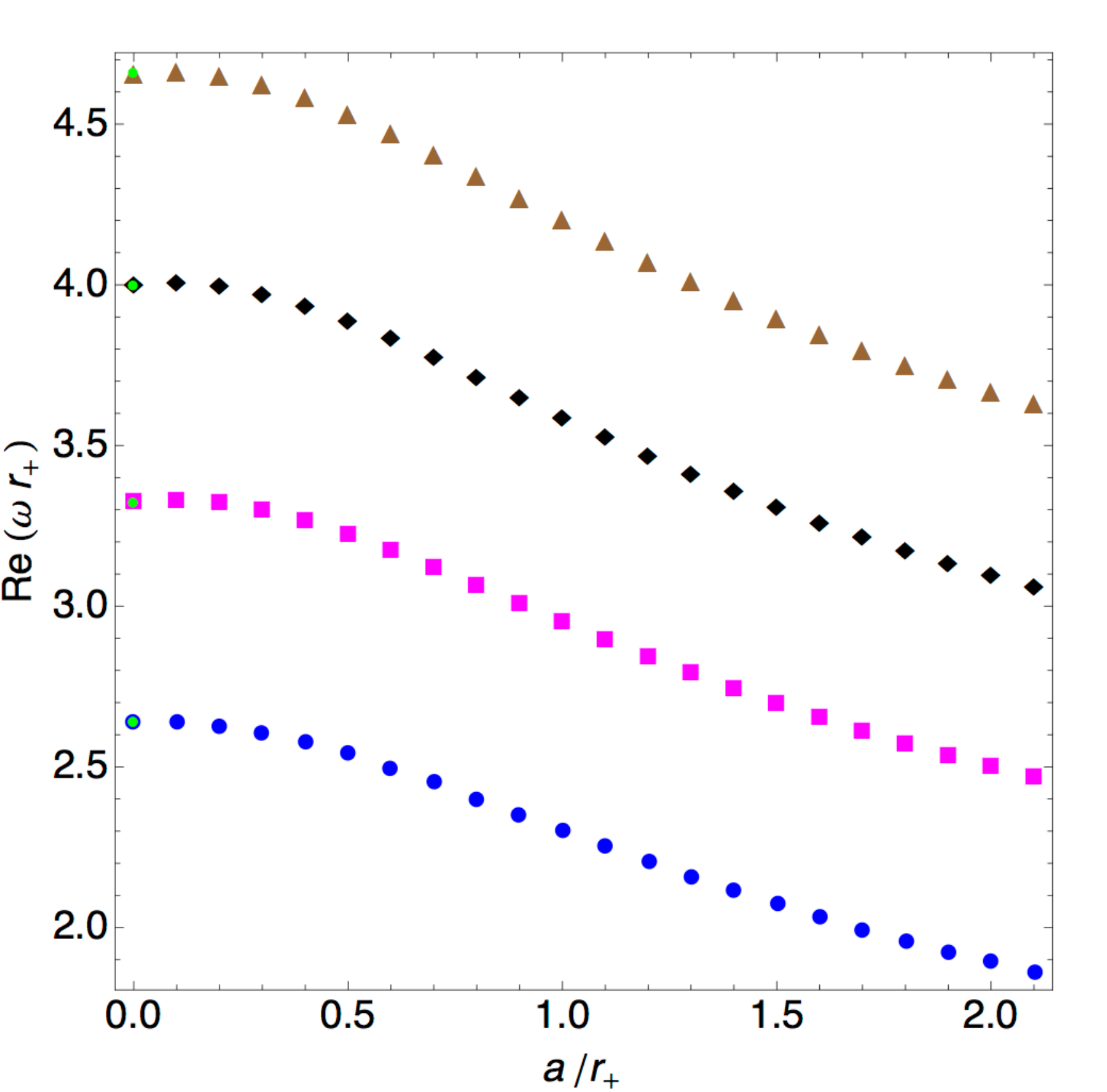}
\hspace{0.5cm}
\includegraphics[width=.47\textwidth]{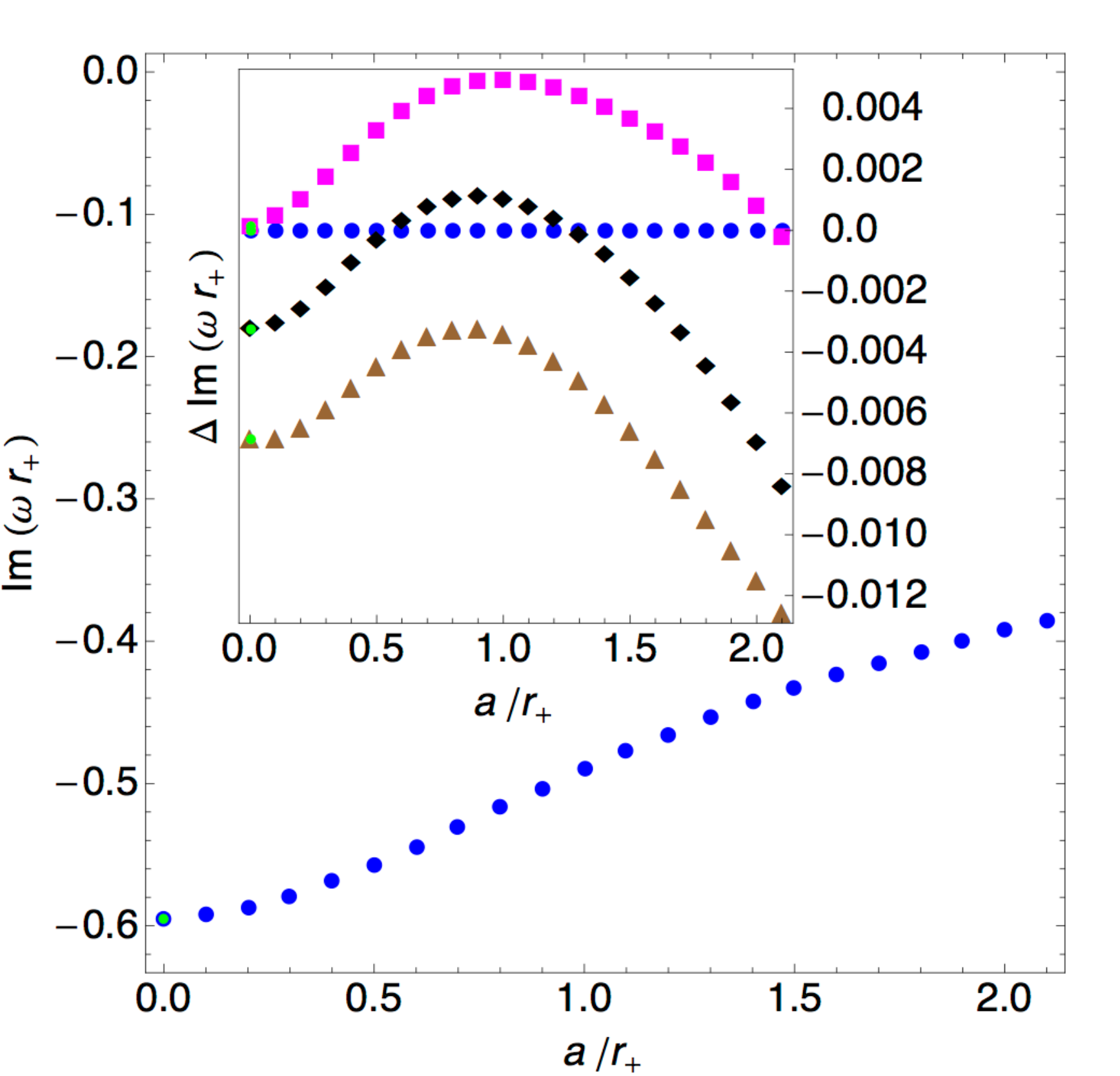}
\caption{{\bf Single MP.} Vector modes in $d=7$ with $m=1$ and: $\ell_V=2$ (blue disks), $\ell_V=3$ (magenta squares), $\ell_V=4$ (black squares) and $\ell_V=5$ (brown triangles). In the {\it Right Panel} the inset plot shows the difference between the imaginary part of the modes with respect to the $\ell_V=2, m=1$ mode. These MP vector modes with  $m=1$ and $\ell_V=2,3,4,5$  respectively connect to  $\widetilde{\ell}_V=3,4,5,6$  KI  vector QNMs when $a/r_+ \to 0$, which are pinpointed as green points.}\label{Fig:vectorm1d7}
\end{figure}  

Vector modes of the singly spinning MP BH exist for $d\geq 6$. As in the scalar case, in the vector sector there is also an infinite family $-$ associated to different radial overtones $-$ of vector QNMs for each pair $(\ell_V,m)$ of vector modes (with $\ell_V\geq 2$ and $|m| \leq \ell_V$).  We will present results only for the lowest-lying QNMs (that have smaller $|{\rm Im}(\omega r_+)|$).
When the rotation vanishes the vector QNMs of the singly spinning MP BH reduce to certain QNMs of the Schwarzschild BH  that are listed in Table \ref{Table:SchwQNMs} and that will be pinpointed as green dots in the plots of this subsection. They will confirm that our spinning QNMs are being correctly computed.   There is a pair of curves (one with positive and the other with negative real part) for each vector mode specified by the angular quantum numbers $(\ell_V,m)$. For the vector modes we will show only the element of the pair that has ${\rm Re}(\omega r_+)>0$, since for $m>0$, only these can go unstable. This is a simple consequence of the area-law argument presented in Section \ref{sec:MPEAM}. Quite often in the inset plots of the imaginary part of the frequency we will present subtracted frequencies with respect to a specified reference mode to better visualise the curves, since the curves would otherwise be very close to each other. 
 
\begin{figure}[ht]
\centering
\includegraphics[width=.47\textwidth]{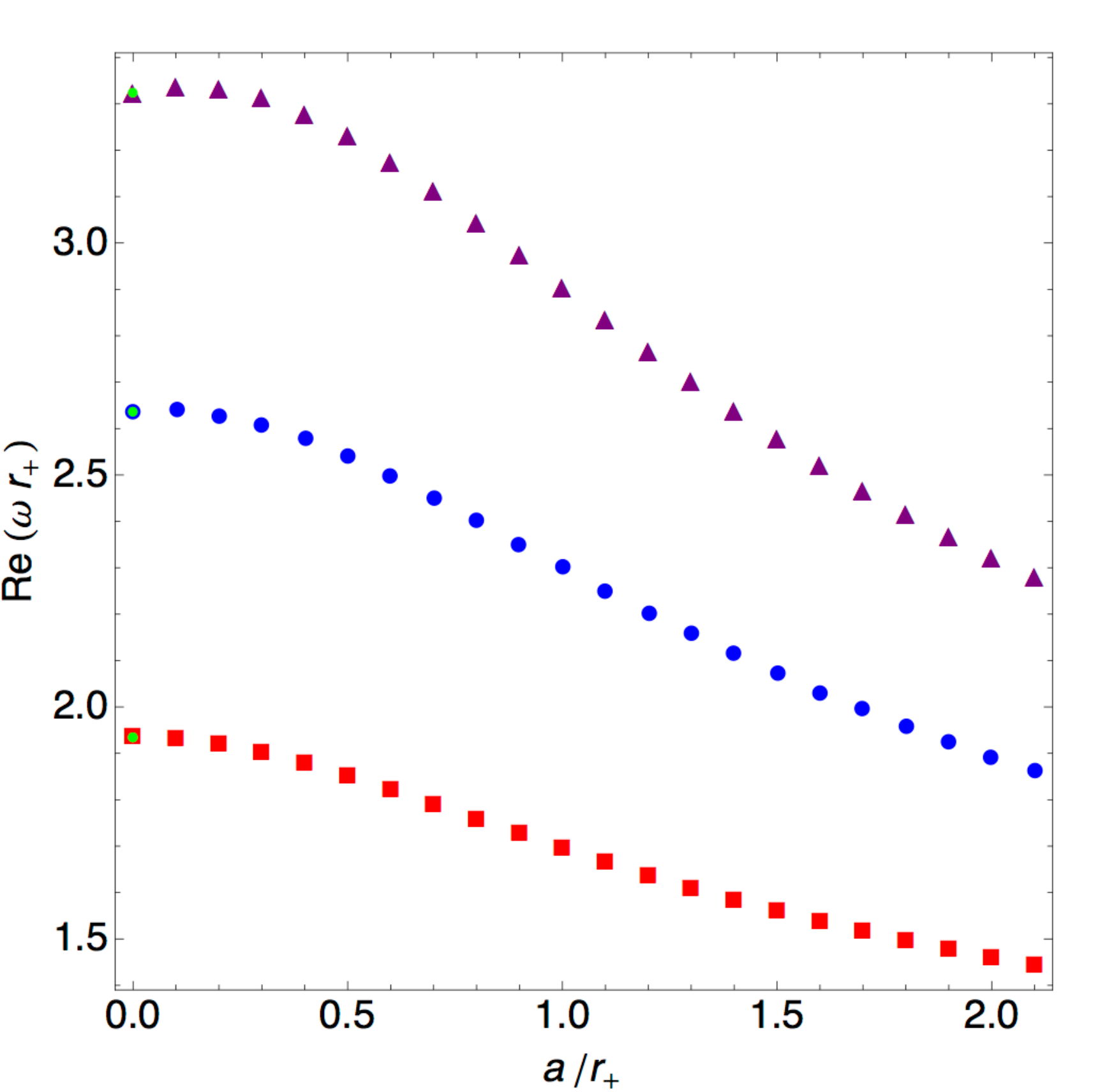}
\hspace{0.5cm}
\includegraphics[width=.47\textwidth]{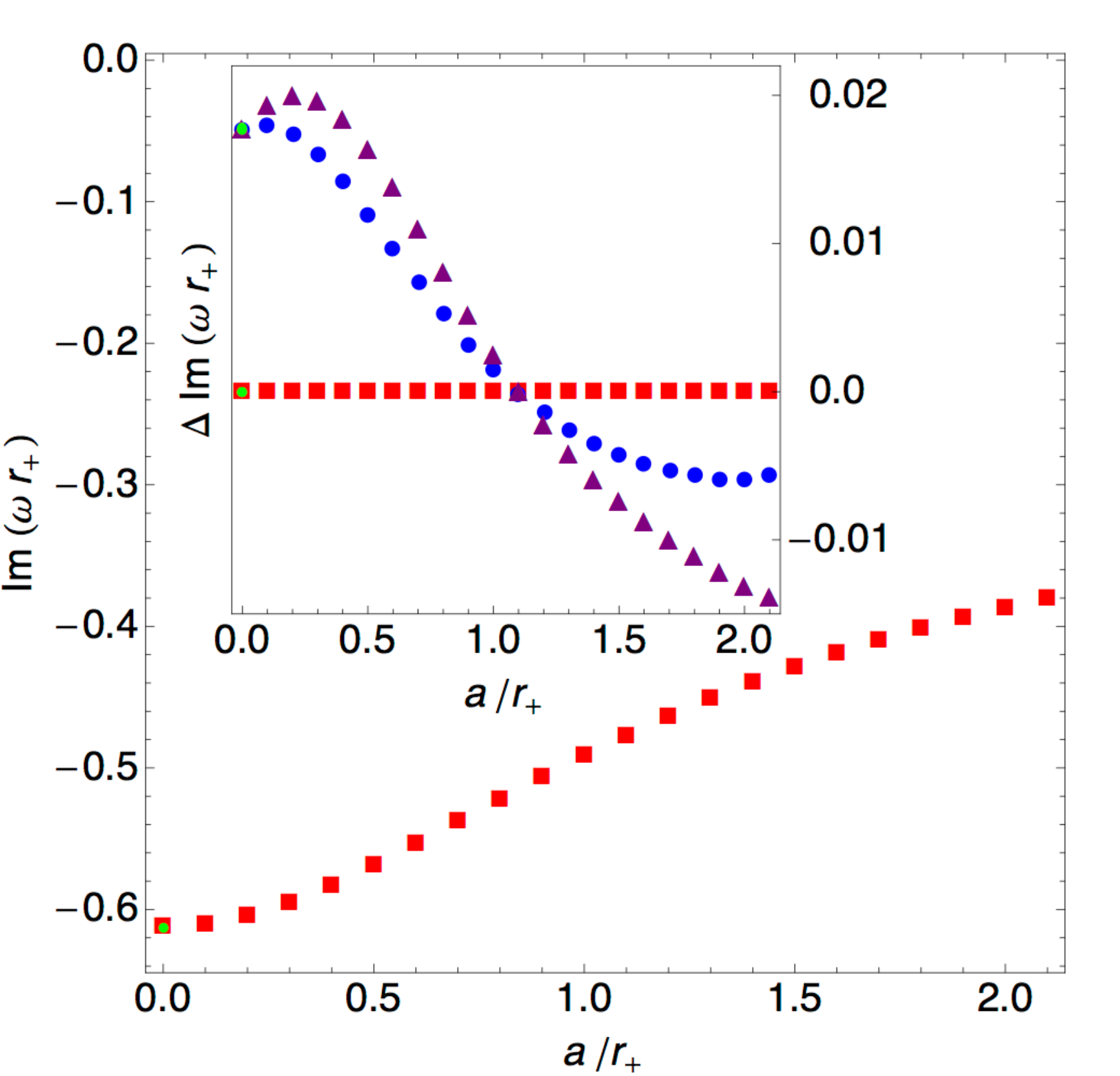}
\caption{{\bf Single MP.}  Vector modes in $d=7$ with $\ell_V=2$ and: $m=0$ (red squares), $m=1$ (blue disks), $m=2$ (purple triangles). In the {\it Right Panel} the inset plot shows the difference between the imaginary part of the modes with respect to the $\ell_V=2, m=0$ mode.
These MP vector modes with $\ell_V=2$ and $m=0,1,2$ respectively connect to  $\widetilde{\ell}_V=2,3,4$  KI  vector QNMs when $a/r_+ \to 0$, which are identified as green points.}\label{Fig:vectorl2d7}
\end{figure}  

 Fig.~\ref{Fig:vectorm1d7} shows the frequencies of the $d=7$ vector modes with fixed $m=1$ and different $\ell_V$'s, namely: $\ell_V=2$ (blue disks), $\ell_V=3$ (magenta squares), $\ell_V=4$ (black squares) and $\ell_V=5$ (brown triangles). These MP vector modes with  $m=1$ and $\ell_V=2,3,4,5$, respectively connect to  $\widetilde{\ell}_V=3,4,5,6$  KI  vector QNMs when $a/r_+ \to 0$, which are pinpointed as green points (see Table \ref{Table:SchwQNMs}).
 We find no instability in this vector spectrum. 

Fig.~\ref{Fig:vectorl2d7} fixes  $\ell_V=2$ and shows the vector modes in $d=7$ with $m=0$ (red squares), $m=1$ (blue disks), $m=2$ (purple triangles). These MP vector modes with $\ell_V=2$ and $m=0,1,2$, respectively connect to  $\widetilde{\ell}_V=2,3,4$  KI  vector QNMs when $a/r_+ \to 0$, which are identified as green points (see Table \ref{Table:SchwQNMs}). Again we find no instability in this spectrum. 

\begin{figure}[ht]
\centering
\includegraphics[width=.47\textwidth]{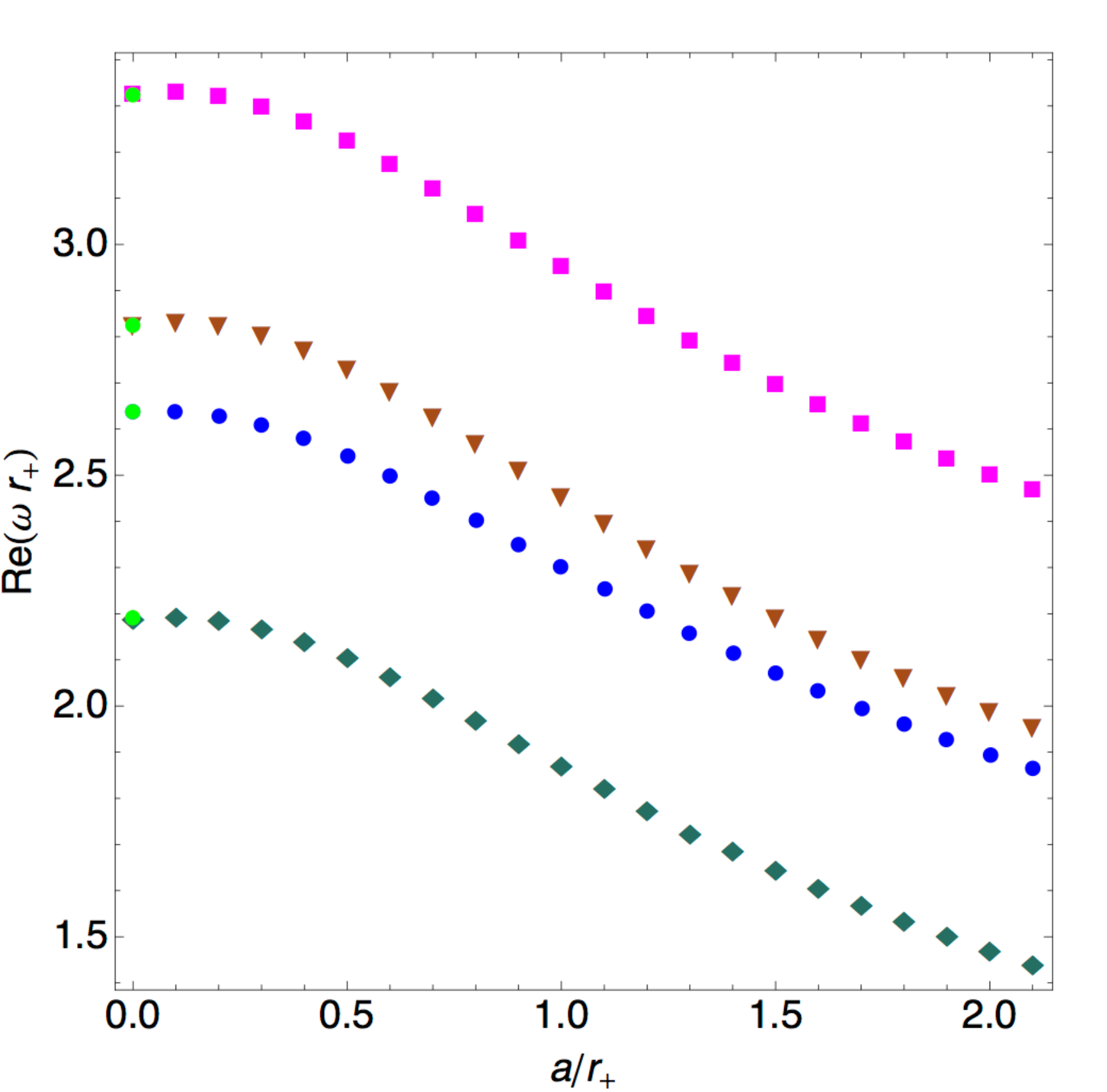}
\hspace{0.5cm}
\includegraphics[width=.47\textwidth]{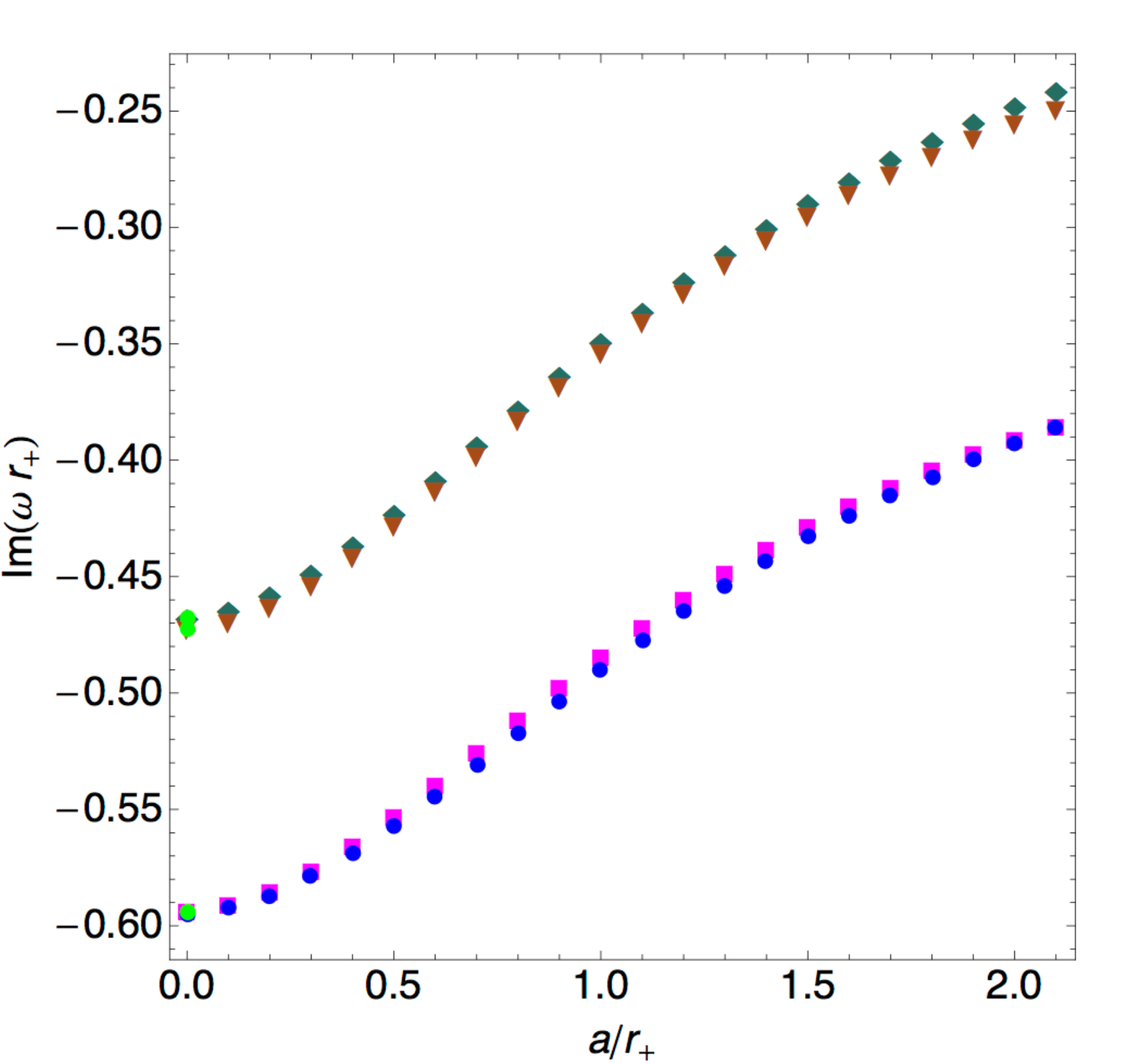}
\caption{{\bf Single MP.} Vector modes in $d=6$ and $d=7$ with $m=1$ (real part in the {\it Left Panel} and imaginary part in the {\it Right Panel}).   In $d=6$ we have the green disks with $\ell_V=2$ and the brown triangles with $\ell_V=3$. In $d=7$ we have the blue disks with $\ell_V=2$ and the magenta squares with $\ell_V=3$. These MP vector modes with  $m=1$ and $\ell_V=2,3$  respectively connect to  $\widetilde{\ell}_V=3,4$  KI  vector QNMs when $a/r_+ \to 0$, which are pinpointed as green points.}\label{Fig:vectorm1d67}
\end{figure}

Finally,  to exemplify that the behaviour of the vector modes is similar for any $d\geq 6$, in Fig.~\ref{Fig:vectorm1d67}  we show again the $(\ell_V,m)=(2,1)$ and  $(\ell_V,m)=(3,1)$ vector modes in $d=7$ but now we compare them with those with same angular quantum numbers in $d=6$.

%%%%%%%%%%%%%%%%%%%%%%%%%%%%%%
%%%%%%%%%%%%%%%%%%%%%%%%%%%%%%
\subsection{Tensor QNMs ($d\geq 7$) \label{sec:singleMPtensor}}
%%%%%%%%%%%%%%

Tensor perturbations of the singly spinning MP background can be expanded in terms of a basis of transverse ($\nabla_i {\bf T}^i_j$) and traceless (${\bf T}^i_i=0$) harmonic tensors ${\bf T}_{ij}$ on the unit sphere $S^{d-4}$ that solve $\left(\Box_{S^{d-4}} +\lambda_T \right) {\bf T}_{ij} =0$. Regular tensor harmonics have eigenvalue $\lambda_T=\ell_T \left(\ell_T+d-5 \right)-2$ with integer $\ell_T\geq 2$.

Tensor-type perturbations have the form 
\begin{equation}\label{SMP:tensorhab} 
   h_{ab}=0 \,, \qquad 
  h_{ai}= 0,\qquad 
  h_{ij} = e^{-i\omega t}e^{i m \phi}  H_T(r,\widetilde{x}) {\bf T}_{ij} \,, 
\end{equation} 
Ref. \cite{Kodama:2009bf} studied this tensor QNM spectrum in great detail so we do not discuss this sector of perturbations further. In particular, \cite{Kodama:2009bf} found no instability in the tensor sector of linear perturbations.

%%%%%%%%%%%%%%%%%%%%%%%%%%%%%%
%%%%%%%%%%%%%%%%%%%%%%%%%%%%%%
\section{Conclusions and Outlook\label{sec:Conc}}
This manuscript makes a thorough study of the QNMs of the two most representative cases of asymptotically flat MP BHs as a function of the spacetime dimension $d$, including in the limit where all rotations are taken to zero, \emph{i.e.} the Schwarzschild BH. For the latter BH, we give particular emphasis to the limit $d\to+\infty$. Also, due to the lack of symmetry of the most general MP BH, we focus on two particular classes of MP BHs: the singly spinning MP BH, and the odd-dimensional equal angular MP BH.

QNMs of the Schwarzschild BH show an interesting structure as $d$ increases. Scalars and vector gravitational perturbations have two distinct sectors of QNMs which exhibit an universal behavior. In the first, which we denominate class $I$, the corresponding QNMs saturate as the spacetime dimension increases. In the second, which we coin class $II$, the QNMs scale with $d$, with (at least some of) the modes scaling as $\mathrm{Im}(\omega r_0) \propto -d^{1/2}$ and $\mathrm{Re}(\omega r_0) \propto d$.  Furthermore, in class $I$, our results allow us to conjecture that the imaginary part is universal, depending on $\ell$ only, and saturates at $\mathrm{Im}(\omega r_0) = -(\ell-1)$, and the real part depends on the sector, being zero for vectors and non vanishing for scalars. Finally, the tensor gravitational perturbations only exhibit QNMs of class $II$. Each of these behaviors should be derivable from a matched asymptotic expansion, such as the one proposed in \cite{Emparan:2014cia,Emparan:2013xia,Emparan:2013moa}.

The structure of QNMs typical of Schwarzschild BHs, seems to persist when rotation is included. The clearest setup where this is evident, in the sense that we are able to increase $d$ to relatively large values, occurs when we study gravitational perturbations of the equal angular momenta MP BHs. For sufficiently high $d$, we have shown that both the bar-mode and ultraspinning instabilities are continuously connected, as the rotation is taken to zero, to QNMs of class $I$. This observation could have been anticipated from the following simple argument. All unstable modes of MP BHs must, at zero rotation, reduce to Schwarzschild modes. These we have found to consist of two classes, those that saturate to finite values in the infinite-$d$ limit, and those that scale with $d$. If an instability of an equal angular momenta BH connected to a class II (scaling) mode, then it would need to move a large distance in the complex $\omega$ plane as the rotation is increased a finite amount in order to turn into an instability (recall that these BHs have extremal limits). This seems unlikely, and thus it is quite natural that instabilities of these large-$d$ MP BHs should stem from saturating Schwarzschild modes. More generally, a generic MP BH with no vanishing angular momenta also has an extremal bound and is expected to suffer from instabilities similar to the equal angular momenta case, and therefore we conjecture that, for sufficiently large $d$, all instabilities of MP BHs with no vanishing angular momenta connect to  class I (saturating) Schwarzschild modes at zero rotation. Singly spinning MP BHs, for $d>5$, do not have extremal limits, and the above argument does not apply. However, for large dimension, one would at least expect that the first mode to go unstable as the rotation is increased would connect to a saturating Schwarzschild mode, since these will be the lowest lying modes.

In addition to these observations concerning perturbations in the large-$d$ limit, we have also continued the study of linear perturbations of the equal angular momenta MP BH. For the case of $d=7$, this study is now complete as we were able to make use of our derivation of the charged vector harmonics on $\mathbb{CP}^2$ to study the vector sector of perturbations. In both cases studied, the equal angular momenta and singly spinning MP BH, only the scalar sector contains QNMs that go unstable. For axisymmetric ($m=0$) modes, this corroborates the near-horizon analysis of \cite{Durkee:2010ea}.

One of the central results of this manuscript is the study of the linear analysis of the bar-mode instability of singly spinning MP BHs. For $d\geq6$ our linear analysis reproduces the key results of the non-linear time evolution of \cite{Shibata:2009ad,Shibata:2010wz}, including the slope of the growth rate as the QNMs become unstable. 
However, we find no evidence of a linear instability in the $d=5$ singly spinning MP, which seems to be at odds with the results found in \cite{Shibata:2009ad,Shibata:2010wz}. In fact, we find that the lowest lying QNMs have an imaginary part that approaches zero exponentially in the rotation parameter $a/r_+$. It would be interesting to try to understand whether this result is amenable to an analytic understanding. We leave this for future work. In $d=5$, it seems that our results are only compatible with those found in \cite{Shibata:2009ad,Shibata:2010wz}, if the bar-mode instability is a non-linear instability. This might well be the case, due to the aforementioned exponential approach.

An extension of our work would be to study the $m=0$ sector of perturbations of the singly spinning MP BH, which describes the ultraspinning instability \cite{Emparan:2003sy,Dias:2009iu,Dias:2010maa}. This would allow to compute the instability timescale, which is missing. This turns out to be a very complicated sector, due to the existence of a zero mode - namely, another MP with a slightly different angular momentum. We did not manage to use Newton's method to disentangle such perturbations, the main reason being that we don't know of a good starting seed for our iteration procedure. For the $m\neq0$, we use the QNMs of the Schwarzschild BH as a starting point. However, for $m=0$, and given what we observe in the equal angular momenta MP, we expect the ultraspinning mode to have zero real frequency. This means we should search for a QNM  of the Schwarzschild BH with zero real part of the frequency, thus lying along the purely imaginary axis, where a branch cut exists. This makes the numerics very challenging.

The work presented in this paper can be extended in many directions. We can use our PDE methods to compute the QNMs of the Kerr-Newman BH, which is a long standing problem in Classical General Relativity. Even though there is no known mechanism to herald an instability in such system, it would be interesting to test some of the conjectures put forward in \cite{Pani:2013ija}. Another extension pertains the study of graybody factors of higher-dimensional asymptotically flat BHs, which is a mild extension of our work with possible implications for LHC physics \cite{Cavaglia:2002si,Kanti:2004nr}. Another longstanding problem that could be tackled by the methods used in this paper, is the stability of the Emparan-Reall black ring found in \cite{PhysRevLett.88.101101}. Recently, using indirect methods, the fat branch of the black ring solutions has been shown to be unstable \cite{Figueras:2011he} (see also \cite{Elvang:2006dd} for a more heuristic argument, that leads to the same conclusion), but the stability of the thin ring remains largely unknown.

%%%%%%%%%%%%%%%%%%%%%%%%%%%%%%
%%%%%%%%%%%%%%%%%%%%%%%%%%%%%%
\begin{acknowledgments}
It is a pleasure to thank Roberto Emparan, Masaru Shibata, Ryotaku Suzuki, Kentaro Tanabe for enlightening discussions and for sharing their results. J.E.S.'s work is partially supported by the John Templeton Foundation. OJCD was supported in part by the ERC Starting Grant 240210 - String-QCD-BH. G.S.H was supported by NSF grant PHY12-05500. The authors thankfully acknowledge the computer resources, technical expertise and assistance provided by CENTRA/IST. Some of the computations were performed at the cluster ``Baltasar-Sete-S\'ois" and supported by the DyBHo-256667 ERC Starting Grant.
\end{acknowledgments}

%%%%%%%%%%%%%%%%%%%%%%%%%%%%%%
%%%%%%%%%%%%%%%%%%%%%%%%%%%%%%
\begin{appendix} 
\section{Spectrum of Charged Vector Harmonics on $\mathbb{CP}^2$} \label{app:vectors}
Here we derive the spectrum of charged vector harmonics on $\mathbb{CP}^2$. We will use the coordinates and conventions of Ref.~\cite{Dias:2010eu}, in which the metric on $\mathbb{CP}^N$ is constructed iteratively starting from the metric on $\mathbb{CP}^1 \simeq S^2$. Letting $d\Sigma_N^2$ and $A_N$ denote the line element of the Fubini-Study metric on $\mathbb{CP}^N$ and the $U(1)$ connection, respectively, for $N=1$ we have
\be d\Sigma_1^2 = \frac{1}{4} d\Omega_2^2 = \frac{1}{4}\left( \frac{dx^2}{1-x^2} + (1-x^2)d\phi^2 \right), \qquad A_1 = \frac{x}{2}d\phi. \ee
The line element and connection for general $N$ can then be determined iteratively,
\begin{eqnarray}
d\Sigma_N^2 &=& \frac{dR_N^2}{\left(1+R_N^2\right)^2} + \frac{1}{4} \frac{R_N^2}{\left(1+R_N^2\right)^2}\left(d\Psi_N + 2A_{N-1}\right)^2+ \frac{R_N^2}{1+R_N^2} d\Sigma_{N-1}^2 \\
A_N &=& \frac{1}{2} \frac{R_N^2}{1+R_N^2} \left(d\Psi_N + 2A_{N-1}\right). \nonumber
\end{eqnarray}
Charged vector harmonics are characterized by the following attributes: they are regular, transverse with respect to $\hat{D}_a = \hat{\nabla}_a - i m A_a$, and eigenvectors of both $\hat{D}^2$ and the complex structure. To analyse the regularity of the vectors, it is important to understand the geometry of $\mathbb{CP}^N$ near the origin. Near $R_N = 0$, the line element takes the form
\be d\Sigma_N^2 \sim dR_N^2 + R_N^2 \left[ \left( \frac{d\Psi_N}{2} + A_{N-1} \right)^2 + d\Sigma_{N-1}^2 \right], \ee
which is easily seen to be $\mathbb{R}^{2N}$ once the quantity in brackets is recognized as $S^{2N-1}$ written in terms of the Hopf fibration with the non-standard periodicity for the $U(1)$ coordinate $\Psi_N$. It is now relatively straightforward to introduce locally Cartesian coordinates. The form of the metric near the origin suggests that these coordinates should correspond to the embedding of $\mathbb{CP}^{N-1}$ in $\mathbb{C}^{N}$. For example, for $N=2$ this is achieved by
\be z_1 = R_2 \sqrt{\frac{1-x}{2}} e^{\frac{i}{2}(\phi-\Psi_2)}, \qquad z_2 = R_2 \sqrt{\frac{1+x}{2}} e^{\frac{i}{2}(\phi+\Psi_2)}. \ee
We are now in a position to derive the charged vector harmonics for the case $N=2$. Consider the following vector,
\be \mathbb{V}^{(0)}_a dx^a = Y_2(R_2) Y_1(x) A_1 e^{i n_2 \Psi_2}. \ee
This can be decomposed into hermitian and anti-hermitian parts,
\be \mathbb{V}_a = \frac{1}{2} \left( \delta_a^{\hspace{3pt} b} + i \epsilon J_a^{\hspace{3pt} b} \right) \mathbb{V}^{(0)}_b, \qquad \epsilon = \pm 1. \ee
Requiring that this vector be transverse yields 
\be Y_1(x) = x^{-1} \left(1-x^2\right)^{-n_2/(2\epsilon)}. \ee
The function $Y_2(R_2)$ and the integer $n_2$ are  determined by requiring that $\mathbb{V}$ be regular. Using the above locally Cartesian coordinates, one can show that the near $R_2 = 0$ geometry reveals that $\mathbb{V}$ displays two types of pathological behaviour. First, even at finite $R_2$, it diverges  along certain 2- and 1-dimensional subspaces, and secondly, it diverges as $R_2^{-2}$. The first problem is resolved by the choice $n_2 = -\epsilon$, and the second will be fixed by the quantization of $\lambda_V$, to which we now turn. 

For the above vector, the eigenvalue equation $(\hat{D}^2 + \lambda_V)\mathbb{V}_a = 0$ reduces to a single ODE,
\be 
Y_2''(R_2) + \frac{Y'(R_2)}{R_2}  - \frac{4 + R_2^2 \left(2 + 2m\epsilon -\lambda_V\right) + R_2^4 (2+m\epsilon)^2 }{R_2^2 (1+R_2^2)^2} .
\ee
It is convenient to introduce the coordinate $z^{-1} = 1+R_N^2$ and to define a new function, $Y_2(z) = (1-z) z^{|2+m \epsilon|/2} w(z)$, which brings the equation into hypergeometric form,
\be z(1-z)w''(z) + \big(c-(a+b-1)z \big) w'(z) - a b w(z) = 0.
\ee
Here $c= 1 + |2 + m \epsilon|$, and $a+b = 3 + |2 + m \epsilon|$. Requiring that the solution is regular will fix the solution and impose a quantization condition on the eigenvalues. The general solution can be written as a sum of hypergeometric functions with argument $z$. Since $c$ is an integer, one of the two independent solutions contains factor of $\ln(z)$, which is not analytic and therefore the coefficient of this solution is set to zero. The solution is then
\be
Y_2(z) = (1-z)z^{|2 + m \epsilon|/2} {}_2F_1(a,b;c;z).
\ee
This is regular at $z=0$ $(R_N = \infty)$. To analyse the regularity near $z=1$, use the identity \cite{abramowitz1964handbook}
\begin{eqnarray} {}_2F_1(a,b;c;z) &  =& \frac{\Gamma(c)\Gamma(c-a-b)}{\Gamma(c-a) \Gamma(c-b)}  {}_2F_1(a,b;a+b-c+1;1-z) \\
&+& \frac{\Gamma(c)\Gamma(a+b-c)}{\Gamma(a)\Gamma(b)} (1-z)^{c-a-b}{}_2F_1(c-a,c-b;c-a-b+1;1-z). \nonumber
\end{eqnarray}
Regularity requires that the coefficient of the second term vanish, since $c-a-b = -2$, so that  both the $(1-z)$ factor is diverging and the hypergeometric function is ill-defined, since the third argument is a non-positive integer. If $b = -\kappa$, where $\kappa = 0, 1, 2,...$, then the second term will vanish and the solution will be manifestly regular at both $z=0,1$. This leads to 
\be \lambda_V^{(N=2)} = 4\kappa \left( \kappa + 3 + |2+m\epsilon| \right) + 6|2+m\epsilon|+6+2m\epsilon.
\ee
For $m = 0$ and $N=2$, this agrees with the result derived in \cite{Durkee:2010ea}.

We have only succeeded in deriving the spectrum of charged vector harmonics for the case $N=2$. The main obstacle to explicitly constructing the harmonics for larger $N$ is finding a tensorial structure that is regular near the origin and satisfies the other criteria. It would be useful if this result could be further generalized to arbitrary $N$.

\end{appendix}

%%%%%%%%%%%%%%%%%%%%%%%%%%%%%%
%%%%%%%%%%%%%%%%%%%%%%%%%%%%%%
\bibliography{refs}{}
\bibliographystyle{JHEP}

\end{document}